\documentclass[a4paper,fleqn,usenatbib]{mnras}

\usepackage{newtxtext,newtxmath}

\usepackage[T1]{fontenc}
\usepackage{ae,aecompl}

\usepackage{graphicx}	
\usepackage{amsmath}	
\usepackage{amssymb}	
\usepackage{multirow}   
\usepackage{subfig}     
\usepackage{bigdelim}   
\usepackage{xspace}     
\usepackage{array}		
 \usepackage{threeparttable}
\newcolumntype{c}[1]{>{\centering\arraybackslash}p{#1}} 

\usepackage{times}
\usepackage[super]{nth} 

\usepackage{xcolor}
\usepackage{wasysym}

\usepackage{enumitem}

\defcitealias{KL14}{KL14}

\newcommand{\myr}{\mbox{${\rm Myr}$}}

\newcommand{\pc}{\mbox{${\rm pc}$}}

\newcommand{\tgas}{\mbox{$t_{\rm gas}$}}

\newcommand{\tstar}{\mbox{$t_{\rm star}$}}
\newcommand{\tover}{\mbox{$t_{\rm over}$}}

\newcommand{\fcl}{\mbox{$f_{\rm cl}$}}
\newcommand{\fgmc}{\mbox{$f_{\rm GMC}$}}

\newcommand{\zetastar}{\mbox{$\zeta_{\rm star}$}}
\newcommand{\zetagas}{\mbox{$\zeta_{\rm gas}$}}

\newcommand{\halpha}{\mbox{${\rm H}\alpha$}\xspace}

\newcommand{\code}{{\sc Heisenberg}\xspace} 

\newcommand{\be}{\begin{equation}}
\newcommand{\ee}{\end{equation}}
\newcommand{\bea}{\begin{eqnarray}}
\newcommand{\eea}{\end{eqnarray}}

\newcommand{\dcrit}{\mbox{$D_{\rm crit}$}}
\newcommand{\dcritlow}{\mbox{$D_{\rm crit, low}$}}
\newcommand{\npin}[1]{\mbox{$N_{\rm peak,#1,input}$}}
\newcommand{\nlambda}{\mbox{$n_{\lambda}$}}
\newcommand{\qcon}{\mbox{$q_{\rm con}$}}
\newcommand{\qconstar}{\mbox{$q_{\rm con, star}$}}
\newcommand{\qcongas}{\mbox{$q_{\rm con, gas}$}}
\newcommand{\qconinfty}{\mbox{$q_{\rm con, \infty}$}}

\newcommand{\fwhm}{\mbox{$G_{\rm FWHM}$}}
\newcommand{\fsignal}{\mbox{$f_{\rm compact}$}}
\newcommand{\fsignalmeasured}{\mbox{$f_{\rm compact,measured}$}}
\newcommand{\fsignaltrue}{\mbox{$f_{\rm compact,true}$}}
\newcommand{\qeta}{\mbox{$q_{\eta}$}}
\newcommand{\qetastar}{\mbox{$q_{\eta, \rm star}$}}
\newcommand{\qetagas}{\mbox{$q_{\eta, \rm gas}$}}

\newcommand{\htwo}{\mbox{H{\sc ii}}}
\newcommand{\hone}{\mbox{H{\sc{i}}}}

\newcommand{\etastar}{\mbox{$\eta_{\rm stars}$}}
\newcommand{\etagas}{\mbox{$\eta_{\rm gas}$}}

\newcommand{\rpeak}{\mbox{$r_{\rm peak}$}}

\newcommand{\npix}[1]{\mbox{$N_{{\rm pix,}#1}$}}

\newcommand{\lpix}{\mbox{$l_{\rm pix}$}}

\newcommand{\phimax}{\mbox{$\phi_{\rm max}$}}
\newcommand{\phimin}{\mbox{$\phi_{\rm min}$}}

\newcommand{\Psiuv}{\mbox{$\Psi(u,v)$}}
\newcommand{\Duv}{\mbox{$D(u,v)$}}

\newcommand{\etaname}[1]{evolutionary phase lifetime adjusted #1filling factor}

\title[On the nature and filtering of diffuse emission]{An uncertainty principle for star formation -- IV. On the nature and filtering of diffuse emission}

\author[A. P. S. Hygate et al.]{Alexander P. S. Hygate,$^{1,2}$\thanks{E-mail: hygate@mpia.de (APSH)}
J.~M.~Diederik Kruijssen,$^{2,1}$
M\'{e}lanie Chevance,$^{2}$
\newauthor
Andreas Schruba,$^{3}$
Daniel T.~Haydon$^{2}$ and
Steven N.~Longmore$^{4}$
\\
$^{1}$Max Planck Institute f\"{u}r Astronomie, K\"{o}nigstuhl 17, 69117, Heidelberg, Germany\\
$^{2}$Astronomisches Rechen-Institut, Zentrum f\"{u}r Astronomie der Universit\"{a}t Heidelberg, M\"{o}nchhofstra\ss e 12-14, 69120 Heidelberg, Germany \\
$^3$Max-Planck Institut f\"{u}r Extraterrestrische Physik, Giessenbachstra\ss e 1, 85748 Garching, Germany \\
$^4$Astrophysics Research Institute, Liverpool John Moores University, IC2, Liverpool Science Park, 146 Brownlow Hill, Liverpool L3 5RF, UK}

\date{Accepted XXX. Received YYY; in original form ZZZ}

\pubyear{2018}

\begin{document}
\label{firstpage}
\pagerange{\pageref{firstpage}--\pageref{lastpage}}
\maketitle

\begin{abstract}
Diffuse emission is observed in galaxies in many tracers across the electromagnetic spectrum, including tracers of star formation, such as H$\alpha$ and ultraviolet (UV), and tracers of gas mass, such as carbon monoxide (CO) transition lines and the 21-cm line of atomic hydrogen (\hone{}). Its treatment is key to extracting meaningful information from observations such as cloud-scale star formation rates. Finally, studying diffuse emission can reveal information about the physical processes taking place in the ISM, such as chemical transitions and the nature of stellar feedback (through the photon escape fraction). We present a physically-motivated method for decomposing astronomical images containing both diffuse emission and compact regions of interest, such as \htwo{} regions or molecular clouds, into diffuse and compact component images through filtering in Fourier space. We have previously presented a statistical method for constraining the evolutionary timeline of star formation and mean separation length between compact star forming regions with galaxy-scale observations. We demonstrate how these measurements are biased by the presence of diffuse emission in tracer maps and that by using the mean separation length as a critical length scale to separate diffuse emission from compact emission, we are able to filter out this diffuse emission, thus removing its biasing effect. Furthermore, this method provides, without the need for interferometry or ancillary spectral data, a measurement of the diffuse emission fraction in input tracer maps and decomposed diffuse and compact emission maps for further analysis.

\end{abstract}



\section{Introduction}

One of the key challenges in observational astronomy is separating populations of interest from other contaminant populations present in data, due to the fact that observations do not take place in controlled laboratory settings. It has been known for a number of decades that there is a significant diffuse component in \halpha observations of ionised gas (known as diffuse ionised gas or DIG) in our own Galaxy (see e.g. \citealt{Reynolds1973}). Detection of diffuse \halpha emission in the discs \citep{Monnet1971} of external galaxies and their haloes (\citealt{Rand1990}; \citealt{Dettmar1990}) confirmed the phenomenon is not restricted to the Milky Way and, as technology has advanced, its ubiquity and significance in other galaxies has been confirmed (e.g. \citealt{Lacerda2018}). The physical origin of the diffuse \halpha{} emission is still debated, but it has been shown that photoionization of the warm neutral medium (WNM) by Lyman continuum leaking out of bright \htwo{} regions \citep[e.g.][]{Mathis1986,Sembach2000,Wood2010} or dust scattering \citep{Seon2012} can provide an important contribution to this diffuse component. The diffuse \halpha{} emission contaminates samples of compact H{\sc ii} regions, where star formation takes place (e.g. \citealt{Liu2011}), but also carries unique astrophysical information. A physically-motivated separation between these components is challenging, but necessary.

Studies of the Warm Ionised Medium with \halpha indicate significant variation in the \halpha diffuse fraction (see e.g. \citealt{Thilker2002}; \citealt{Oey2007}; \citealt{Lacerda2018}) between different galaxies. For example, \citet{Oey2007} find diffuse \halpha fractions of between 12\% and 100\%, by comparing the total \halpha flux of a sample of 109 galaxies to the flux of their H{\sc ii} regions as identified by the automated H{\sc ii} region photometry package {\sc HIIPHOT} \citep{Thilker2000}. In addition, \citet{Lacerda2018} measure distributions of the \halpha diffuse fraction that shift with Hubble type, observing that the contribution of the DIG is generally smaller in later-type galaxies. Within galaxies, the diffuse \halpha fraction is observed to vary between different environments. For example, \citet{Blanc2009} observe higher \halpha diffuse fractions in fainter \htwo{} regions in M51 and \citet{Kreckel2016} find higher \halpha{} diffuse fractions in spiral-arm \htwo~regions than in inter-arm \htwo~regions.

Due to the high optical depth of the Earth's atmosphere in the UV part of the spectrum, it was not until the advent of space telescopes that studies of diffuse UV could be made. Studies of the UV diffuse fraction indicate variation, as with \halpha. For example, \citet{Hoopes2001} find that the diffuse emission not associated to \halpha-selected H{\sc ii} regions contributes 72 - 91\% of the total FUV flux of 10 spiral galaxies. This could reflect the difference in the lifetimes of the \halpha{} and UV-bright phases of young stellar populations, with the UV-bright phase lifetime being roughly a factor of 5 longer \citep{haydon18}. Moreover, comparisons indicate that there are differences between diffuse fractions in different phases and within galaxies themselves. For example, \citet{Thilker2005} find that the ratio of FUV diffuse fraction to \halpha diffuse fraction changes radially. Consequently, the question arises as to whether these variations can be linked to galaxy properties or the physical processes driving the star formation process.

Furthermore, diffuse emission has also been found across a wide range of tracers used as direct probes of star formation rate or in composite star formation prescriptions, including infrared (see for e.g. in 8 $\umu m$ \citealt{Crocker2013}); diffuse x-ray in starbursts \citep{Fabbiano1990} and normal galaxies \citep{Bregman1994} (see also \citealt{Strickland2004} for a larger sample of galaxies with diffuse x-ray emission); and [\ion{C}{ii}] \citep{Kapala2015}, which is also used as a tracer of molecular gas mass.

Taking account of the presence of diffuse emission is important for comparisons between integrated galaxy-scale measurements and small scale measurements of star formation rate and for properly interpreting the source of observed ionising radiation (see e.g \citealt{Blanc2009} and \citealt{Leroy2012}). Diffuse emission also impacts other measurements such as line-ratios and therefore the interpretation of diagnostic diagrams (such as the BPT diagram, see \citealt{Baldwin1981}) and measurements of metallicity  \citep{Zhang2017}. Thus correctly dealing with diffuse emission is key to proper interpretations of observations.

Gas tracers may also include diffuse emission. Comparison of the $^{12}$CO to $^{13}$CO ratio using single dish and interferometric observations of regions in M33 by \citet{WilsonWalker1994} indicated that diffuse molecular clouds contribute up to 60\% of the $^{12}$CO emission. More recently, comparisons between interferometric and single-dish CO line observations across the discs of galaxies have revealed that observed molecular gas consists of both a compact clumpy component (molecular clouds) and a component spread-out over large scales (diffuse emission) (\citealt{CalduPrimo2013}; \citealt{CalduPrimo2015}). Furthermore, this diffuse component can be a significant fraction of the emission, with for example $\sim$50\% of observed CO(2-1) emission in M51 originating from scales in excess of 1.3 kpc \citep{Pety2013}. The nature of diffuse CO emission is unknown. It may consist of unresolved, low-mass clouds, or be truly diffuse in nature.

A number of different approaches have been taken to separate diffuse emission from compact emission. Examples of definitions are: the flux lost in an interferometric map compared to the flux from a single dish measurement (e.g. \citealt{Pety2013}), the flux of identified regions of interest versus the total flux in an image (e.g. application of automated H{\sc ii} region photometry package {\sc HIIPHOT} \citealt{Thilker2000}, by \citealt{Thilker2002}), use of additional spectral information available with integral field unit (IFU) datacubes to divide pixels into diffuse and non-diffuse pixels (e.g~an \halpha equivalent width criterion employed by \citealt{Lacerda2018}), and the use of diagnostic line ratios to separate the contribution of diffuse emission (e.g. [S{\sc ii}]/\halpha ratio to separate DIG from H{\sc ii} region emission by \citealt{Blanc2009}). Very few of these methods are physically motivated and they all depend on properties of the observations. This, in part, explains why current measurements in the literature exhibit significant differences and that the role of this diffuse emission for the star formation process remains unclear. A general, physically motivated approach to separate diffuse emission from compact emission is therefore desirable in order to make progress on better understanding the nature and origin of diffuse emission.

Finally, we have recently developed a statistical method for constraining the evolutionary timeline of star formation and mean separation length between compact star-forming regions with galaxy-scale observations of (molecular) gas and young stellar emission \citep{KL14,kruijssen18}. This method (`an uncertainty principle for star formation', hereafter \citetalias{KL14} principle) has been prepared for observational applications in the form of the {\sc idl} code \code \citep{kruijssen18}. The method fundamentally assumes that all emission contained in the gas and young stellar maps belongs to individual regions evolving on an underlying evolutionary timeline. However, in practice, the maps generally contain both the regions of interest (e.g. molecular clouds or \htwo{} regions) and diffuse emission that does not emanate from these regions. This requires the diffuse component of each map to be isolated and separated out. Because the \code code is suitable for applications across a wide range of tracers, including \hone{}, CO, H$\alpha$ and UV, we require an objective method that is not instrument dependent and can be applied uniformly to different tracers.

This paper presents a method for the separation of diffuse emission from tracer maps through filtering in Fourier space implemented into the \code code.
The structure of the paper is as follows. We outline the \citetalias{KL14} principle in Section~\ref{sec:heisenberg}. We then describe our new method of diffuse emission filtering in Section~\ref{sec:method} and its implementation into the \code code in Section~\ref{sec:implementation}. Following this, we detail the steps taken to generate simulated data, on which we test the method in Section~\ref{sec:test_images}. We show the results of this testing in Section~\ref{sec:signal_loss}, focusing on how the compact regions of interest can be affected by the filtering of large-scale emission, and in Section~\ref{sec:testing}, focusing on how the quantities measured by {\sc Heisenberg} are improved by our filtering method and how well it retrieves the fraction of emission within an image that is diffuse. Finally, we present a summary of our results and conclusions in Section~\ref{sec:conclusions}.

\section{Uncertainty principle for star formation}
\label{sec:heisenberg}

This section provides a brief summary of the `uncertainty principle' for star formation within which the method in this paper is implemented. For a full description, the reader is referred to \citet{KL14} for the theoretical basis, \citet{kruijssen18} for the complete method and its realisation in the form of the {\sc Heisenberg} code, and \citet{haydon18}  for the calibration of the obtained evolutionary timelines through the use of characteristic `reference' time-scales of star formation rate tracers such as H$\alpha$ or UV emission.

In short, the \citetalias{KL14} principle describes how variations in the spatially resolved flux ratio between two tracers of successive phases of a given evolutionary sequence are controlled by the relative lifetimes of emission peaks in these tracers, together with the mean separation length between these regions. The {\sc Heisenberg} code applies the \citetalias{KL14} principle such that it can measure the evolutionary timeline describing different phases of the star formation process from observed emission maps tracing these phases. It places circular apertures of different sizes on emission peaks in both maps, calculates the enclosed flux ratio difference relative to the average across the entire field of view, and fits a statistical model to this relative flux ratio difference as a function of the aperture size to measure the peak lifetimes and separation length.

An example of such a timeline is as follows: CO traced gas clouds live for \tgas{} and evolve into young stars which are visible in \halpha for \tstar{} and both co-exist for \tover{}. In this case, we measure $\lambda$, $\tgas/\tstar$ and $\tover/\tstar$. Given a priori knowledge of \tstar, based on stellar population synthesis models by \citet{haydon18} for \halpha and UV emission, we can convert these relative lifetimes to the absolute timescales \tgas{} and \tover{}.

In this paper, we will investigate how these three key parameters, $\lambda$, $\tgas$ and $\tover$ are impacted by the presence of additional flux from diffuse emission present in tracer maps and present a modification of the \code code that removes the diffuse emission with Fourier filtering. By using (a multiple of) the region separation length, $\lambda$, as the critical wavelength for defining the separation between diffuse and compact emission, this method presents a physically-motivated way of decomposing the tracer maps into diffuse and compact emission images, thus allowing the fractions of diffuse and compact emission in these images to be measured.

\section{Method for decomposing emission maps into diffuse and compact components}
\label{sec:method}

We exploit the differing spatial distribution between regions of interest (for example molecular clouds), which are relatively small and concentrated, and the diffuse emission, which spans large scales. While these two emission components are co-spatial in a map, in `real space', they are more easily distinguished in Fourier space where their emission is primarily located in different Fourier frequency regions. We therefore use filtering in Fourier space to separate these two types of emission.

For a given two-dimensional astronomical image $f(m,n)$ of size $\npix{x} \times \npix{y}$, we transform the image into Fourier space using a discrete Fourier transform:

\begin{equation}
\label{eq:Fourier_transform}
F(u,v) = \frac{1}{\npix{x}\npix{y}} \sum_{m,n} f(m,n) e^{-i2\pi (um + vm) } ,
\end{equation}

\noindent
where $\npix{x}$ and $\npix{y}$ are the number of pixels in the x and y dimensions, respectively. The Fourier frequencies ($u,v$) are defined as:

\begin{equation}
u =  \frac{ p -  \left\lfloor\frac{\npix{x} -1 }{2}\right\rfloor}{\npix{x}}	 \\
\end{equation}

\begin{equation}
v =  \frac{ q -  \left\lfloor\frac{\npix{y} -1 }{2}\right\rfloor}{\npix{y}}	 \\
\end{equation}

\noindent
with $p \in \mathbb{N}  \ | \ 0 \leqslant p \leqslant \npix{x} -1 $ and $ q \in \mathbb{N}  \ | \ 0 \leqslant q \leqslant \npix{y} -1 $. We note that $\left\lfloor x \right\rfloor$ represents the floor function of $x$, i.e. the value obtained when rounding $x$ down to nearest integer. These Fourier frequencies define the frequency axes of the Fourier space into which we transform the image. For these frequencies, all values are such that $ -0.5 < u,v \leqslant 0.5 $. Each frequency axis also has a value at zero, with the $u = 0$, $v = 0$ component, known as the `DC' component, representing the mean flux of the image. Within Fourier space, the emission characterised by long-wavelength Fourier modes, i.e. that from extended large-scale emission, is concentrated in the low frequency part of space, whereas the emission characterised by short-wavelength Fourier modes, i.e. that from compact regions in real space, is concentrated in the high frequency part of space.

The transformed image $F(u,v)$ is then multiplied with a mask, \Psiuv{}, with values between 0 and 1 (corresponding to 0\% and 100\% transmission respectively), defined by one of the filters defined in Section~\ref{sec:Filters} and transformed back into the image domain with an inverse Fourier transform:

\begin{equation}
\label{eq:Fourier_filter}
 f'(m,n) = \Omega(m,n) \sum_{u,v}  \Psiuv F(u,v)  e^{i2\pi (um + vm)} ,\\
\end{equation}

\noindent
where $\Omega(m,n)$ is a post-processing mask applied to the result of the inverse Fourier transform (see Section~\ref{sec:post_processing}) and $f'(m,n)$ is the filtered image with diffuse emission removed or reduced. An illustration of this process can be seen in Figure~\ref{fig:fourier_space_illustration}.

\newcommand{\procwidth}{0.49\columnwidth}
\begin{figure}
	\centering
	\subfloat[original image]{\includegraphics[width=\procwidth]{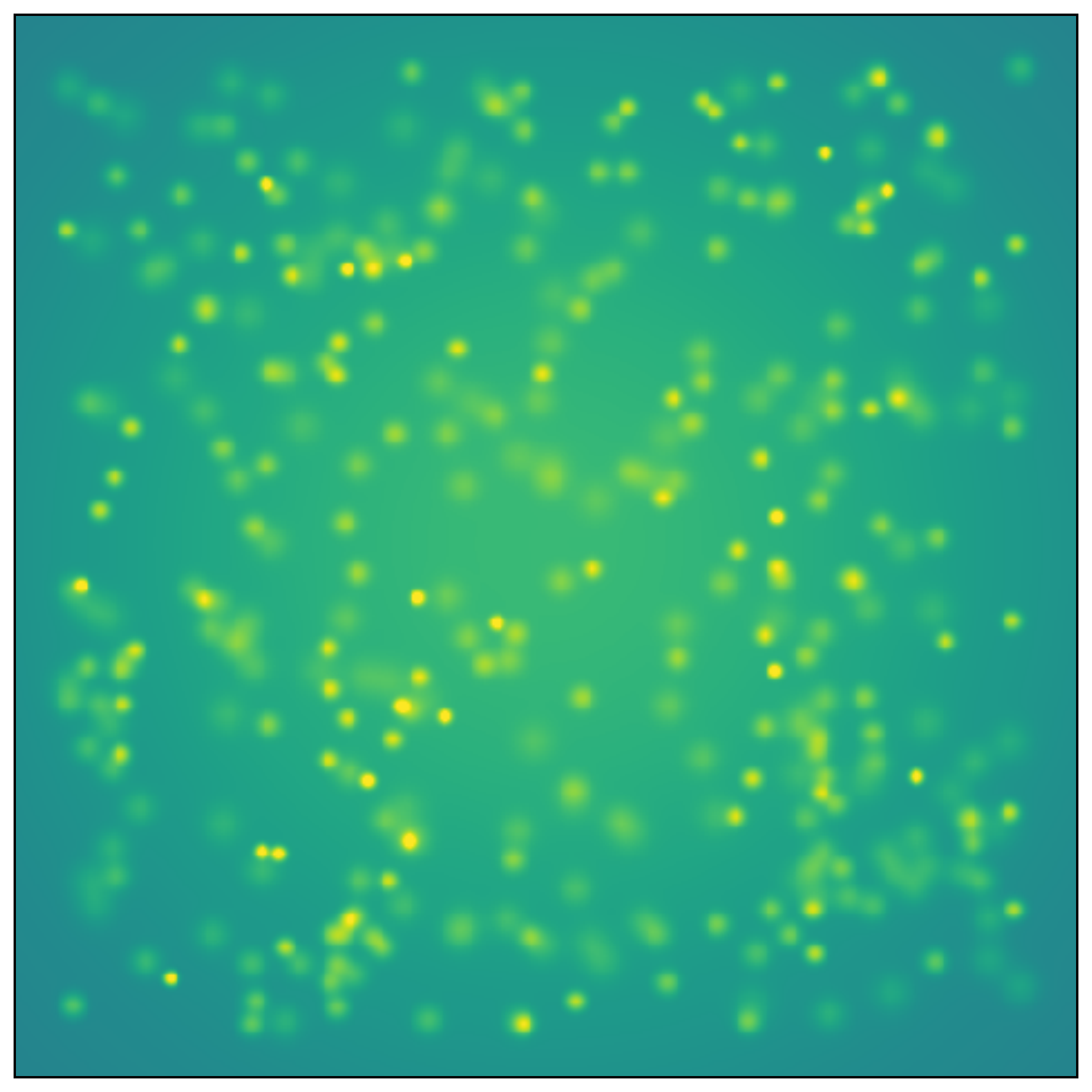} \label{fig:fsi_0}}
	\subfloat[power spectrum]{\includegraphics[width=\procwidth]{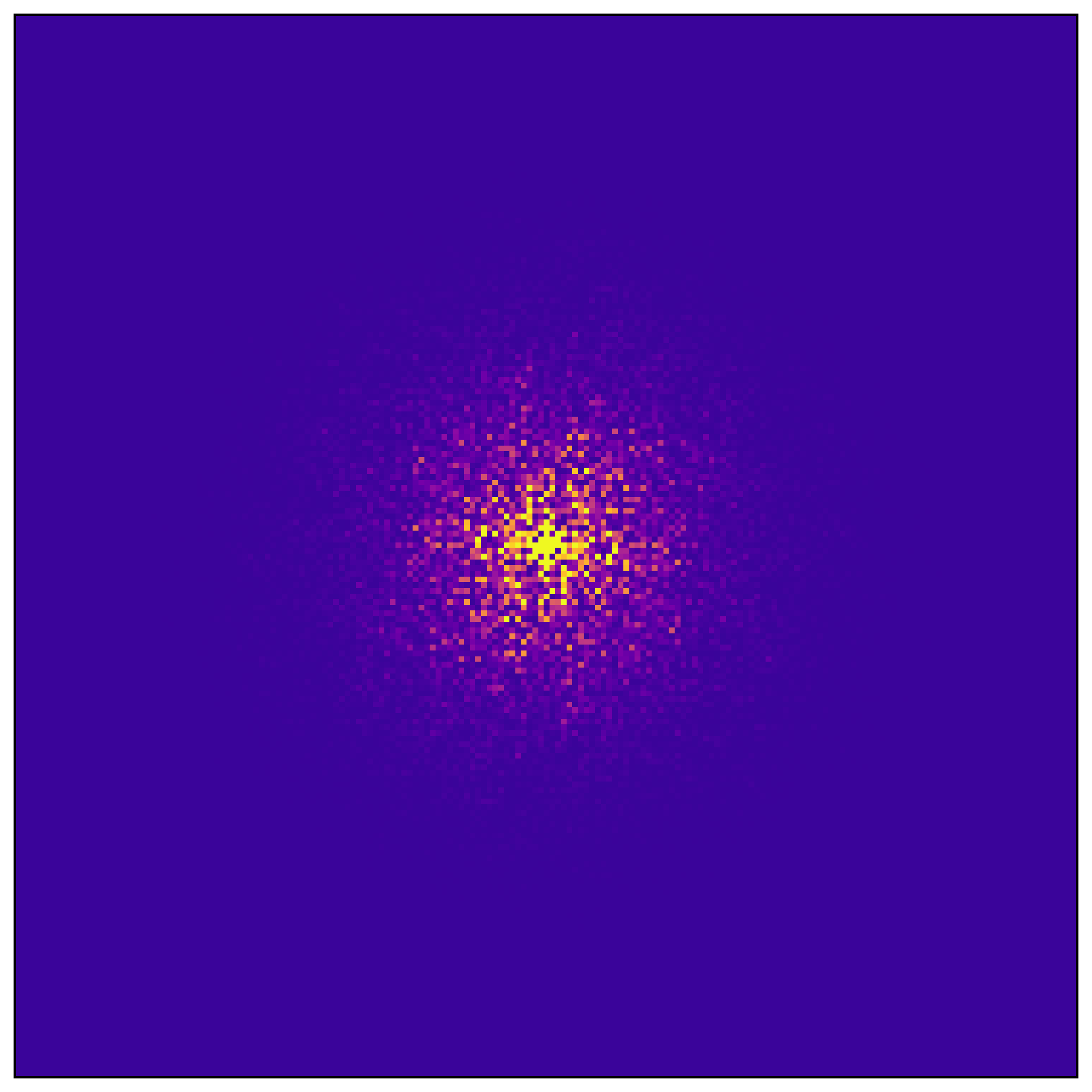}}\\
	\subfloat[filtered power spectrum]{\includegraphics[width=\procwidth]{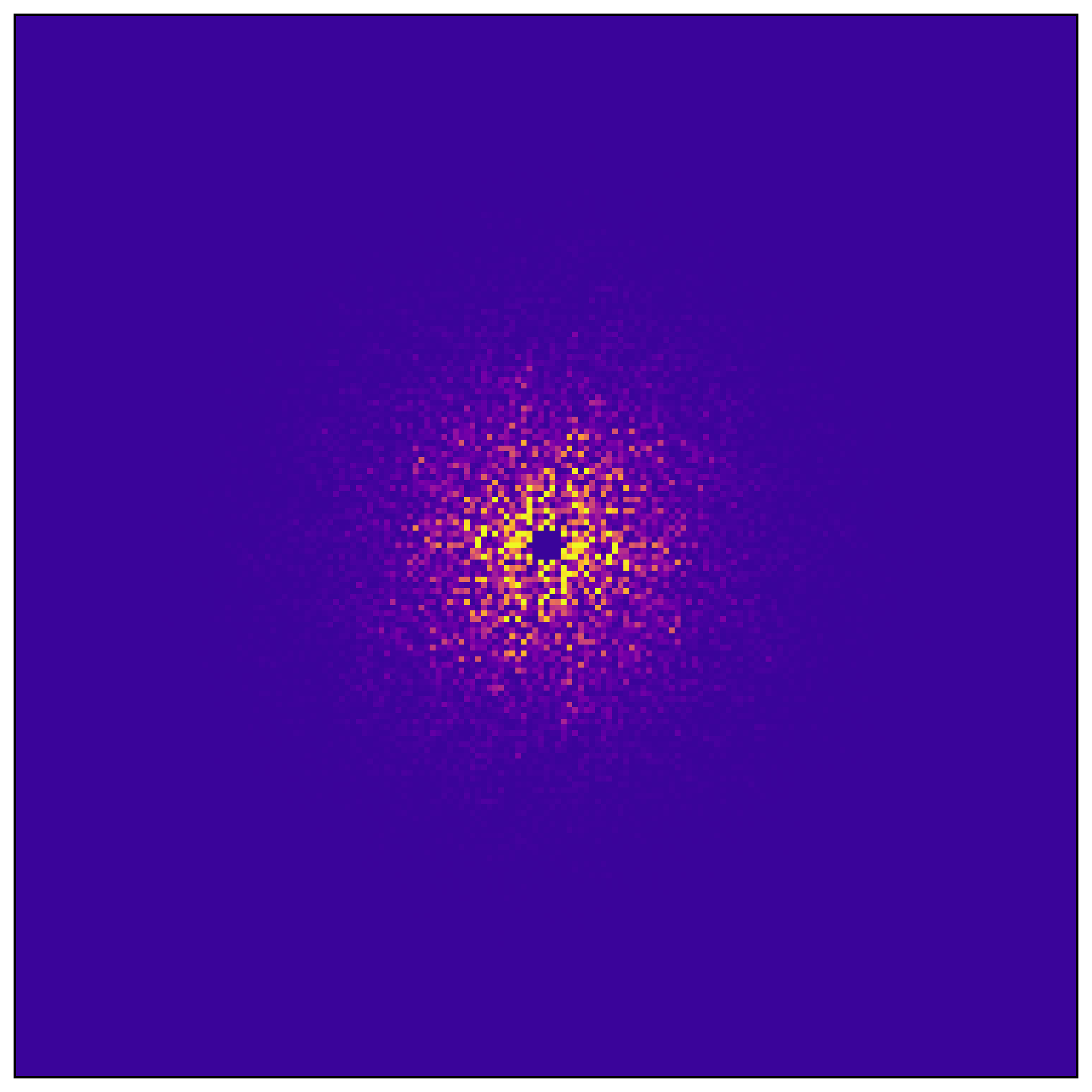}}
	\centering
	\subfloat[filtered image]{\includegraphics[width=\procwidth]{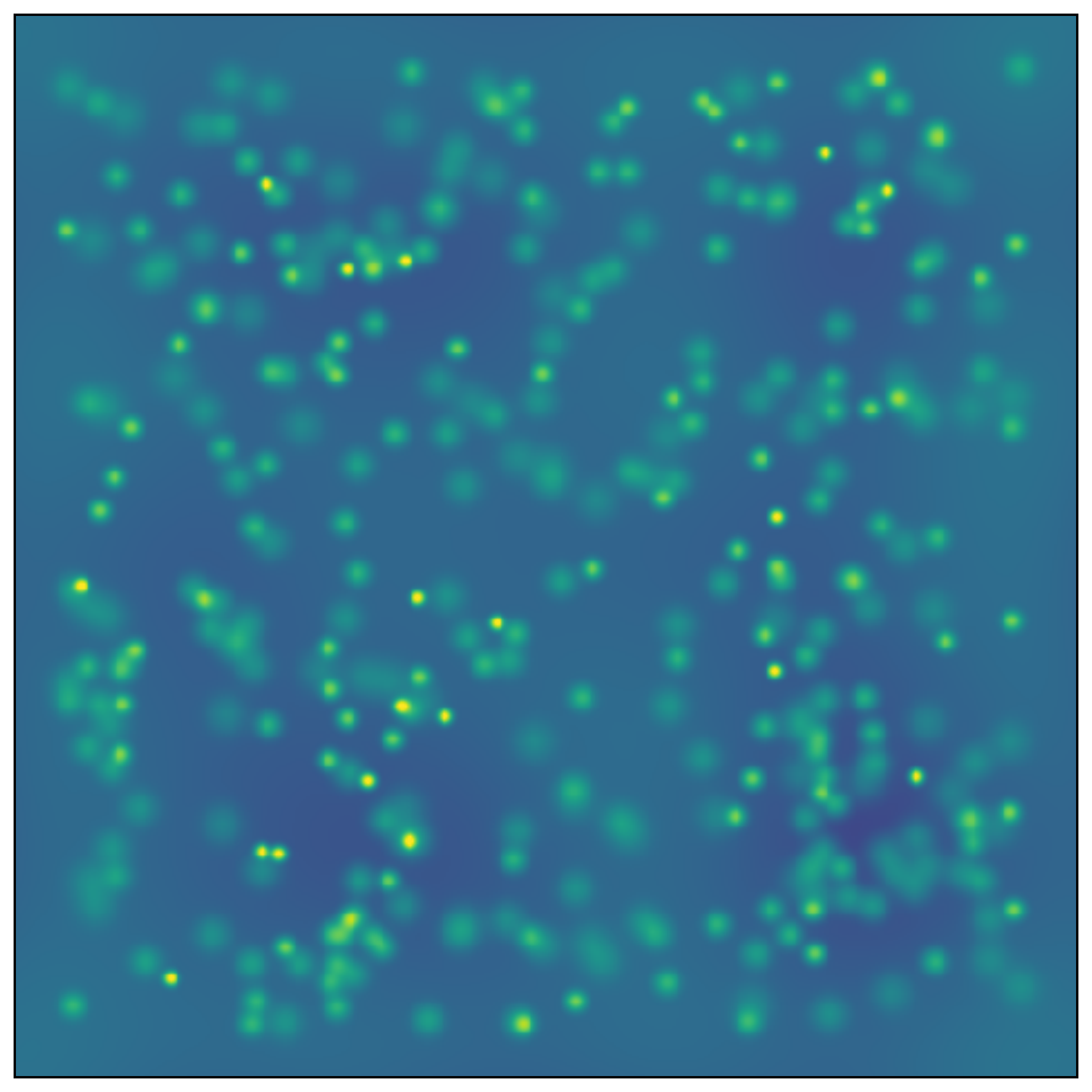}} \\
	\subfloat[positive filtered image]{\includegraphics[width=\procwidth]{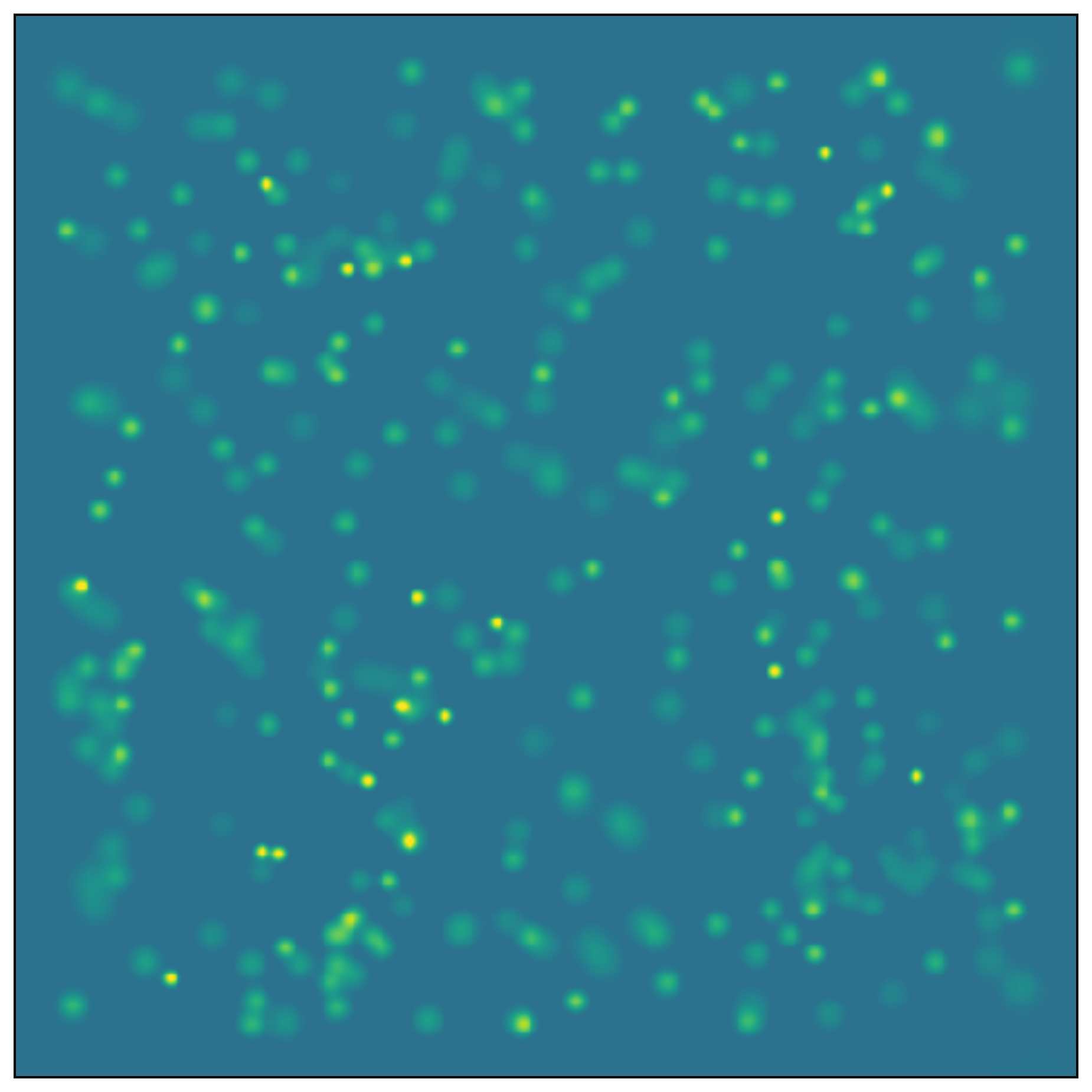}}
	\subfloat[diffuse image]{\includegraphics[width=\procwidth]{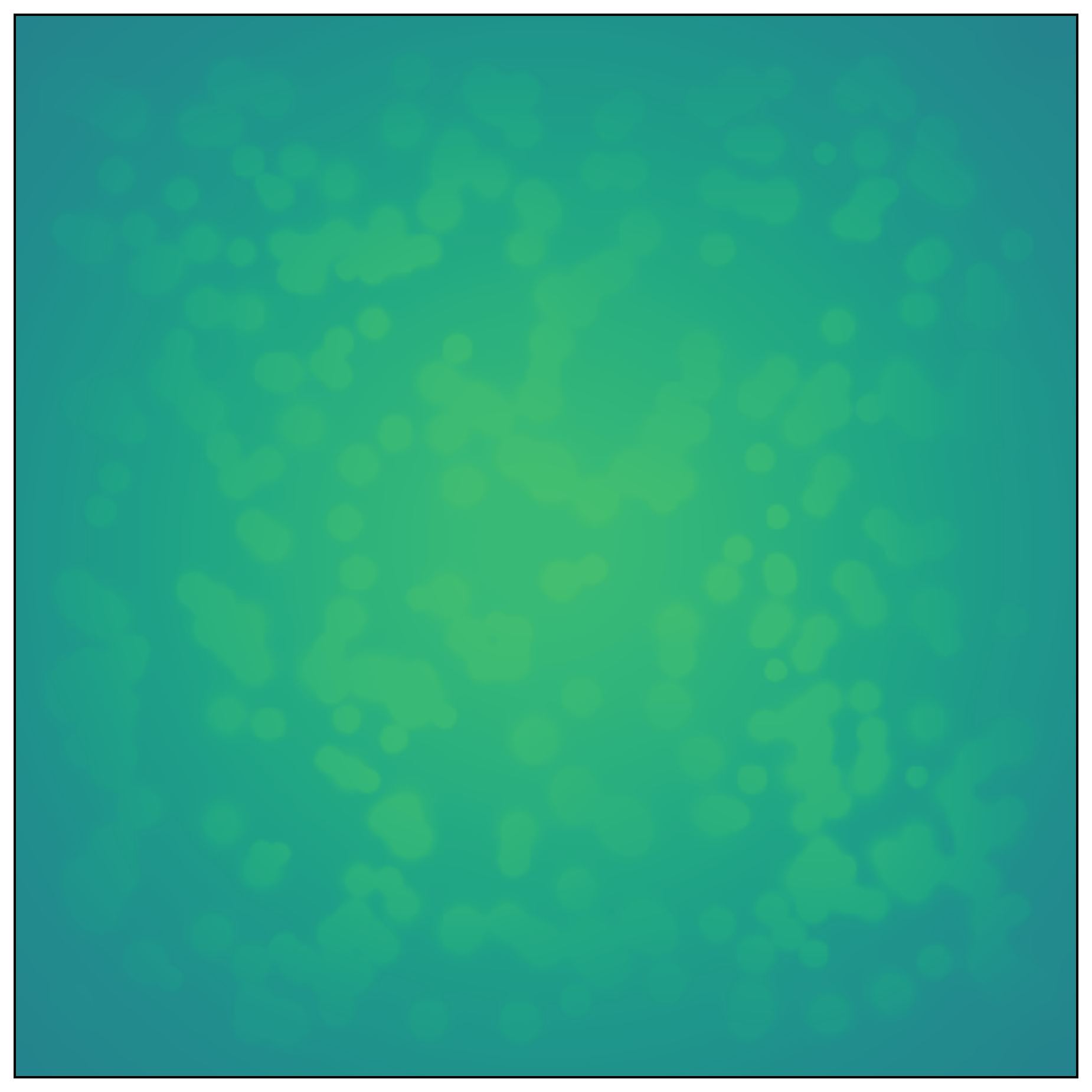}}

	\caption{An illustration of the process of filtering diffuse emission in Fourier space. a) $f(m,n)$ the original unfiltered image, which is a combination of compact regions and a diffuse background model. b) $ \left| F(u,v) \right|^2 $, the Fourier power spectrum of the original image. c) The Fourier power spectrum of the image with an Ideal filter applied, $   \left| \Psiuv F(u,v) \right|^2 $, masking a portion of the low frequency, large-spatial wavelength region of Fourier space. d) The image after the application of the highpass Ideal filter in the Fourier domain and transformation back to the image domain. This has removed the large-scale diffuse emission from the image. As the `DC' component of the Fourier spectrum has been set to zero the image has zero mean flux and thus parts of the image are negative, seen in the image as the dark blue patches. e) The final filtered image, $f'(m,n)$ after the application of a mask to all negative pixels (see equation~\ref{eq:Fourier_filter}). f) An image of the diffuse emission in the original image, obtained by subtracting the filtered image from the original image ($f(m,n) - f'(m,n)$).  }
	\label{fig:fourier_space_illustration}
\end{figure}

\subsection{Filters}
\label{sec:Filters}

As a basis for describing frequency-based filters, we define two key quantities. Firstly, we define a distance for each component of the Fourier spectrum, in two-dimensional frequency space, from the zero-frequency $(u=0,v=0)$ `DC' component:

\begin{equation}
 \Duv  = \sqrt{u^2 + v^2} .
\end{equation}

\begin{figure}
	\centering
	\includegraphics[width=\columnwidth]{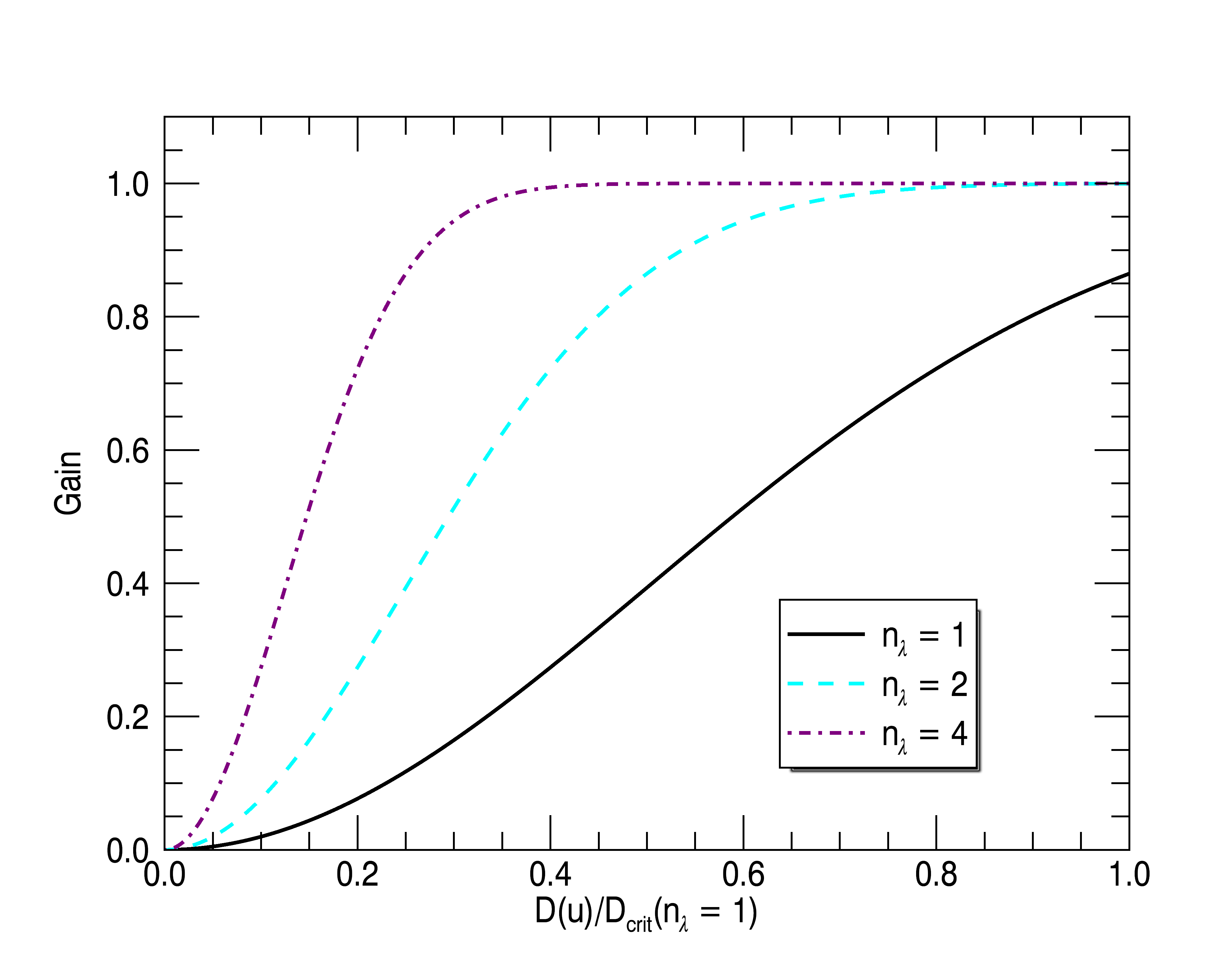}
	\caption{An illustration of the impact of choosing different values of the filtering-to-region separation length scale ratio, \nlambda{}. The frequency response (gain) of  three 1D Gaussian filters are shown are shown against the 1D frequency distance $D(u)$ normalised to the critical frequency, as calculated with $\nlambda{} = 1 $: $\dcrit(\nlambda{} =1)$. It can be seen that higher values of \nlambda{} have the effect of lowering the critical frequency and reducing the amount of attenuation from the filter. }
	\label{fig:n_lambda_comparison}
\end{figure}

\noindent
Secondly, we define a critical distance in frequency space:

\begin{equation}
\label{eq:d_crit}
\dcrit  = \frac{\lpix}{\nlambda \lambda} ,
\end{equation}

\noindent
where $\lambda$ is a physical size scale such that Fourier modes with wavelengths above this value are considered to be diffuse. We use the region separation length, $\lambda$, as measured by the \code code to define this distance. The filtering-to-region separation length scale ratio, \nlambda{}, is a multiplicative factor of $\lambda$ that softens the cut, reducing the value of \dcrit~and thus reducing the amount of attenuation from the filter. An illustration of its impact on the shape of a filter can be seen in Figure~\ref{fig:n_lambda_comparison}. Finally,  $\lpix$ is the length of one pixel in the same physical units as $\lambda$, where the pixels must be square.

We require a filter that removes diffuse emission while leaving the emission from the populations of interest (e.g. gas clouds and regions of star formation) intact. As these populations are compact, the Fourier transform of their emission will be concentrated in the low wavelength/high spatial-frequency part of Fourier space. Conversely, large scale diffuse emission, such as that from a sheet of gas or a galaxy scale Gaussian, will be primarily located in the high wavelength/low spatial-frequency part of Fourier space. For this reason, we only consider highpass filters, which allow high spatial frequencies to pass through the filter. We consider three different highpass filters:

\begin{enumerate}[leftmargin=*]
    \item Ideal highpass filter
        \begin{equation}
	        \label{eq:Ideal_filt}
            \Psiuv = \begin{cases}
            0 &  \Duv \leq \dcrit \\
            1 & \Duv > \dcrit\\
            \end{cases}
        \end{equation}

    \item Gaussian highpass filter
        \begin{equation}
	        \label{eq:Gaussian_filt}
            \Psiuv = 1-  \exp \left( - \frac{(\Duv)^{2}}{2 \dcrit^{2}}     \right)
        \end{equation}

    \item Butterworth highpass filter
        \begin{equation}
	        \label{eq:butterworth_filt}
            \Psiuv =\frac{1}{1+ \left(\frac{\dcrit}{\Duv}\right) ^{2n_{b}} }
        \end{equation}
        \noindent
        with the Butterworth order $n_{b} \in \mathbb{N}  \ | \ n_{b} \geqslant 1 $.

\end{enumerate}

\noindent

\begin{figure}
	\centering
	\includegraphics[width=\columnwidth]{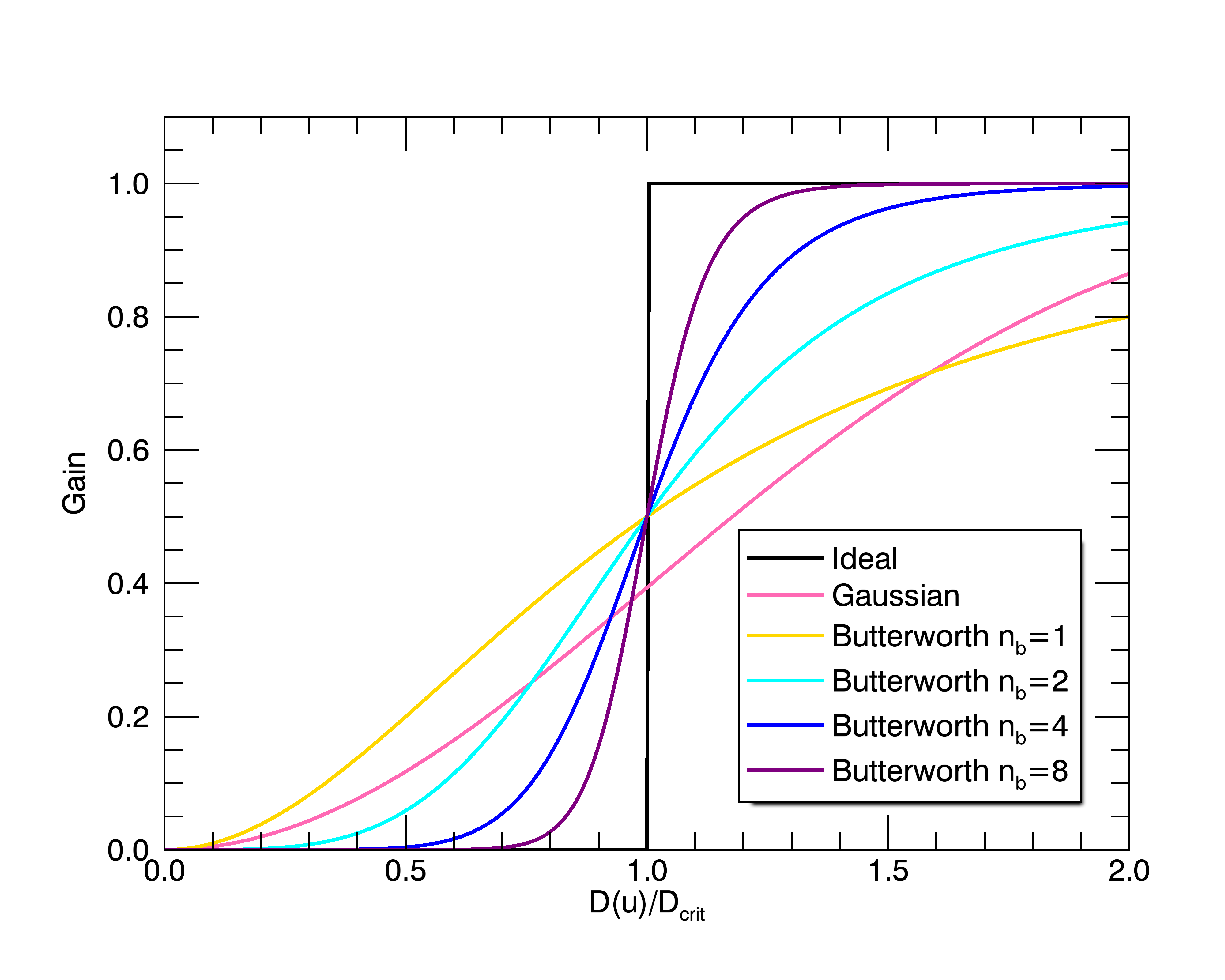}
	\caption{A comparison of the attenuation of the 1D counterparts of the filters described in Section~\ref{sec:Filters}. The frequency response (gain) of each of the filters is shown against the 1D frequency distance $D(u)$ normalised to the critical frequency \dcrit.}
	\label{fig:filter_comparison}
\end{figure}

A visual comparison of the Ideal, Gaussian, and Butterworth filters can be seen in Figure~\ref{fig:filter_comparison}. For a given value of \dcrit, the Ideal filter is the sharpest filter, fully attenuating frequencies lower than \dcrit~(the diffuse component) and allowing frequencies higher than \dcrit~(the compact component) to pass with no attenuation. On the other hand the Gaussian filter has a smoother roll-off and thus attenuates the compact part of frequency-space while not fully removing the diffuse part of frequency space. The Butterworth filter has a tunable sharpness that approaches a fully sharp Ideal filter as $n_{b} \to \infty $. However, the advantage, of smoother roll-off is that application of a smoother filter will create less `ringing' or distortion in the resultant filtered image than a sharp filter does. We assess the suitability of these filters for use in filtering diffuse emission from input images for the \code code in Section~\ref{sec:testing}.

\subsection{Image post-processing}
\label{sec:post_processing}

Due to the mathematical definition of the three highpass filters that we consider, the zero-frequency $(u=0,v=0)$ `DC' component, which represents the mean flux of the image, is always set to zero (i.e. $\Psi(0,0) = 0 $). This necessarily results in a filtered image with zero mean flux and thus roughly zero total flux. As flux remains in the filtered image corresponding to the parts of frequency space that were unattenuated or partially attenuated by the filter, this leads to negative pixels in the filtered image. As we do not expect negative emission from a tracer map, this problem may be solved simply by applying a binary mask that sets all pixels to zero wherever the reverse Fourier transform is negative:

\begin{equation}
\label{eq:negative_mask}
\Omega(m,n) = \begin{cases}
0 & \sum_{u,v}  \Psiuv F(u,v)  e^{i2\pi (um + vm)} < 0 \\
1 & \sum_{u,v}  \Psiuv F(u,v)  e^{i2\pi (um + vm)} >0\\
\end{cases} .
\end{equation}

\noindent
This results in a filtered image $f'(m,n)$ with a positive total flux. This is done in order to avoid an image with approximately zero flux after the removal of the `DC' component. In the case where an image contains noise, this has the effect of biasing the image with positive noise flux. We investigate this effect in more detail in Appendix~\ref{sec:noise}.

\subsection{Diffuse fraction measurement}

The method that we employ to filter diffuse emission has the added benefit of decomposing an input tracer map into a map of compact emission and a map of diffuse emission. This allows us to quantify the fraction of emission within the map that is diffuse. The fraction of the flux in an image that is compact is simply the ratio of the total flux in the filtered (compact emission) image $f'(m,n)$ to the total flux in the original unfiltered image $f(m,n)$:

\begin{equation}
\label{eq:signal_fraction}
\fsignal{} = \frac{\sum_{m,n} f'(m,n)}{\sum_{m,n} f(m,n)} .
\end{equation}

\noindent
The relationship of $f'(m,n)$ to $f(m,n)$ is described by equations~\ref{eq:Fourier_transform} and~\ref{eq:Fourier_filter}. The diffuse fraction is then simply $f_{\rm diffuse} = 1-  \fsignal{}$, as all flux must be either diffuse or compact. This gives us a physically motivated measure of the relative contribution of the diffuse and compact emission components in the tracer map. We use the cloud-separation length, $\lambda$, as the dividing line between diffuse and compact emission , with emission at spatial wavelengths greater than $ \nlambda{} \lambda$ being diffuse emission and that at spatial wavelengths less than $ \nlambda{} \lambda$ being compact emission. For example, in the case where $f(m,n)$ is a tracer map of the molecular gas phase in a galaxy, the compact emission fraction represents the fraction of emission that comes from compact structures such as molecular clouds, whereas the diffuse fraction represents the fraction of the emission coming from large scale diffuse emission such as a reservoir of diffuse molecular gas or unresolved, low-mass molecular clouds.

\section{Implementation within the \code code}
\label{sec:implementation}

We present a method, using iterative filtering in Fourier space, to remove diffuse emission present in tracer maps to be used with the \code code \citep{kruijssen18}. As with the \code code, we have implemented this method in the Interactive Data Language ({\sc IDL})\footnote{\url{http://www.harrisgeospatial.com/SoftwareTechnology/IDL.aspx}}, in part using routines from the {\sc IDL} Astronomy User's Library\footnote{\url{https://idlastro.gsfc.nasa.gov/}} and the {\sc IDL Coyote} Library.\footnote{\url{http://www.idlcoyote.com/}}

\begin{figure}
	\centering
	\includegraphics[width=\columnwidth]{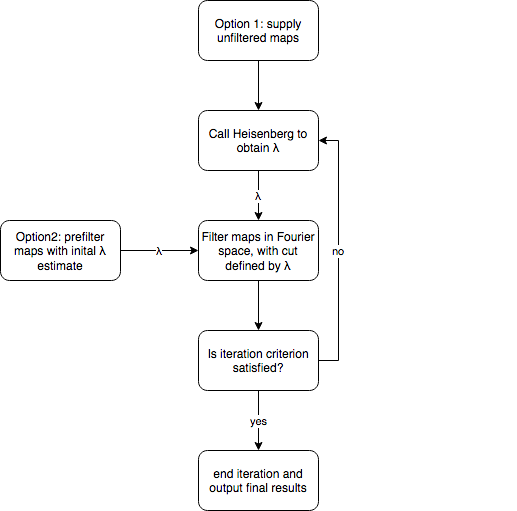}
	\caption{Flowchart summarising the iterative diffuse filtering method.}
	\label{fig:flowchart}
\end{figure}

A flowchart summarising the steps taken in the method can be seen in Figure~\ref{fig:flowchart}. A user has two options of how to start the filtering process: they can either supply the original images alone, in which case \code is called to obtain a value $\lambda$ for the initial calculation of \dcrit; or, if the user has calculated a value of $\lambda$ externally, this value may also be supplied to the code for the initial calculation of \dcrit, in which case the images are pre-filtered, before the first call to \code is made. After that, the \code code is fitted to the filtered maps, and a new value of $\lambda$ and thus \dcrit{} is calculated. The code then checks for convergence and either repeats the process of fitting \code in the case of no convergence or terminates and outputs the final values of all output parameters.

\subsection{Input parameters}

In order to run the presented method, a user has to specify a number of input flags, shown in Table~\ref{tab:flags}, and input parameters, shown in Table~\ref{tab:parameters}, in addition to those flags and parameters described in Tables 1 and 2 of \citet{kruijssen18}.

Both the flags and input parameters are divided into two groups separated by white space. The first of these comprises flags and parameters added to \code itself and the second comprises flags and parameters that control the iterative diffuse filtering process. We consider most of these to be well described by reference to Tables~\ref{tab:flags} and~\ref{tab:parameters}. We provide some further explanation for some of the flags and parameters:

\begin{enumerate}[leftmargin=*]
	\item If the flag {\tt peak\_find\_tui} is set, the user is presented with a text-user-interface allowing them to adjust peak selection (see section 3.2.7 of \citealt{kruijssen18} for details). The user is also presented with a DS9 window\footnote{This optional feature requires the user to have the DS9 software package, which can be obtained from: \url{http://ds9.si.edu}}, showing the input images with their identified peaks overlaid. The user may then adjust the parameters controlling peak selection through the text-user-interface and repeat the peak-finding process until they are satisfied with the identified peaks. This feature is of particular use during the iterative Fourier filtering process, as it allows the user to alter their peak selection parameters to adjust peak-finding after the removal of the diffuse background.

	\item Together, the parameters $q_{\rm crit}$, a fractional tolerance between the measured value of $\lambda$ in the current iteration and previous iteration and $r_{\rm crit}$, the number of iterations over which the criterion should be satisfied, specify the iteration's stopping condition:

	\begin{equation}
	q_{\rm crit} <  \left| \frac{ \lambda_{i-r} - \lambda_{i} }{\lambda_{i}} \right| ,
	\end{equation}
	for $r \in \mathbb{N}  \ | \ 1 \leqslant r \leqslant r_{\rm crit} $. i.e. if the relative difference between value of $\lambda$ calculated in the current iteration step and all values of $\lambda$ calculated in the $r_{\rm crit}$ previous steps is less than $q_{\rm crit}$, the iteration is considered to have converged and is terminated.

	\item In addition,  $r_{\rm max}$ sets the maximum number of iterative steps that will be taken ($r_{\rm max}$  + 1) before the code will terminate even if convergence has not been reached.

\end{enumerate}

\subsection{Code output}

\begin{figure}
	\centering
	\subfloat{\includegraphics[width=\columnwidth]{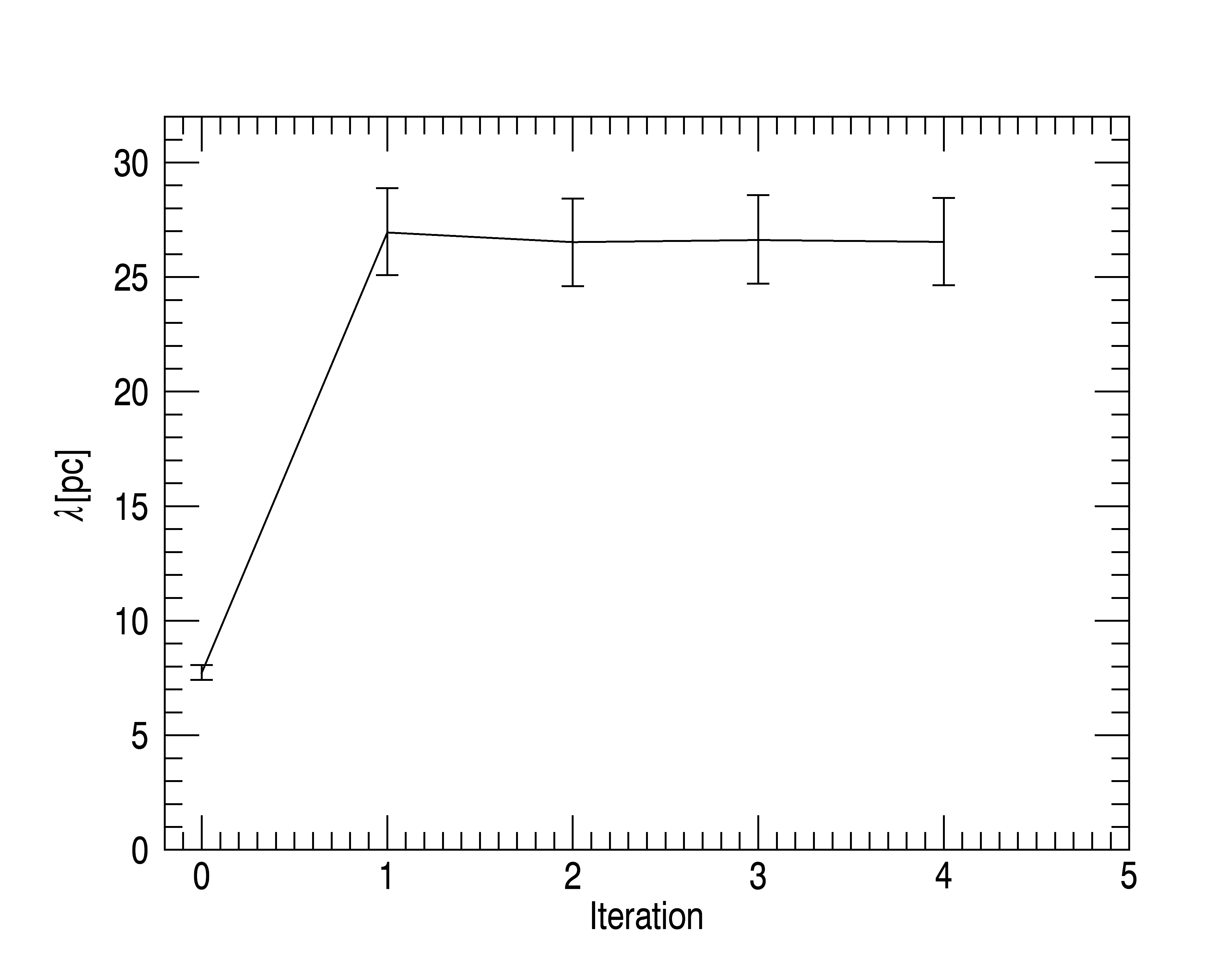}} \\
	\centering
	\subfloat{\includegraphics[width=\columnwidth]{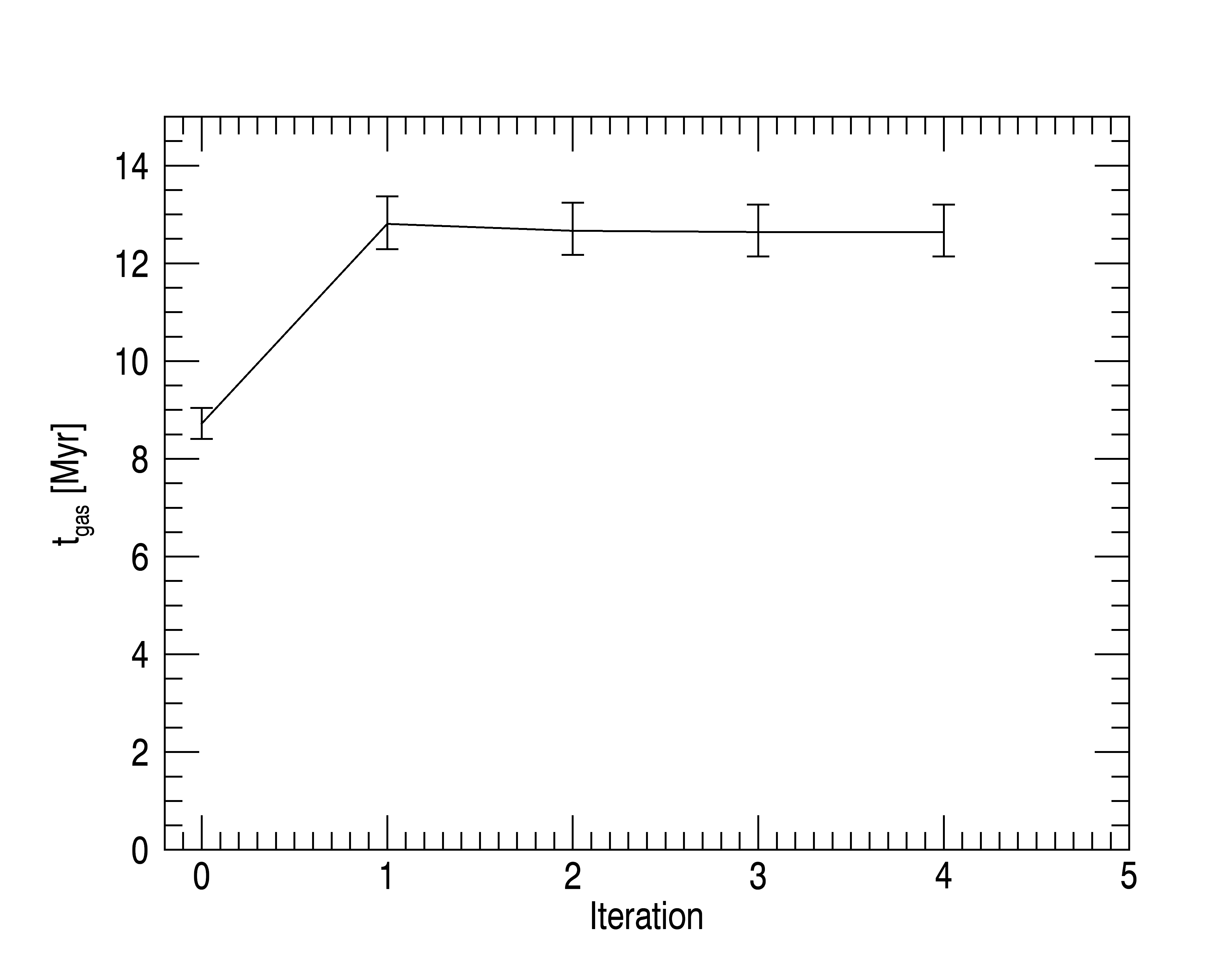}} \\
	\subfloat{\includegraphics[width=\columnwidth]{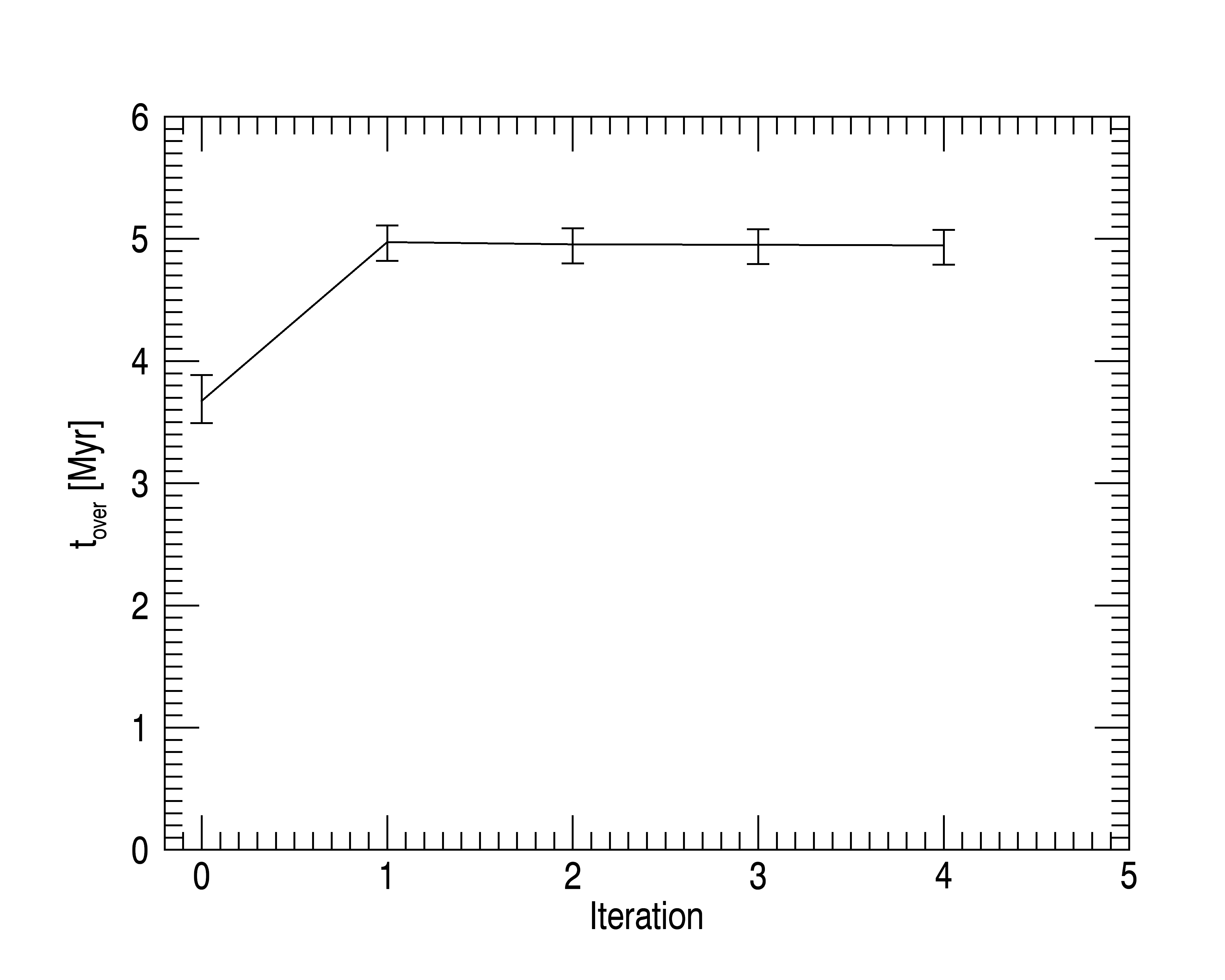}}
	\caption{Example output from a successful run, in which convergence has been reached, showing how the quantities $\lambda$, \tgas{} and \tover{} vary with iteration number. A change in the values of all three quantities can be seen from the leftmost data-point in each plot, where no Fourier filtering has been applied, and all subsequent points after the application of Fourier filtering. }
	\label{fig:quantity_iter_fig}
\end{figure}

Each iterative step applies the \code code and the quantities described by \citet{kruijssen18} in their table 4 are measured and returned, in addition to the image compact emission fractions, as described in Table~\ref{tab:output_quantities}. The quantities are tracked and figures are made for the user showing how they change with each iterative step. Figure~\ref{fig:quantity_iter_fig} shows examples of these figures for the fundamental physical quantities, $\lambda$, \tgas~and \tover, which are measured by \code. This allows the user to quickly visually inspect that convergence in the value of $\lambda$, on which the method's iteration condition is based, has also resulted in satisfactory convergence of the other quantities. In addition, the filtered map (of compact emission) and the diffuse emission map produced in each step are saved to disk. This makes these maps available for further analysis on the two components and allows the user to visually inspect the filtered map for distortions. An example of these maps can be seen in panels e and f of Figure~\ref{fig:fourier_space_illustration}.

\begin{table*}
	\caption{Flags to be set for the presented analysis additional to those described in table 1 of \citet{kruijssen18}}
\begin{tabular}{
		p{\dimexpr.13\linewidth-2\tabcolsep}
		c{\dimexpr.14\linewidth-2\tabcolsep}
		p{\dimexpr.73\linewidth-2\tabcolsep}}
			\hline
			Flag & Values (\textbf{default}) & Description \\
			\hline
			{\tt diffuse\_frac} & 0/\textbf{1} & Calculate diffuse fraction in images (off/\textbf{on}) \\
			{\tt peak\_find\_tui} &  \textbf{0}/1  & Use interactive peak identification interface to refine initial peak selection (\textbf{off}/on) \\
			\multirow{2}{*}{\tt f\_filter\_type} &
				\hspace{-10pt}\ldelim\{{3}{10pt} 0 & Use an Ideal filter for the determination of diffuse fractions (see equation~\ref{eq:Ideal_filt}) \\
				& \textbf{1} & Use a Gaussian filter for the determination of diffuse fractions (see equation~\ref{eq:Gaussian_filt}) \\
				& 2 & Use a Butterworth filter for the determination of diffuse fractions (see equation~\ref{eq:butterworth_filt})  \\
			\multirow{2}{*}{\tt use\_sds} &
				\hspace{-10pt}\ldelim\{{2}{10pt} 0 & Automatically calculate map sensitivity limits as described in  section 3.2.6 of  \citet{kruijssen18} \\
				& \textbf{1} & Pass in externally calculated values of map sensitivity limits to the code \\
			\\
			\multirow{2}{*}{\tt use\_guess} &
				\hspace{-10pt}\ldelim\{{2}{10pt} \textbf{0} & Calculate \dcrit~for the first filtering step by fitting \code  to the unfiltered maps (option 1 in Figure~\ref{fig:flowchart})\\
				& 1 & Use an initial estimated value of $\lambda$ to determine \dcrit~for the first filtering step  (option 2 in Figure~\ref{fig:flowchart}) \\

			\hline
		\end{tabular}
	\label{tab:flags}
\end{table*}

\newcommand{\two}{2.0}
\begin{table*}
	\caption{Input parameters to be set for the presented analysis additional to those described in table 2 of  \citet{kruijssen18}}
	\begin{tabular}{
			p{\dimexpr.10\linewidth-2\tabcolsep}
			c{\dimexpr.10\linewidth-2\tabcolsep}
			p{\dimexpr.80\linewidth-2\tabcolsep}}
		\hline
		Parameter & Default & Description \\
		\hline
		$n_{b,{\rm diffuse}}$ & 2  & The order of the Butterworth filter used to calculate map diffuse fractions (only used if {\tt f\_filter\_type} = 0) \\
		$n_{\lambda,{\rm diffuse}}$ &  1.0  & Multiplicative factor used in calculating map diffuse fractions for calculating \dcrit~that reduces filter attenuation for a given value of $\lambda$ (see equation ~\ref{eq:d_crit})  \\
		$\sigma_{\rm sens,star}$  &  -  & The measured sensitivity limit of the star formation tracer map in units of the map \\
		$\sigma_{\rm sens,gas}$ & - & The measured sensitivity limit of the gas tracer map in units of the map  \\
		\\
		$\lambda_{\rm initial}$ & -  &    Length in pc of the initial estimate of lambda (only used if {\tt use\_guess} = 1) \\
		$q_{\rm crit}$ &  0.05  & Maximum fractional difference between $\lambda$  and previous $\rm\lambda$ value(s) that triggers the iteration process to terminate \\
		$r_{\rm crit}$ &  2  & Number of steps over which the iteration criterion, $q_{\rm crit}$, must be true for iteration to terminate \\
		$r_{\rm max}$  &    10   &  Maximum number of steps in the iterative diffuse filtering process, before the iteration is terminated \\
		$n_{b,{\rm iter}}$ &  2  & The order of the Butterworth filter used in the iterative diffuse filtering process  (only relevant if {\tt f\_filter\_type} = 0 )\\
		$n_{\lambda,{\rm iter}}$ &  1.0  & Multiplicative factor used in the iterative diffuse filtering process for calculating \dcrit~that reduces filter attenuation for a given value of $\lambda$ (see equation ~\ref{eq:d_crit}) \\

		\hline
	\end{tabular}
	\label{tab:parameters}
\end{table*}

\begin{table*}
	\caption{Quantities constrained by \code additional to those described in table 4 of \citet{kruijssen18} }
	\begin{tabular}{
			p{\dimexpr.10\linewidth-2\tabcolsep}
			c{\dimexpr.10\linewidth-2\tabcolsep}
			p{\dimexpr.80\linewidth-2\tabcolsep}}
		\hline
		Quantity & Equation & Description \\
		\hline
		\fcl{} & \ref{eq:signal_fraction}  & The fraction of emission in the stellar tracer map that is not diffuse \\
		\fgmc{} &  \ref{eq:signal_fraction} & The fraction of emission in the gas tracer map that is not diffuse  \\
		\qconstar & \ref{eq:qcon} & The relative flux loss from compact stellar regions due to filtering in Fourier space (see Section~\ref{sec:signal_loss}) \\
		\qcongas{} & \ref{eq:qcon} & The relative flux loss from compact gas regions due to filtering in Fourier space (see Section~\ref{sec:signal_loss}) \\
		\etastar{} & \ref{eq:eta_star} & The \etaname{star formation } (see Section~\ref{sec:signal_loss})  \\
		\etagas{} & \ref{eq:eta_gas} & The \etaname{gas } (see Section~\ref{sec:signal_loss}) \\
		\qetastar{} & \ref{eq:qeta} & The relative flux loss from stellar regions due to overlap between regions in the stellar map (see Section~\ref{sec:signal_loss}) \\
		\qetagas{} & \ref{eq:qeta} &  The relative flux loss from gas regions due to overlap between regions in the gas map (see Section~\ref{sec:signal_loss}) \\

		\hline
	\end{tabular}
	\label{tab:output_quantities}
\end{table*}

\section{Generation of test images}
\label{sec:test_images}

In order to test and validate the presented iterative diffuse filtering method we generate test input datasets. Each test dataset consists of a pair of test images, of size $\npix{x} \times \npix{x}$, such that both images in the pair are square and of the same size. We note that these choices are made for simplicity and are not restrictions of the method. Each image $f(m,n)$ is made up of two components: a compact component $s(m,n)$ and a background component $b(m,n)$:

\begin{equation}
\label{eq:basic_image_model}
f(m,n) = s(m,n) + b(m,n) .
\end{equation}

\noindent
An example image generated in this manner is shown in Figure~\ref{image_componets}.

\newcommand{\imgenwidth}{0.245}

\newcommand{\imgenwidththree}{0.32}
\begin{figure*}
	\centering
	\subfloat{\includegraphics[width=\imgenwidththree\textwidth]{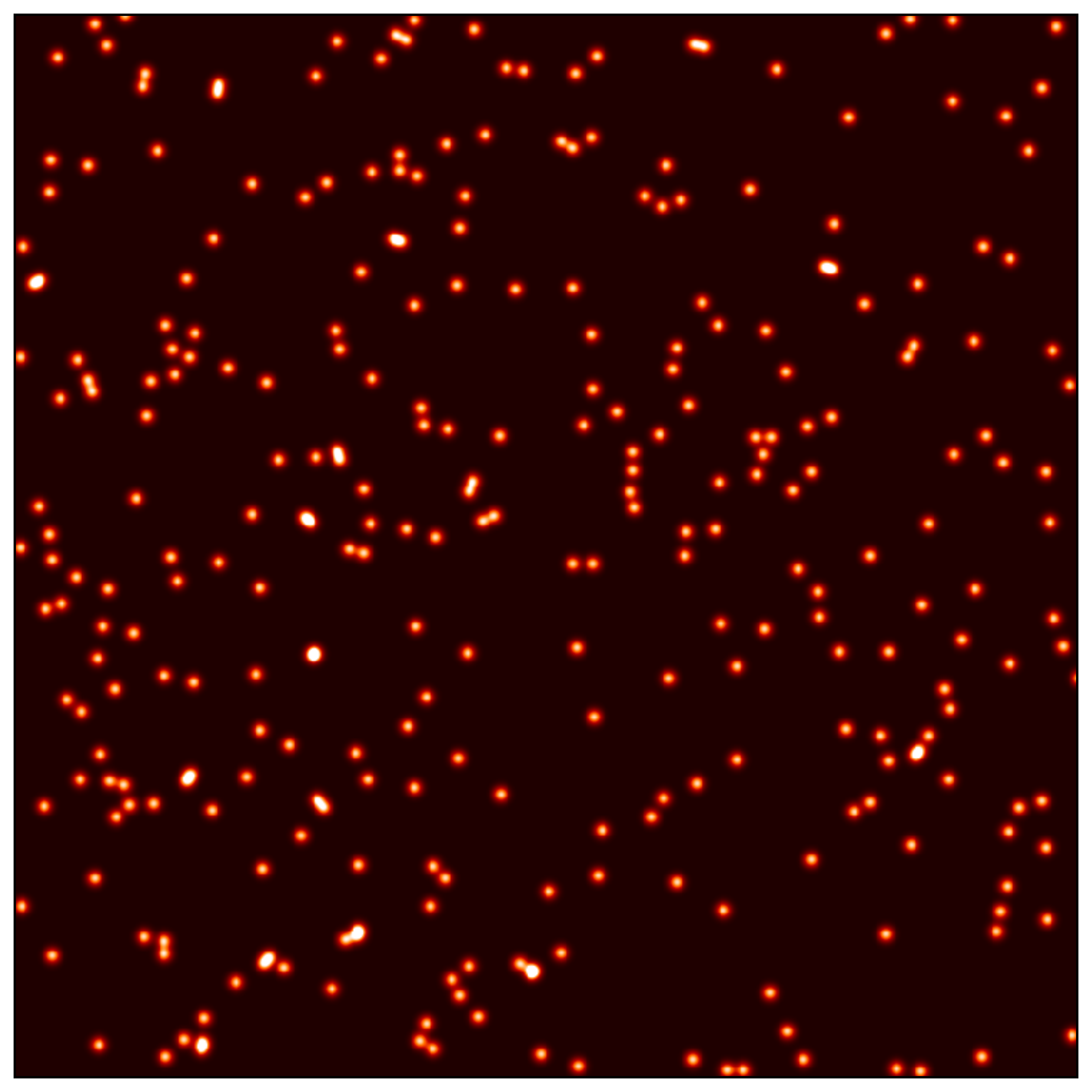}}
	\hfill
	\subfloat{\includegraphics[width=\imgenwidththree\textwidth]{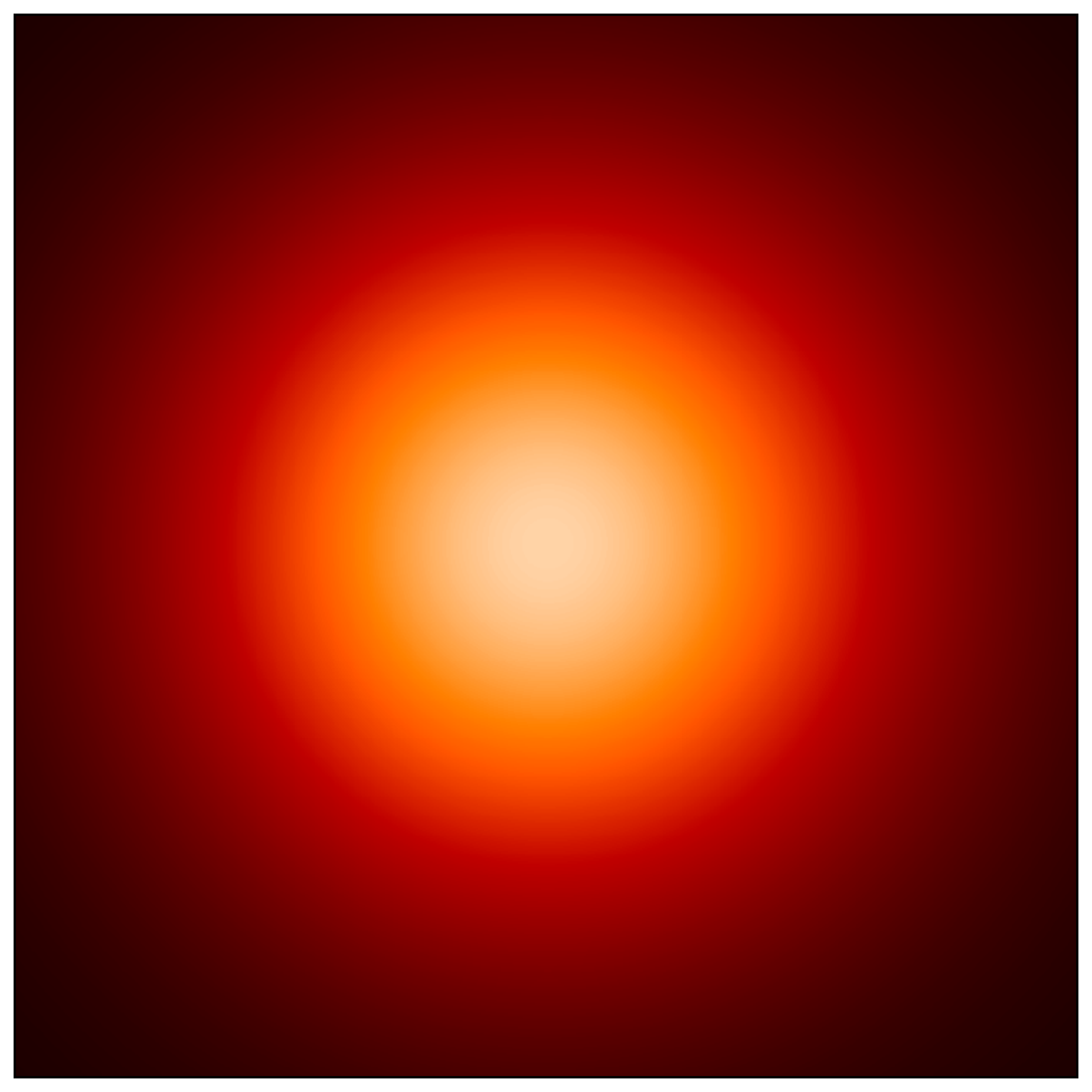}}
	\hfill
	\subfloat{\includegraphics[width=\imgenwidththree\textwidth]{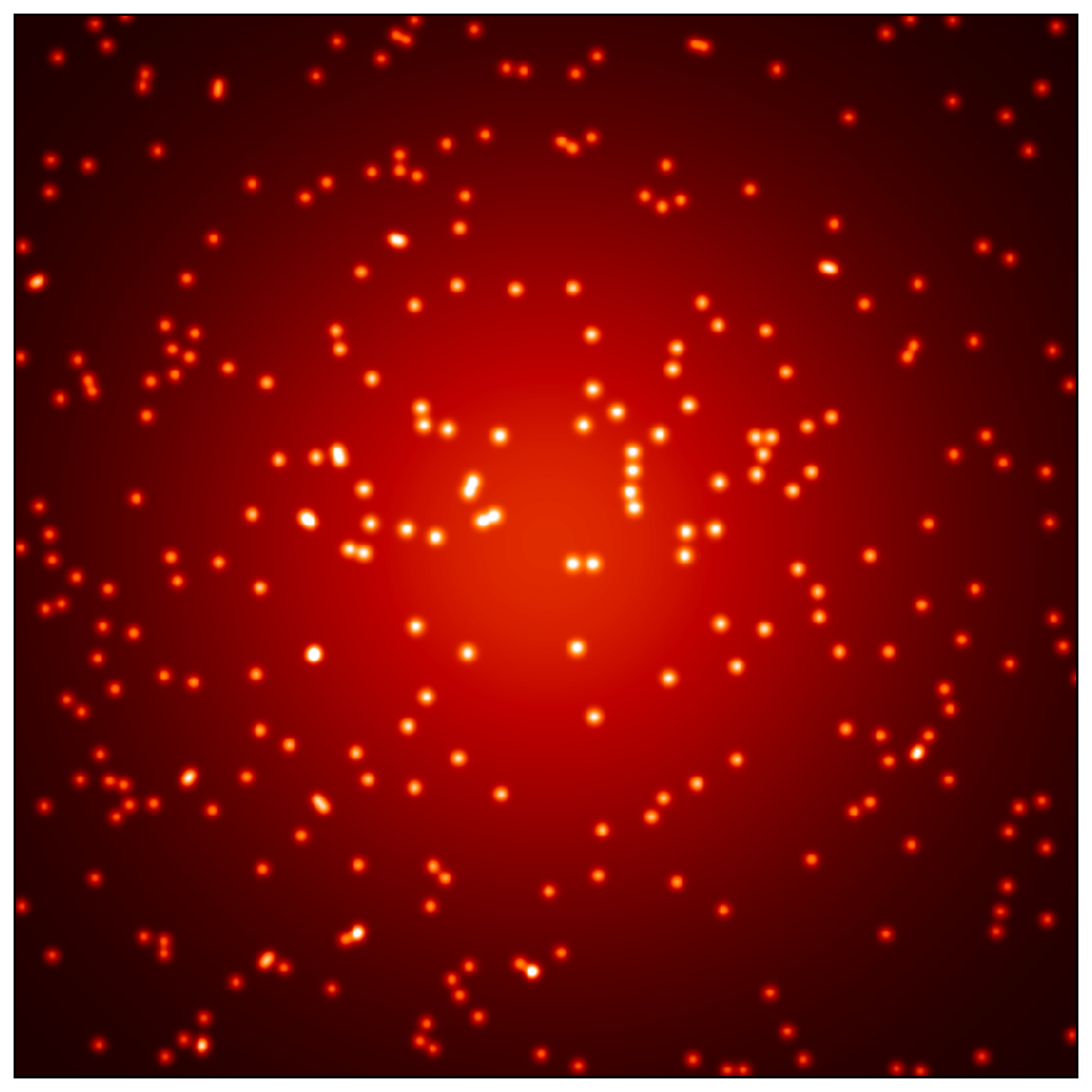}}

	\caption{Example generated image and its components.
		Left panel: the compact component. Note that the placement of Gaussians is entirely random such that they may overlap partially or entirely with other Gaussians, leading to pixels brighter than the peak brightness of a single Gaussian.
		Middle panel: the background component, here a single large `galaxy scale' Gaussian function.
		Right panel: the final generated image with both components summed together.}\label{image_componets}

\end{figure*}

\subsection{The signal component}
\label{sec:signal_component}

For the \code code to be applied we require two maps: one representing a progenitor phase and one representing a descendent phase in an evolutionary sequence. In order to create the compact component for these maps, we generate three compact populations: a population that is visible only in the progenitor map, one that is visible only in the descendant map and an `overlap' population that is visible in both maps. Hereafter, we will assume that our progenitor population is made up of gas clouds and that our descendent population is made up of young stellar regions (\htwo~regions) and thus refer to them as `gas' and `stars'. In this scenario, our overlap population represents the phase of star formation where young stars have been formed and are still co-spatial with their natal cloud. This is the phase during which feedback operates to eventually cause the cloud to cease be visible in the gas tracer map (for example through photodissociation or physical transportation of the gas away from the young stars). We note, however, that the \code code may be applied applied more generally, to any scenario characterised characterised by an evolution from a population visible in one tracer map to a descendent population visible in another tracer map. Likewise, the presented method of removing diffuse emission from a tracer map may be used in other situations, in which one wishes to remove an extended diffuse component from a tracer map containing both this diffuse component and a population of compact regions.

To simulate the gas, stellar and overlap populations, we generate Gaussian functions and position them randomly within our maps. We determine the position $(m_{\rm gc,i},n_{\rm gc,i})$ of the central pixel of the $i^{th}$ Gaussian function by drawing random numbers from a uniform distribution:

\begin{equation}
\left\lbrace m_{\rm gc,i} \in \mathbb{N} \ | \ 0 \leqslant m_{\rm gc,i} \leqslant N_{\rm pix,x} -1 \right\rbrace
\end{equation}

\begin{equation}
 \left\lbrace n_{\rm gc,i} \in \mathbb{N} \ | \ 0 \leqslant n_{\rm gc,i} \leqslant N_{\rm pix,y} -1 \right\rbrace .
\end{equation}

We set the characteristics of the compact component with the following parameters: \npin{gas}, the number of Gaussian functions appearing in the gas map only; \npin{star}, the number of Gaussian functions appearing in the stellar map only; \npin{over}, the number of peaks appearing in both the gas and stellar maps; and $G_{\rm FWHM, star}$ and $G_{\rm FWHM, gas}$, the full width at half maximum (FWHM) of the Gaussian functions seeded into the stellar and gas maps, respectively. We consider a number of different models for the FWHMs of the regions:

\begin{enumerate}[leftmargin=*]
	\item Uniform FWHM model: \\
	In this model, all the Gaussian regions in each pair (stellar and gas) of images have the same FWHM:
	\begin{equation}
		 G_{\rm FWHM, star, i} =  G_{\rm FWHM, gas, i}  = \phi
	\end{equation}
	where $\phi$ is a uniform random number chosen for each experiment such that $\phi \in \mathbb{R}  \ | \ \phimin{}  < \phi  < \phimax{} $. In this model, the total flux of each region is equal to that of each other region.

	\item Asymmetric uniform FWHM model: \\
	In this model, we set a common FWHM value for all Gaussian regions in the stellar map, and another one for all the Gaussian regions in the gas map:
	\begin{equation}
		G_{\rm FWHM, star, i} = \phi_{\rm star}
	\end{equation}
	\begin{equation}
		G_{\rm FWHM, gas, i}  = \phi_{\rm gas} ,
	\end{equation}
	where $\phi_{\rm star}$ and $\phi_{\rm gas}$ are individually chosen uniform random numbers chosen for each experiment such that $\phi_{\rm star} \in \mathbb{R}  \ | \ \phimin{}  < \phi_{\rm star} < \phimax{} $ and $\phi_{\rm gas} \in \mathbb{R}  \ | \ \phimin{}  < \phi_{\rm gas} < \phimax{} $.As with the previous model, the flux of each region is equal to that of each other region.

	\item FWHM spread model: \\
	In this model, each individual Gaussian region in the stellar and gas maps has a randomly selected FWHM, selected within the same range of values for both maps

	\begin{equation}
		\label{eq:spread_fwhm_stars}
		G_{\rm FWHM, star, i}  \in \mathbb{R}  \ | \ \phimin{} < G_{\rm FWHM, star, i} < \phimax{}
	\end{equation}
	\begin{equation}
		\label{eq:spread_fwhm_gas}
		G_{\rm FWHM, gas, i}   \in \mathbb{R}  \ | \ \phimin{} < G_{\rm FWHM, gas, i} < \phimax{} .
	\end{equation}
	\noindent
	As with the previous models, the flux of each region is equal to that of the other regions, irrespective of the randomly selected FWHM of a region.

	\item FWHM flux spread model: \\
	The FWHM of regions in this model is selected in the same way as the `FWHM spread model' (i.e. according to equations~\ref{eq:spread_fwhm_stars} and~\ref{eq:spread_fwhm_gas}). However, in this model the flux of regions are not equal to that of each other region, instead the flux of a region is proportional to its size, i.e. $\mathcal{F}_{i} \propto \fwhm_{i}$.

\end{enumerate}

For each of these models, regions in the overlap phase are selected from the same size distribution as the gas and stellar regions. For the asymmetric uniform FWHM model, an overlap region has size $\phi_{\rm star}$ in the stellar map and size $\phi_{\rm gas}$ in the gas map.

In order to test the effectiveness of the method, we compare measured values of the three key quantities measured by the \code code (the gas cloud lifetime, \tgas, the overlap phase lifetime, \tover, and the mean separation length between regions, $\lambda$; recall that the young stellar lifetime $t_{\rm star}$ is assumed to be known a priori) to values known from the generation of the test datasets. The relative visibility lifetimes of each population are linked to the relative number of regions in each population. Therefore, for \tgas~and \tover~in relation to \tstar~we have:

\begin{equation}
\label{eq:tgas}
\tgas =\tstar \frac{\npin{gas}+\npin{over}}{\npin{star}+\npin{over}}
\end{equation}

\noindent
and

\begin{equation}
\label{eq:tover}
\tover = \tstar \frac{\npin{over}}{\npin{star}+\npin{over}} ,
\end{equation}

\noindent
where \tstar~is used as the reference time-scale. As we are not directly simulating a particular tracer, the actual value of the reference time-scale is not relevant. We choose $\tstar{} = 10\, \myr{}$, which is close to typical values found for visibility lifetimes of star formation tracers by \citet{haydon18}.

For each image, we select a value of the  filling factor, $\zeta$, the region size-to-mean separation ratio. Given our selected value of \fwhm{} and  $\zeta$, we calculate a value of the mean separation length between regions, $\lambda$, following the definition of $\zeta$ in equations 136 and 137 in \citet{kruijssen18}, which we reproduce here: 

\begin{equation}
\label{eq:zeta_star}
\zeta_{\rm star} = \frac{G_{\rm FWHM, star}/\sqrt{2 \ln{2}} }{\lambda}
\end{equation}

\begin{equation}
\label{eq:zeta_gas}
\zeta_{\rm gas}  = \frac{G_{\rm FWHM, gas}/\sqrt{2 \ln{2}} }{\lambda}
\end{equation}

\noindent
where $\fwhm/2\sqrt{2 \ln{2}} = \rpeak$ from the original definition in \citet{kruijssen18}. This filing factor represents the amount of blending between peaks and is a key quantity in determining the applicability of the \code code. Testing performed by \citet{kruijssen18} shows that the quality of measurements obtained  declines as $\zeta$ increases above $0.5$. We will therefore assess the quality of the presented diffuse filtering method at different filling factors based on that criterion.

Lastly, we determine the number of pixels, \npix{x}, of the image to be generated, by using the total number of peaks in the image set, the mean separation length, $\lambda$, and the pixel length scale, \lpix{}:

\begin{equation}
\label{eq:pred_lambda}
\npix{x} =  \sqrt{\frac{\pi \npin{total}}{\lpix{}^2} \left(\frac{\lambda}{2}\right)^2  },
\end{equation}

\noindent
where $\npin{total} = \npin{gas} + \npin{star} + 2 \npin{over}$. As \npix{x} must be an integer number, this calculation introduces a small rounding error. We therefore recalculate the effective value of $\lambda$ from the definition of the geometric mean separation length for the total number of regions in the total area of a map \citep[appendix A2 of][]{KL14}, which in practice is an inversion of equation~(\ref{eq:pred_lambda}).

\begin{table*}
	\centering
	\begin{minipage}{\hsize}
		\centering
		\caption{A summary of the properties of the experiment sets used in  Section~\ref{sec:signal_loss} (\qeta{} points and  \qeta{} Gaussians) and in Section~\ref{sec:testing} (Main set 1 and Main set 2). For a full explanation of these properties see Section~\ref{sec:test_images}}.
		\label{tab:main_parameter_space}
		\label{tab:all_parameter_space}\vspace{-1mm}
		\begin{tabular}{l| l | l | l| l | l}
			\hline
			{} & \multicolumn{3}{l}{Value range} & {}  & {} \\
			Parameter & \qeta{} points & \qeta{} Gaussians & Main set 1 & Main set 2 & Origin  \\
			\hline
			$\zeta$ & N/A$^{\rm a}$ & $0.2-0.5$ & $0.2-0.7$ & $0.2-0.7$ & input  \\
			\fwhm & N/A$^{\rm a}$ & $63 - 135$ \pc $^{\rm a}$  & $45-99$ \pc $^{\rm b}$ & $45-99$ \pc $^{\rm b}$ & input  \\
			\npin{star} & $250$ & $100 - 400$  & $200$ & $0 - 120$ & input  \\
			\npin{gas} & $250$  & $100- 1000$  & $100 - 1200$  & $540- 1200$ & input  \\
			\npin{over} & $100$  & $100 - 200$ & $0 - 400$  & $120 - \npin{star}$ & input \\
			\fsignal{} & $12-100\%$ & $15-100\%$ & $10-100\%$ & $10-100\%$ & input \\
			\tstar{} & 10 \myr{} & 10 \myr{} & 10 \myr{} & 10 \myr{}  &  Equation~\ref{eq:tgas} \\
			\tgas{} & 10 \myr{} & $10 -50$ \myr{} & $5 - 60$ \myr{} & $45 -100$ \myr{}  &  Equation~\ref{eq:tgas} \\
			\tover{} & $2.9$ \myr{} & $2.9 - 5$ \myr{} & $0 - 10$ \myr{} & $0 - 10$ \myr{} & Equation~\ref{eq:tover}  \\
			$\lambda$ & $177$ \pc $^{\rm b,c}$  & $107- 573$ \pc $^{\rm b}$  & $55 - 420$ \pc $^{\rm b}$  & $55 - 420$ \pc $^{\rm b}$ & Equation~\ref{eq:zeta_star} \vspace{3mm} \\

		\end{tabular} \\
		$^{\rm a}$ The definitions of $\zeta$ and \fwhm{} do not apply to single pixel regions \\
		$^{\rm b}$  For all quantities in units of \pc{} we have applied a pixel length scale of $\lpix{} = 9 \pc{}$ \\
		$^{\rm c}$ For images containing point sources, we calculate lambda with equation \ref{eq:pred_lambda} by specifying the image size \npix{x} \\
	\end{minipage}
\end{table*}

\subsection{The background component}
\label{sec:backround_component}

We consider a number of different models of the diffuse background, with increasing levels of complexity:

\begin{enumerate}[leftmargin=*]
	\item Constant background: \\
	The simplest model that we consider is that of a spatially uniform constant diffuse background:

	\begin{equation}
		b(m,n) = \beta_{\rm const} ,
	\end{equation}
	\noindent
	with $\beta_{\rm const} \in \mathbb{R}  \ | \ \beta_{\rm const} > 0 $ selected such that the fraction of the diffuse background may vary in significance relative to the compact component.

	\item Extended Gaussian and constant background: \\
	This background model consists of both a constant background and a large `galaxy-scale' Gaussian function.
	\begin{equation}
	b(m,n) = \beta_{\rm const} + \beta_{\rm Gauss} \exp{\left( \frac{x^{2} + y^{2}}{ \beta_{\rm FWHM} /2 \sqrt{2 \ln{2}}  } \right)}  ,
	\end{equation}
	\noindent
	with the normalisation factors, $\beta_{\rm const} \in \mathbb{R}  \ | \ \beta_{\rm const} > 0 $ and $\beta_{\rm Gauss} \in \mathbb{R}  \ | \ \beta_{\rm Gauss} > 0 $ varying such that the two components may vary in significance in relation to each other and that the significance of the background component may vary relative to the compact component. The FWHM of the `galaxy-scale' Gaussian, $\beta_{\rm FWHM}$, is chosen such that $\beta_{\rm FWHM} \lesssim N_{\rm pix, x}$ so as to represent a large galactic background that decreases smoothly with radius.

	\item Envelopes, Extended Gaussian and constant background: \\
	In addition to the constant and galaxy scale Gaussian background, this background model includes Gaussian envelopes with centres co-spatial to the compact Gaussian regions seeded into the maps. The FWHM of the envelopes is set to be a multiple of the region separation length, $\lambda$, where we consider $ {\rm FWHM}_{\rm envelope} = 0.5 \lambda, 1.0 \lambda, 2.0 \lambda $. As with the previous models, the normalisation of each of the components is randomly chosen for each experiment such that the significance of each component varies and the significance of the total background varies in relation to the compact component.

\end{enumerate}

The true value of the compact emission fraction (\fsignaltrue) in an image $f(m,n)$ is calculated as the total flux in the compact component divided by the total flux in the image (including the background component):

\begin{equation}
 \fsignaltrue = 1 - \frac{\sum_{m,n} s(m,n)}{\sum_{m,n} \left(s(m,n) + b(m,n)\right)} .
\end{equation}

\noindent
Overlap between regions causes flux loss from compact regions after the application of Fourier filtering.
In Section~\ref{sec:signal_loss}, we use the method presented here to generate simulated datasets to investigate this effect. In Section~\ref{sec:testing}, we again use the method to generate datasets to evaluate the performance of the method at removing diffuse emission. A summary of the parameter spaces used for each each section is presented in Table~\ref{tab:parameters}.

\section{The impact of filtering on signal regions}
\label{sec:signal_loss}

The value of \fsignal~given by equation~\ref{eq:signal_fraction} assumes that filtering removes none of the flux of the compact regions. However, the flux from a Gaussian region is spread out in Fourier space, such that the application of a filter in Fourier space removes some of the flux from compact regions. This flux loss should be quantified, so that it can be corrected for.

\subsection{Flux loss for a single Gaussian region}
\label{sec:qcon}

We define the fraction of flux remaining in an image containing a single Gaussian function after filtering as:

\begin{equation}
\qcon{} =  \frac{f'(m,n)}{f(m,n)} .
\end{equation}

\noindent
Figure~\ref{fig:single_gauss_part1} shows \qcon{} after the application of Gaussian filters, where $D_{\rm crit}$ is set with variable values of $\nlambda \lambda$. For the smallest, $\fwhm = 1$ pixel region, which we do not show, we do not see a smooth function between the flux fraction and critical filter length-scale, as the function is insufficiently resolved.

In the right-hand panel of Figure~\ref{fig:single_gauss_part1}, we see that \qcon{} can be well fitted by a simple analytical function of the relative width of the Gaussian filter over the width of the compact regions, $\nlambda \lambda / \fwhm$, for all the Gaussian functions with  $\fwhm \geqslant 2$ pixels.

\newcommand{\cwidth}{0.49\textwidth}

\begin{figure*}
	\centering
	\subfloat{\includegraphics[width=\cwidth]{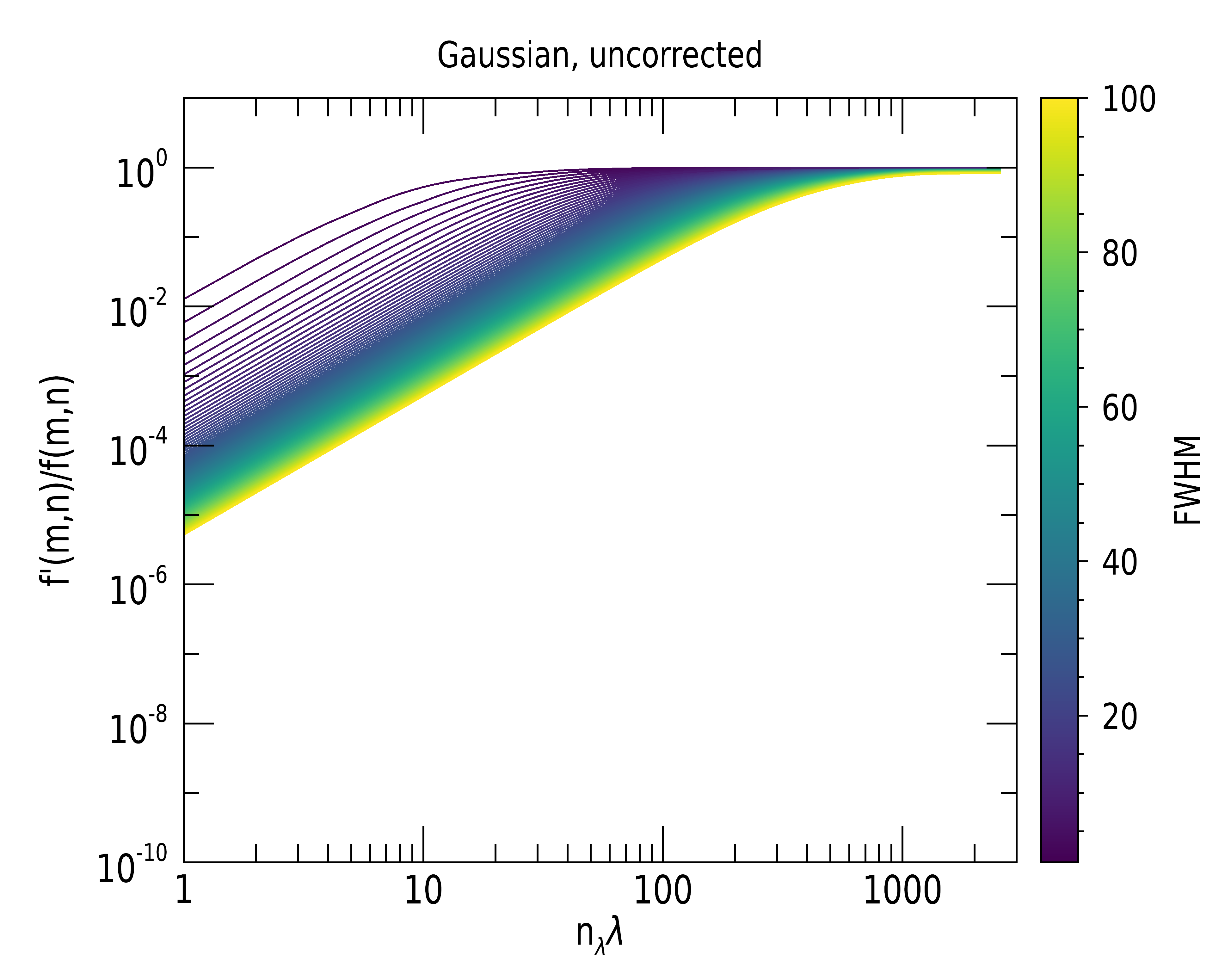}}
	\subfloat{\includegraphics[width=\cwidth]{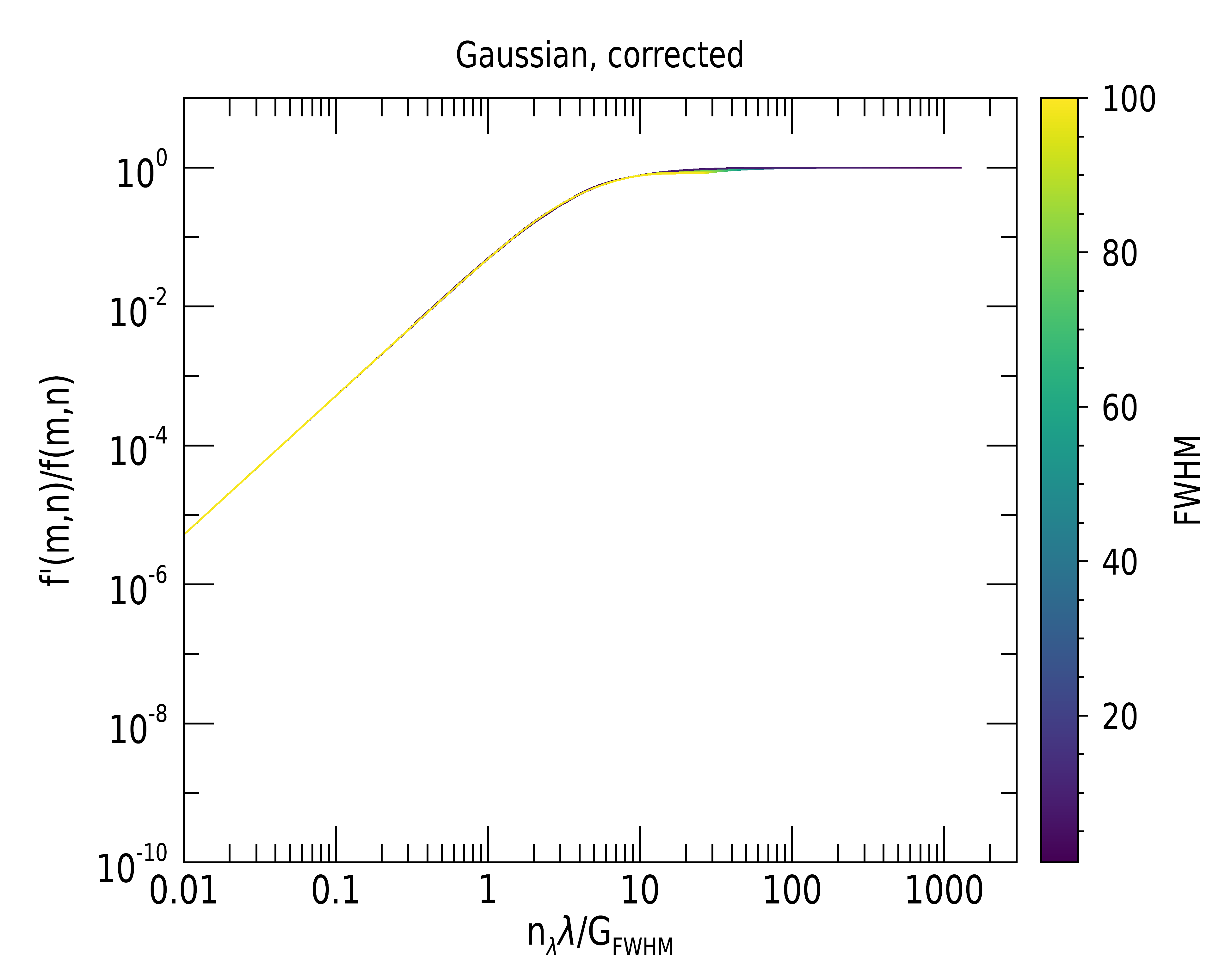}} \\

	\caption{The fraction of flux remaining in an image containing a single Gaussian function of varying FWHM (as indicated by the colour bar) after the application of a  Gaussian filter. Left panel: the remaining flux fraction as a function of $\nlambda{} \lambda $. Right panel: the remaining flux as a function of $\nlambda{} \lambda / G_{\rm FWHM}$, i.e. where the x-axis has been normalised by the FWHM of the Gaussian function. After application of this correction we recover a simple analytic relationship for all the images where $\fwhm \geqslant 2 $. }

	\label{fig:single_gauss_part1}

\end{figure*}

\begin{figure*}
	\centering
	\subfloat{\includegraphics[width=\cwidth]{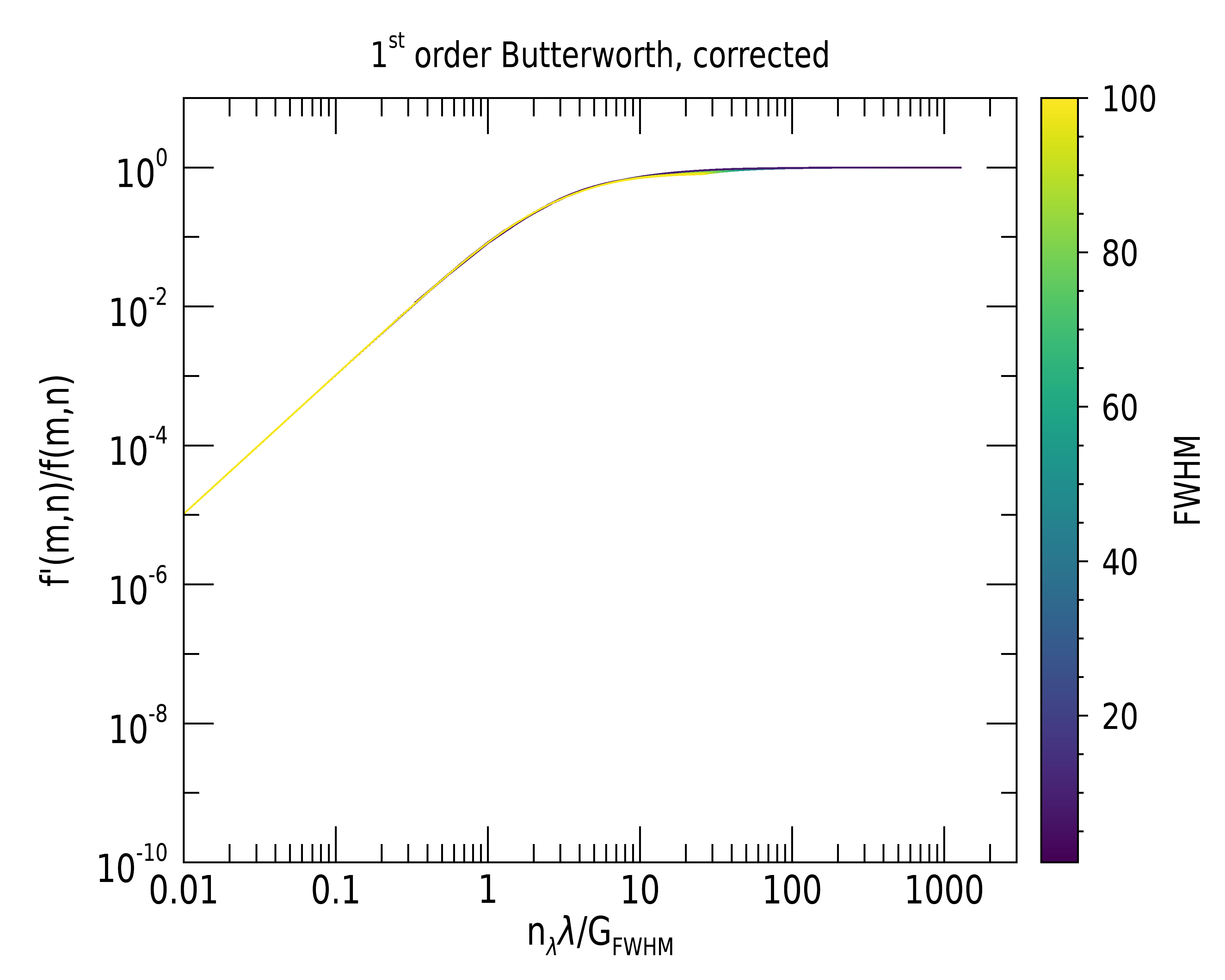}}
	\subfloat{\includegraphics[width=\cwidth]{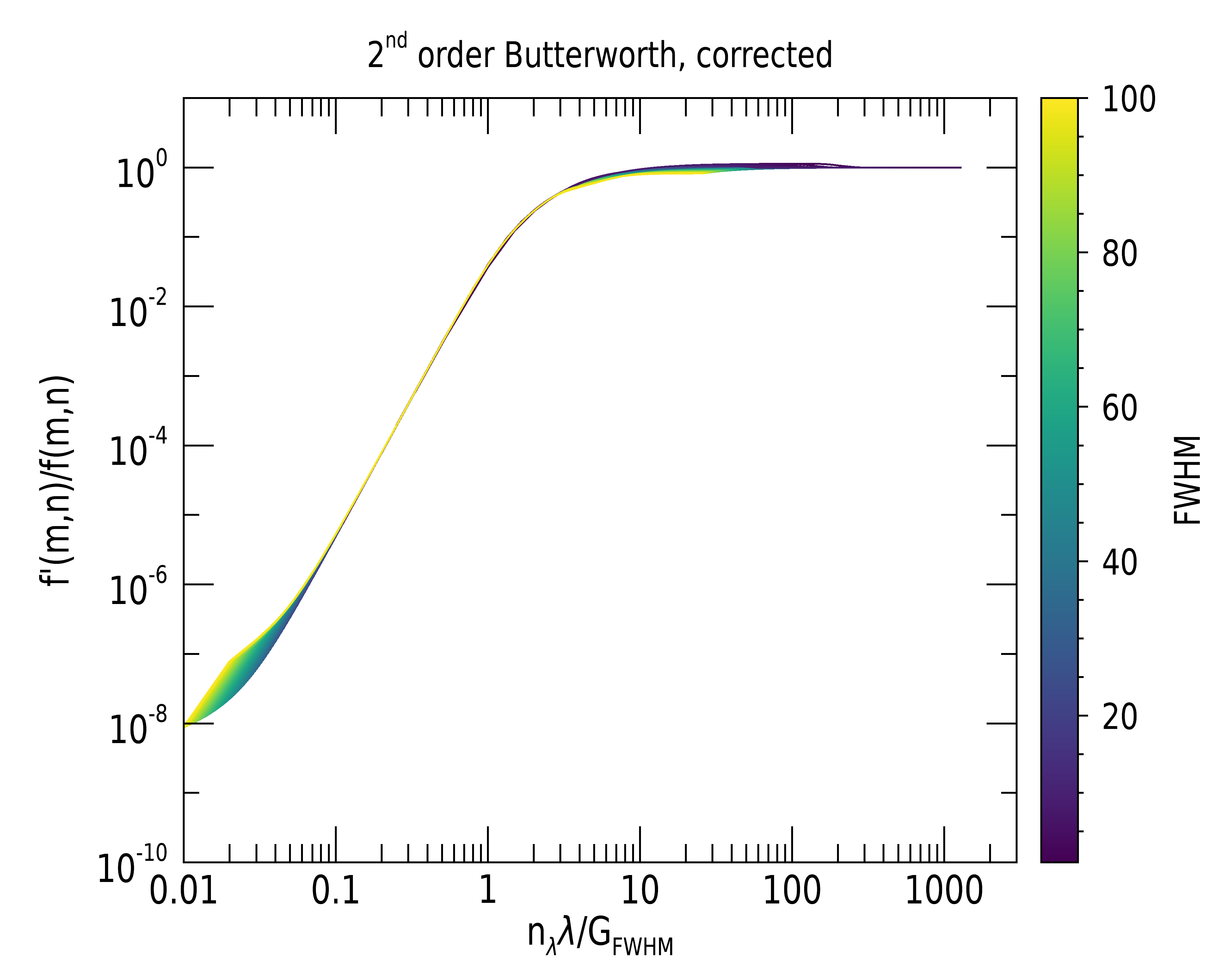}} \\
	\centering
	\subfloat{\includegraphics[width=\cwidth]{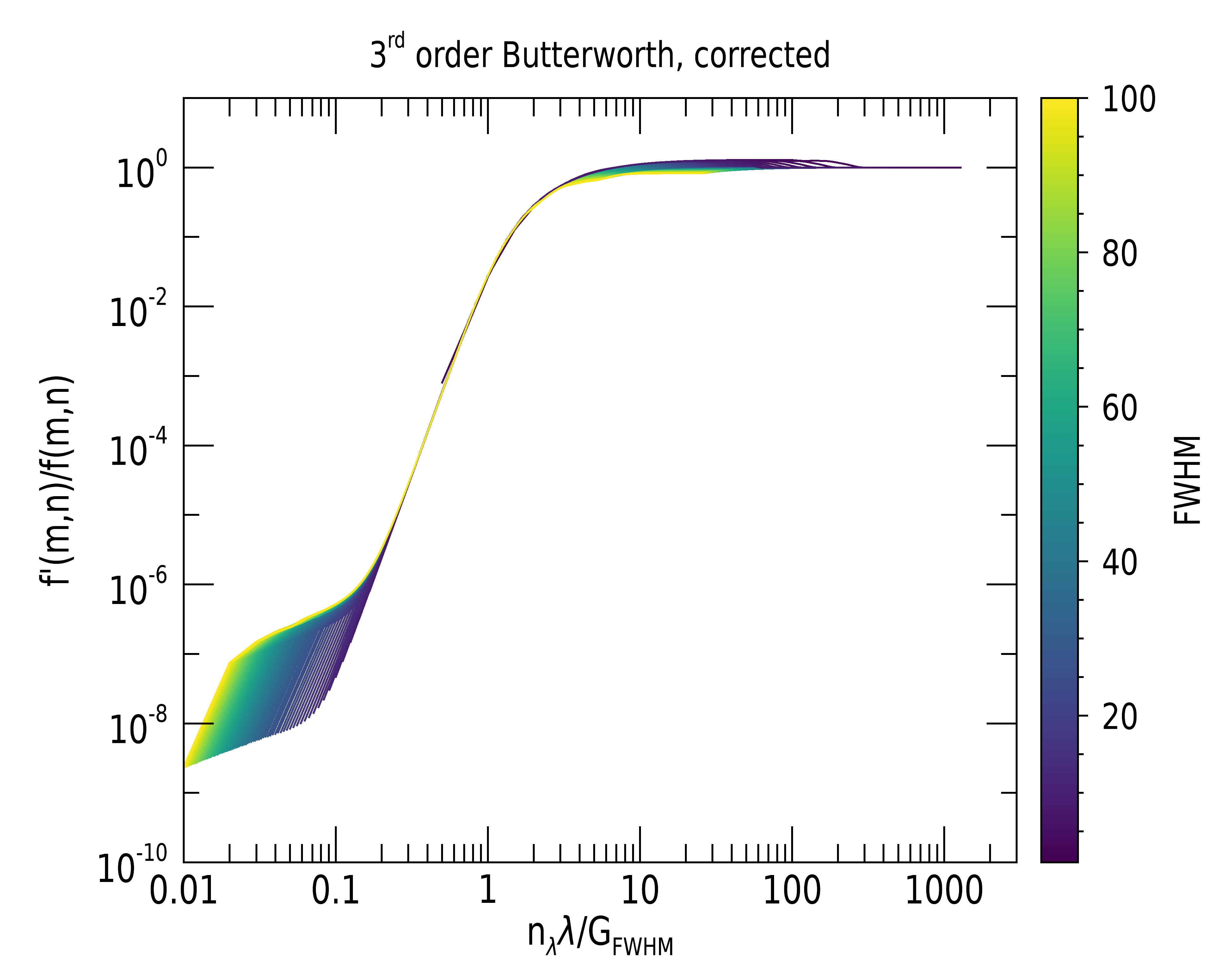}}
	\subfloat{\includegraphics[width=\cwidth]{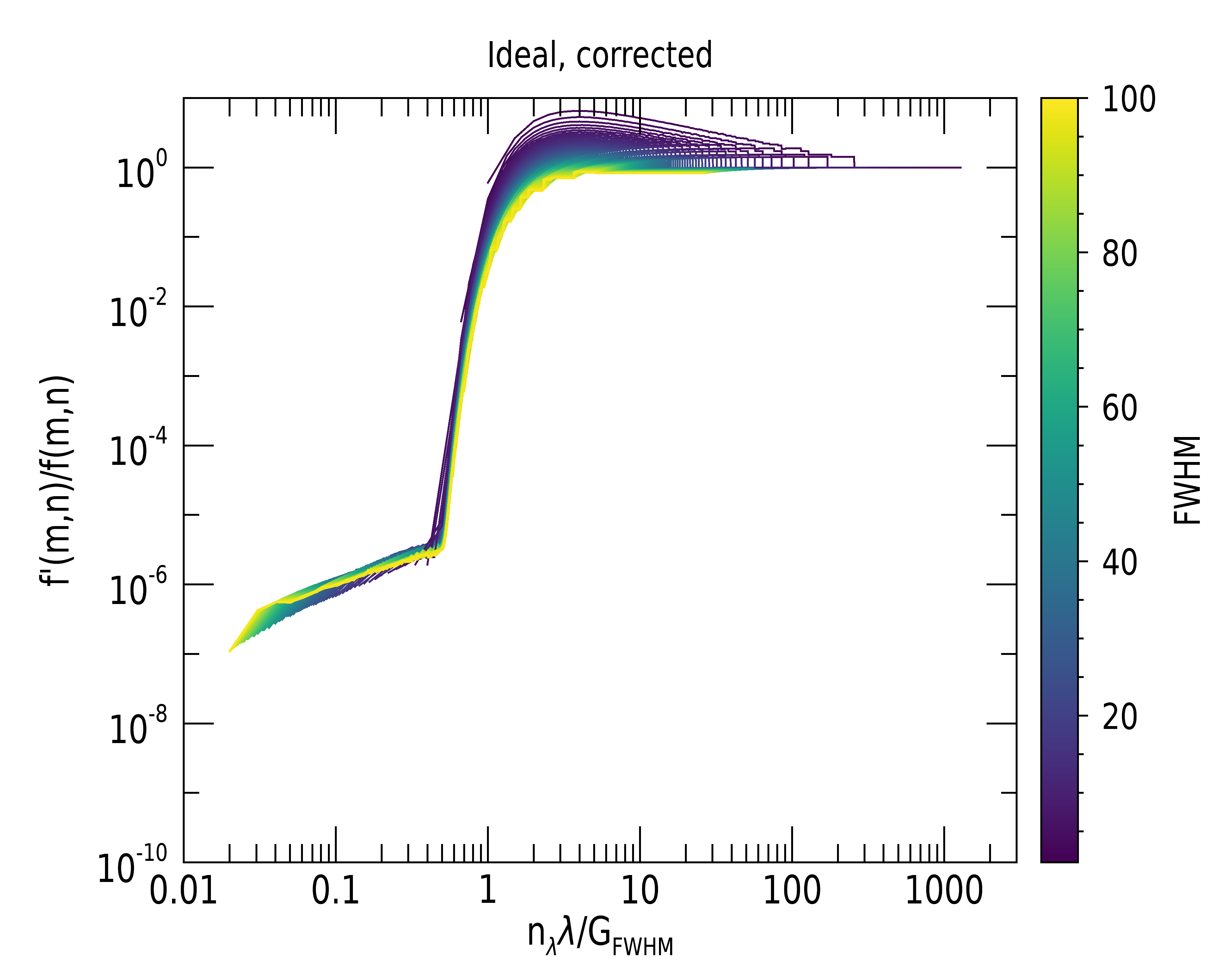}} \\

	\caption{The fraction of flux remaining in an image containing a single Gaussian function of varying FWHM (as indicated by the colour bar) as a function of $\nlambda{} \lambda / G_{\rm FWHM}$ after the application of a \nth{1} order Butterworth filter (top left), a \nth{2} order Butterworth filter (top right), a \nth{3} order Butterworth filter (bottom left) and an Ideal filter (bottom right). As with the Gaussian filter, the remaining flux fraction for the \nth{1} order Butterworth filter is strictly less than or equal to unity and well fitted with a simple analytic function. By contrast, the remaining flux after the application of an Ideal filter significantly exceeds unity in some cases, due to the significant distortions introduced into the image. Butterworth filters with order greater than one also lead to remaining flux fractions greater than unity with the magnitude of the excess increasing with order.}

	\label{fig:single_gauss_part2}

\end{figure*}

\begin{figure}
	\centering
	\subfloat{\includegraphics[width=\columnwidth]{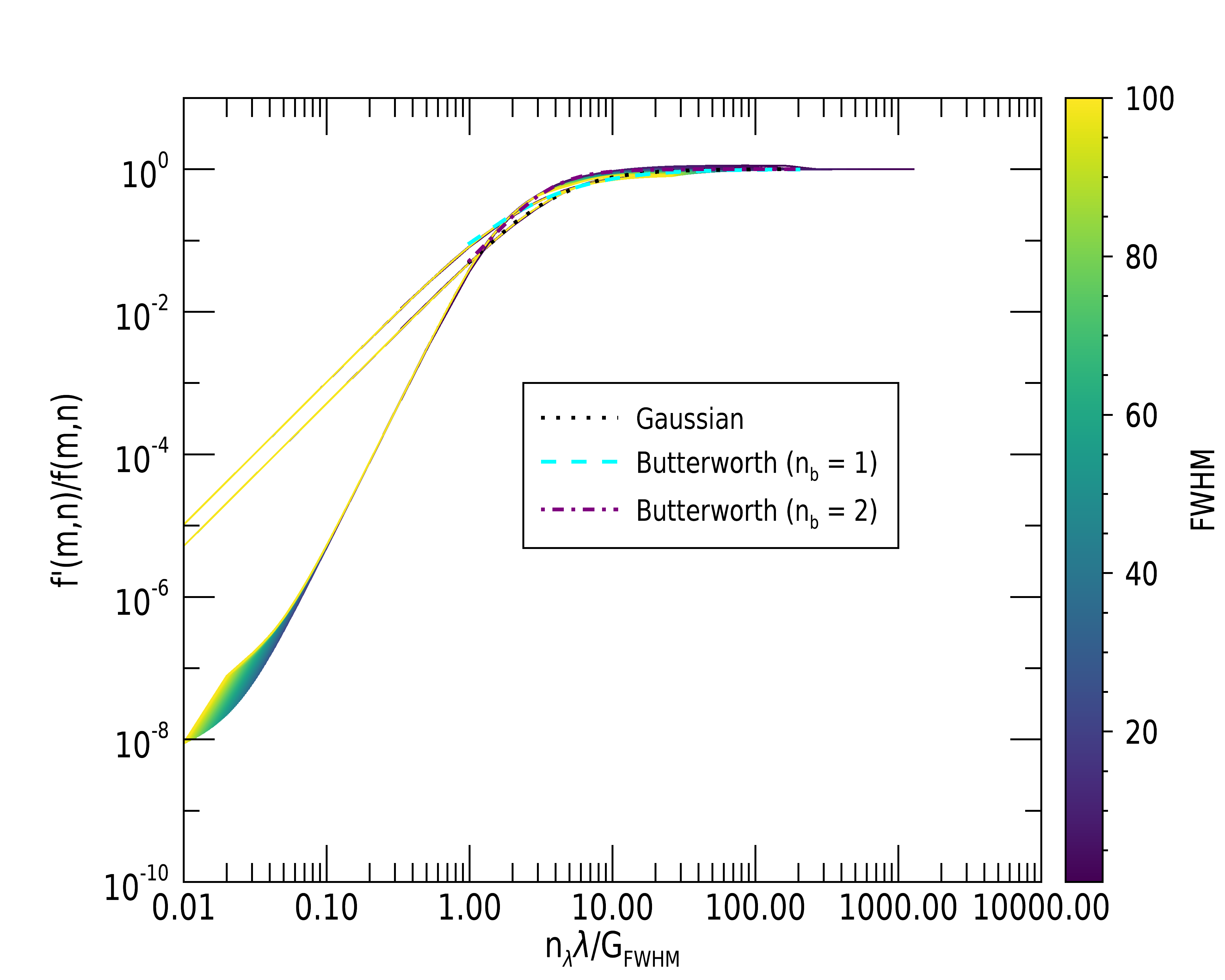}}

	\caption{A comparison of the fitted remaining flux fraction, $\qcon{} = f'(m,n)/f(m,n)$, after the application of a Gaussian filter and a \nth{1} and \nth{2} order Butterworth filter as a function of the ratio between the cut length, $\nlambda \lambda$, over the region size, $\fwhm$, from experimental datasets (solid coloured lines) and from fits to this data with a functional form shown in Equation~\ref{eq:qcon} (dotted and dashed lines). It can be seen that the \nth{2} order Butterworth filter offers the steepest response curve of those considered, followed by the Gaussian and \nth{1} order Butterworth respectively.}
	\label{fig:gaussian_gcon_fit}
\end{figure}

\begin{figure}

	\centering
	\subfloat{\includegraphics[width=\columnwidth]{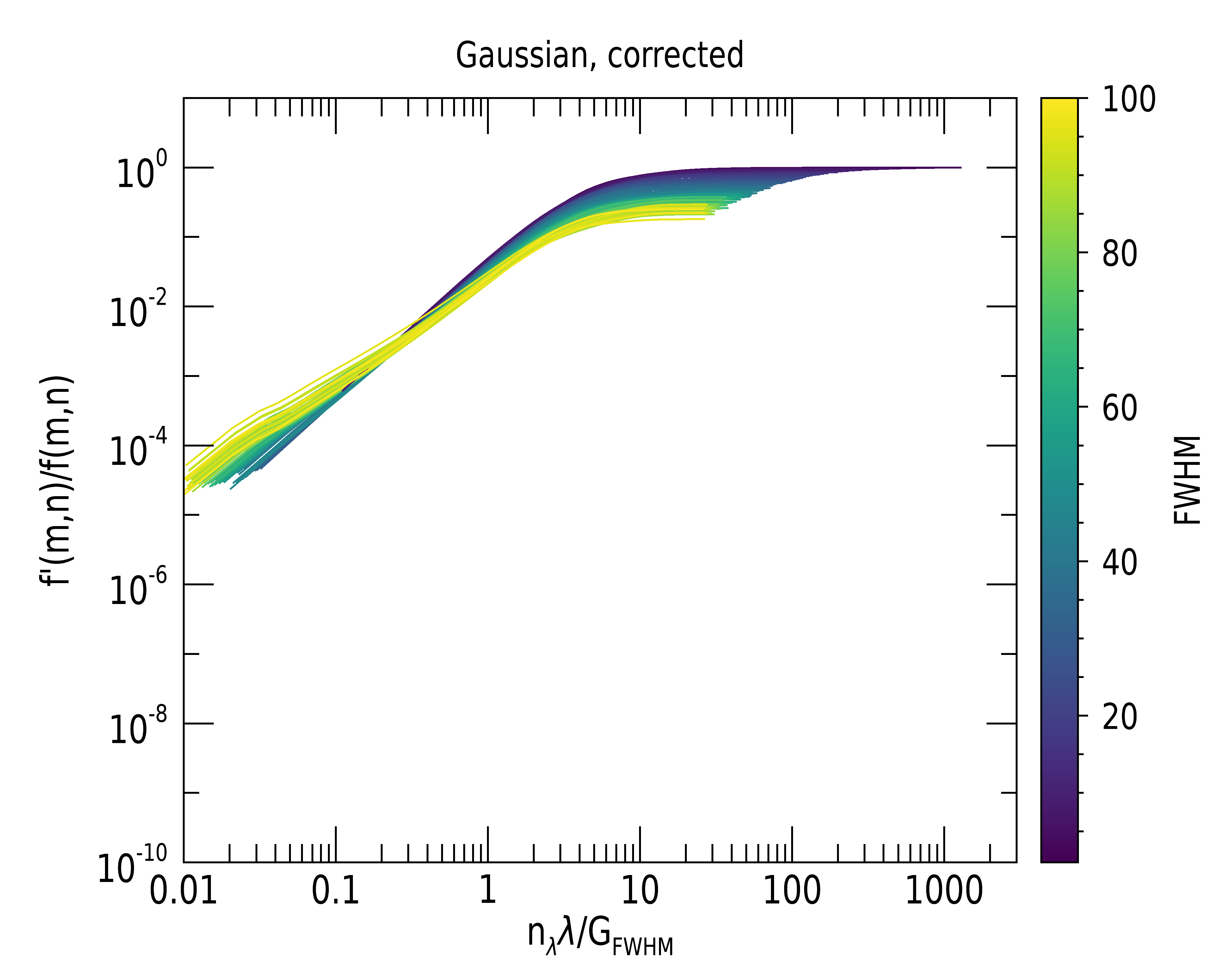}}  \\
	\subfloat{\includegraphics[width=\columnwidth]{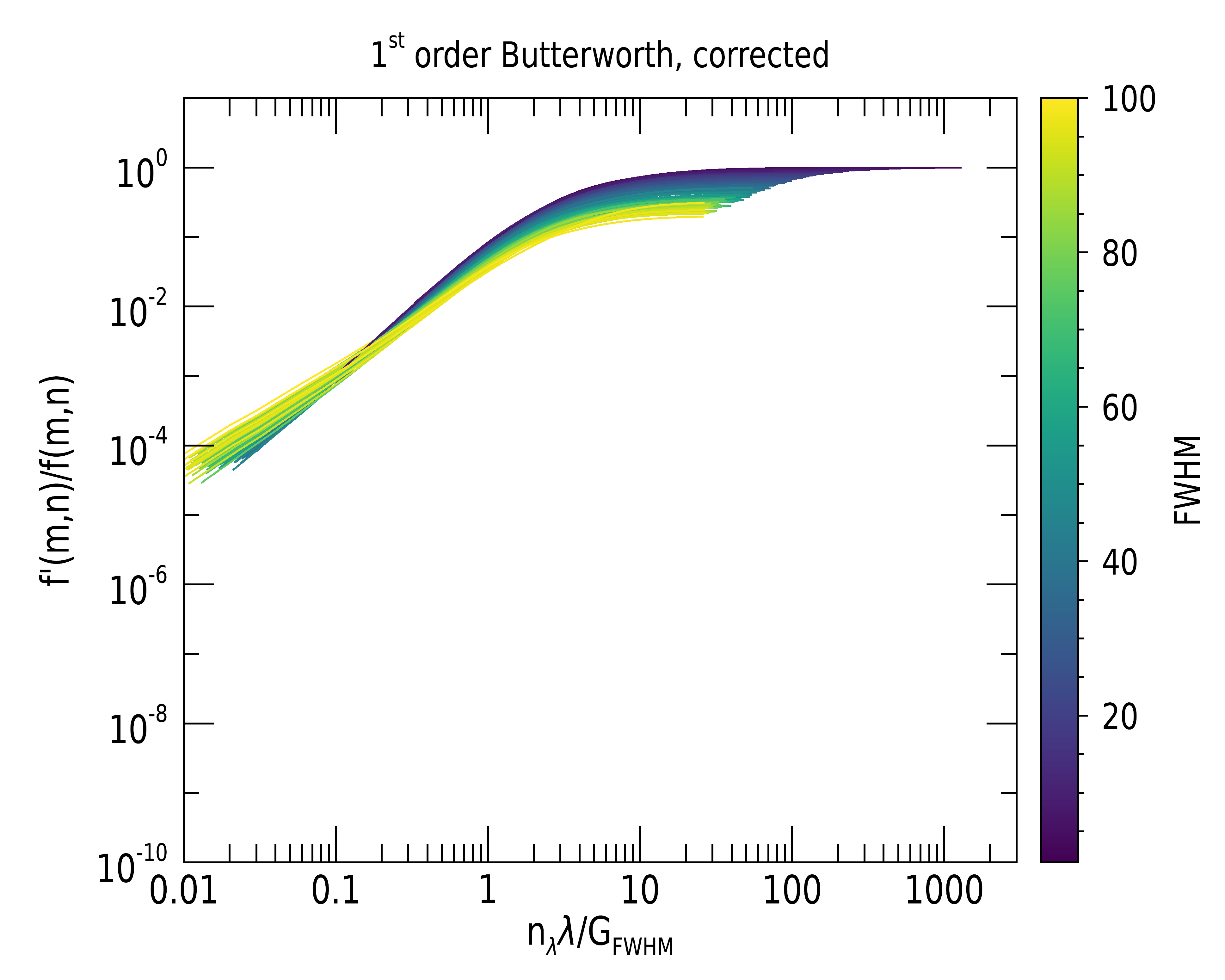}} \\

	\subfloat{\includegraphics[width=\columnwidth]{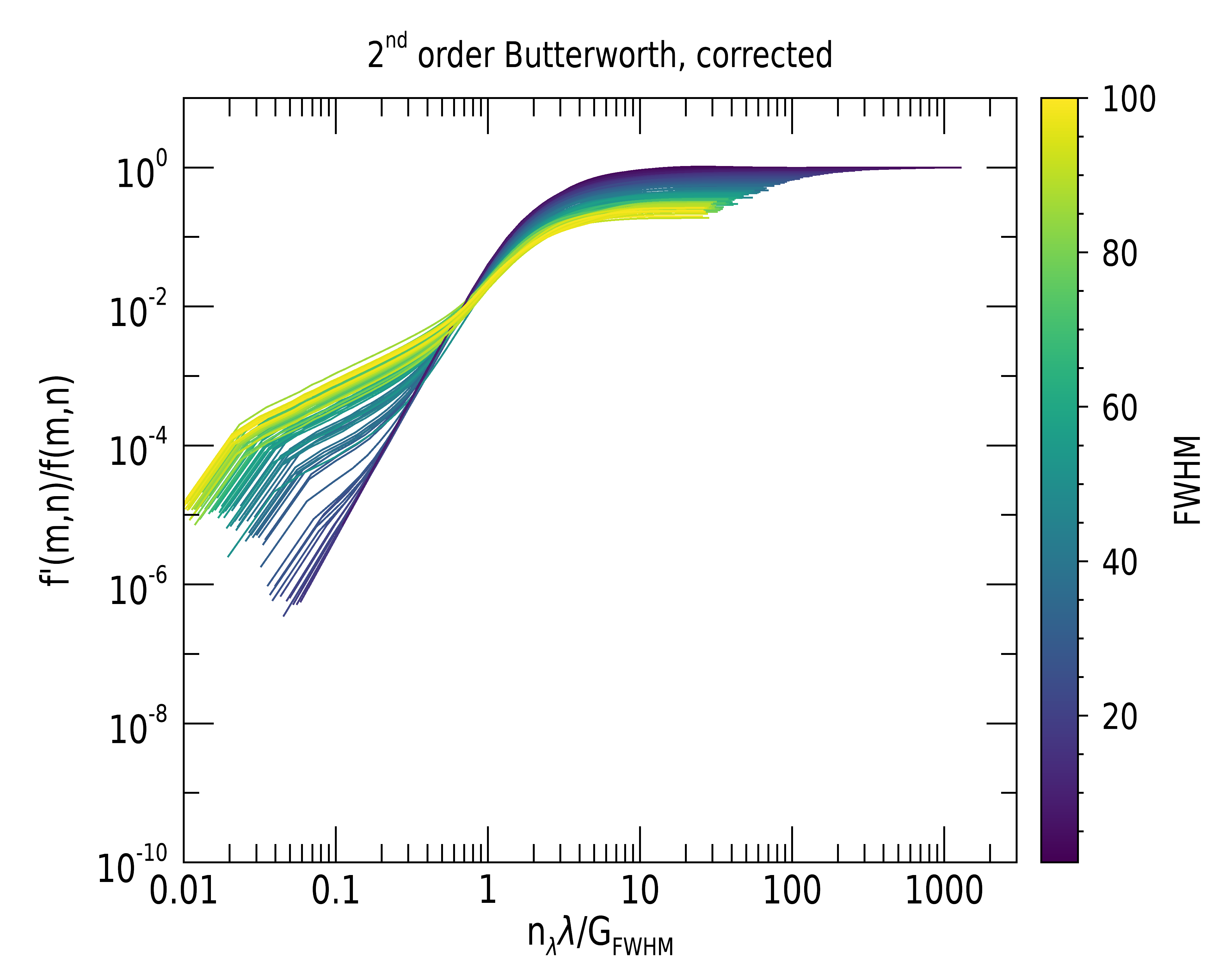}}
	\caption{The fraction of flux remaining in an image containing multiple Gaussian functions of uniform FWHM after the application of a Gaussian filter (top), a \nth{1} order Butterworth filter (middle) a \nth{2} order Butterworth filter (bottom) against $\nlambda{} \lambda / \fwhm{}$. In contrast to the case with only a single Gaussian region, shown in Figures~\ref{fig:single_gauss_part1} and ~\ref{fig:single_gauss_part2}, where tight relationships are obtained, there is still a significant trend of the remaining flux with FWHM, due to the increasing amount of overlap between regions with increasing region size.}

	\label{fig:multi_gauss_part1}

\end{figure}

By again plotting \qcon{} against $\nlambda \lambda / \fwhm$ for the other considered filters, the Ideal filter and Butterworth filters, we obtain relationships of varying quality, which are shown in Figure~\ref{fig:single_gauss_part2}. We again find the smallest ($\fwhm = 1$ pixel) Gaussian functions to be insufficiently resolved and thus we impose the condition that the FWHM of the Gaussian regions in the image $\fwhm \geqslant 2$ pixels for the successful application of the method. While the remaining flux fraction is strictly less than or equal to unity after the application of the Gaussian filter, it is above one (even up to $\sim 5$) in some cases after the application of the Ideal filter, due to the significant distortions introduced into the image. For this reason, we discard the Ideal filter from further consideration. The Butterworth filter is tunable, where at low values of $n_{b}$ the filter is smooth, akin to the Gaussian filter, but becomes sharper, approaching an Ideal filter as $n_{b} \to \infty$. For the \nth{1} order Butterworth filter,  we again obtain a simple analytical function, where $\qcon \leqslant 1$. For $n_{b} \geq 2$, however, we see that \qcon~is greater than 1 in some instances, with the magnitude of the effect greater as  $n_{b}$ increases.

We fit a sigmoidal function for $q_{con}$, as a function of $\nlambda{} \lambda / \fwhm $:

\begin{equation}
\label{eq:qcon}
q_{\rm con} = \qconinfty + \frac{\left(a - \qconinfty \right)}{\left(1 + \left(\frac{\nlambda \lambda}{\fwhm}/c\right)^b\right)} ,
\end{equation}

\noindent
where we set $\qconinfty=1$ for all filters, so that as $\nlambda{} \lambda / \fwhm \to \infty $, $q_{\rm con} \to 1 $, as this represents 0\% flux loss from the Gaussian region in the image. The value of $a$ sets the zero point of the function, such that $\qcon = a$ for $\nlambda{} \lambda/G_{\rm FWHM} =0$. The value of $b$ sets the steepness of the transition between $\qcon = \qconinfty$ and $\qcon = a$, with higher values of $b$ representing a steeper transition. The value of $c$ is the value of $\nlambda{} \lambda / \fwhm$ at which $\qcon = a + \left(\left(a-1\right)/2\right) $. The best fitting values of these parameters are shown in Table~\ref{tab:qcon_table} and a comparison of the fitted functions is shown in Figure~\ref{fig:gaussian_gcon_fit}.

We first compare the Gaussian and 1st order Butterworth filters, which both have the property that the flux remaining after their application in the single Gaussian region is strictly less than or equal to 100\% of the original flux. Between the two, the Gaussian filter offers a steeper response curve that removes less flux from the compact regions than the \nth{1} order Butterworth filter. Butterworth filters of \nth{2} order and greater offer a significantly steeper response curve than the Gaussian filter and thus remove less flux from the compact regions than the Gaussian filter. However, due to introduced distortions, application of the filter does not necessarily result in an image that has $ \leqslant 100\% $ of the flux of the original image. As the Gaussian filter shows few distortions we chose to focus on this filter for the remainder of our analysis.

We can now correct the measured value of \fsignal{} for the lost emission from a single Gaussian region as follows:

\begin{equation}
\label{eq:signal_fraction_qcon}
\fsignal{} = \frac{1}{\qcon} \frac{\sum_{m,n} f'(m,n)}{\sum_{m,n} f(m,n)}  .
\end{equation}

\newlength\qconlength
\setlength\qconlength{\dimexpr.25\columnwidth-2\tabcolsep-0.5\arrayrulewidth\relax}

\begin{table}
	\centering
	\begin{minipage}{\columnwidth}
	\centering
	\caption{The best fitting parameters for the sigmoidal function the flux loss from a single Gaussian region due to application of a filter in Fourier space, \qcon{} (see equation~\ref{eq:qcon}), fitted to the response curves of the Gaussian and \nth{1} and \nth{2} order Butterworth filters (shown in Figures~\ref{fig:single_gauss_part1}~and~\ref{fig:single_gauss_part2}).}
	\label{tab:qcon_table}
	\begin{tabular}{l|l|l|l}
		\hline
		Parameter & Gaussian & \nth{1} order BW & \nth{2} order BW \\
		\hline
		$a$ & $-0.016$ & $-0.038$ & $0.0019$  \\
		$b$ & $1.69$ & $1.30$   &  $2.38$ \\
		$c$ & $4.86$ & $4.45$ & $3.43$ \\
		\hline
	\end{tabular} \\
	BW = Butterworth\vspace{-1mm}
	\end{minipage}
\end{table}

\subsection{Flux loss for a set of overlapping Gaussian regions}
\label{sec:qeta}

Astronomical images do not consist of only single isolated Gaussian functions. Instead, there are a number of overlapping regions distributed within each image. As Figure~\ref{fig:multi_gauss_part1} demonstrates, in the case of an image made up of a field of multiple Gaussian functions, we no longer recover a simple analytical relationship between the remaining flux fraction, \qcon, and $\nlambda{} \lambda / \fwhm{}$. We can characterise the effect of overlap between regions in terms of in terms of the the \etaname{}, $\eta$, which we define as:

\begin{equation}
\label{eq:eta_star}
\etastar = \sqrt{\frac{\tstar}{\tau}} \zeta_{\rm star} = \sqrt{\frac{\tstar}{\tau}}~\frac{G_{\rm FWHM, star}/\sqrt{2 \ln{2}} }{\lambda},
\end{equation}

\noindent
for the stellar map, and

\begin{equation}
\label{eq:eta_gas}
\etagas = \sqrt{\frac{\tgas}{\tau}} \zeta_{\rm gas} =  \sqrt{\frac{\tgas}{\tau}}~\frac{G_{\rm FWHM, gas}/\sqrt{2 \ln{2}} }{\lambda},
\end{equation}

\noindent
for the gas map, where $\tau = \tstar + \tgas- \tover$ is the total duration of evolutionary timeline. Thus, $\eta$ is given by the global filling factor, $\zeta$,\footnote{For the definition of this filling factor, see Equations~\ref{eq:zeta_star}~and~\ref{eq:zeta_gas}.} multiplied by $\sqrt{\tstar / \tau}$ and $\sqrt{\tgas / \tau}$ for the stellar and gas maps, respectively, in order to weight the global filling factor by the number of regions in each of the two tracer maps, as expected from the evolutionary timeline.

\begin{figure}
	\centering
	\subfloat{\includegraphics[width=\columnwidth]{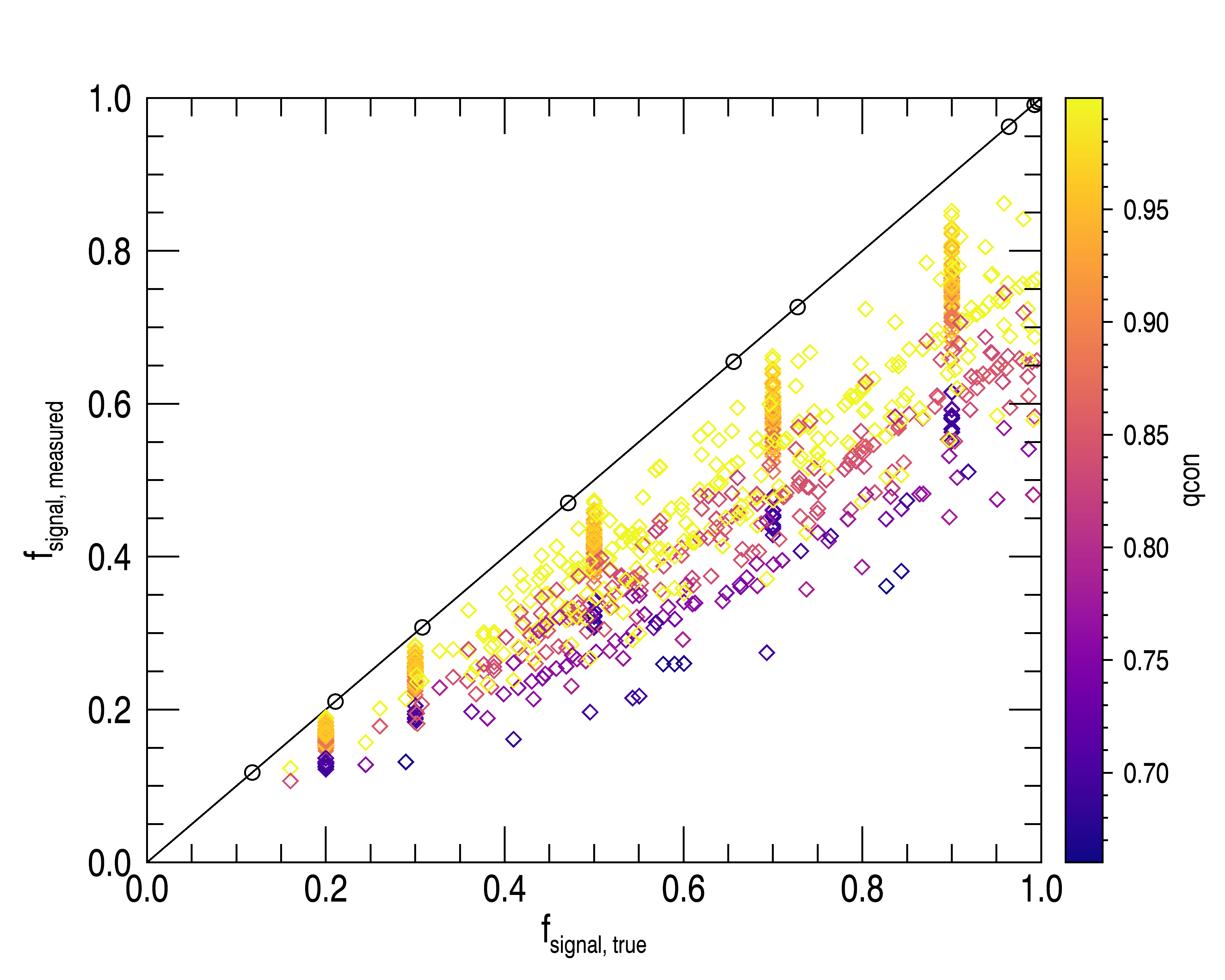}}

	\caption{The value of the fraction of emission in an image that is  compact, \fsignal{}, measured with the application of a Gaussian filter, against the true value for a number of images. The solid black line shows the 1:1 relationship between these two quantities. The black circles show the results from generated map sets where compact regions are simulated with single-pixel points. In this case, there is no effect on \fsignalmeasured{} from increasing amounts of the wings of compact regions overlapping with increasing $\eta$ or of flux loss from spatially extended regions and we recover \fsignaltrue{} without needing to apply any corrections to the measured result. The coloured diamonds show results from generated map sets where compact regions are simulated with Gaussian Functions. The diamonds are coloured by the calculated value of \qcon, as calculated from equation~\ref{eq:qcon} with the appropriate values of $\nlambda \lambda$ and \fwhm.}

	\label{fig:qcon_uncorrected_plot}

\end{figure}

\begin{figure}
	\centering
	\subfloat{\includegraphics[width=\columnwidth]{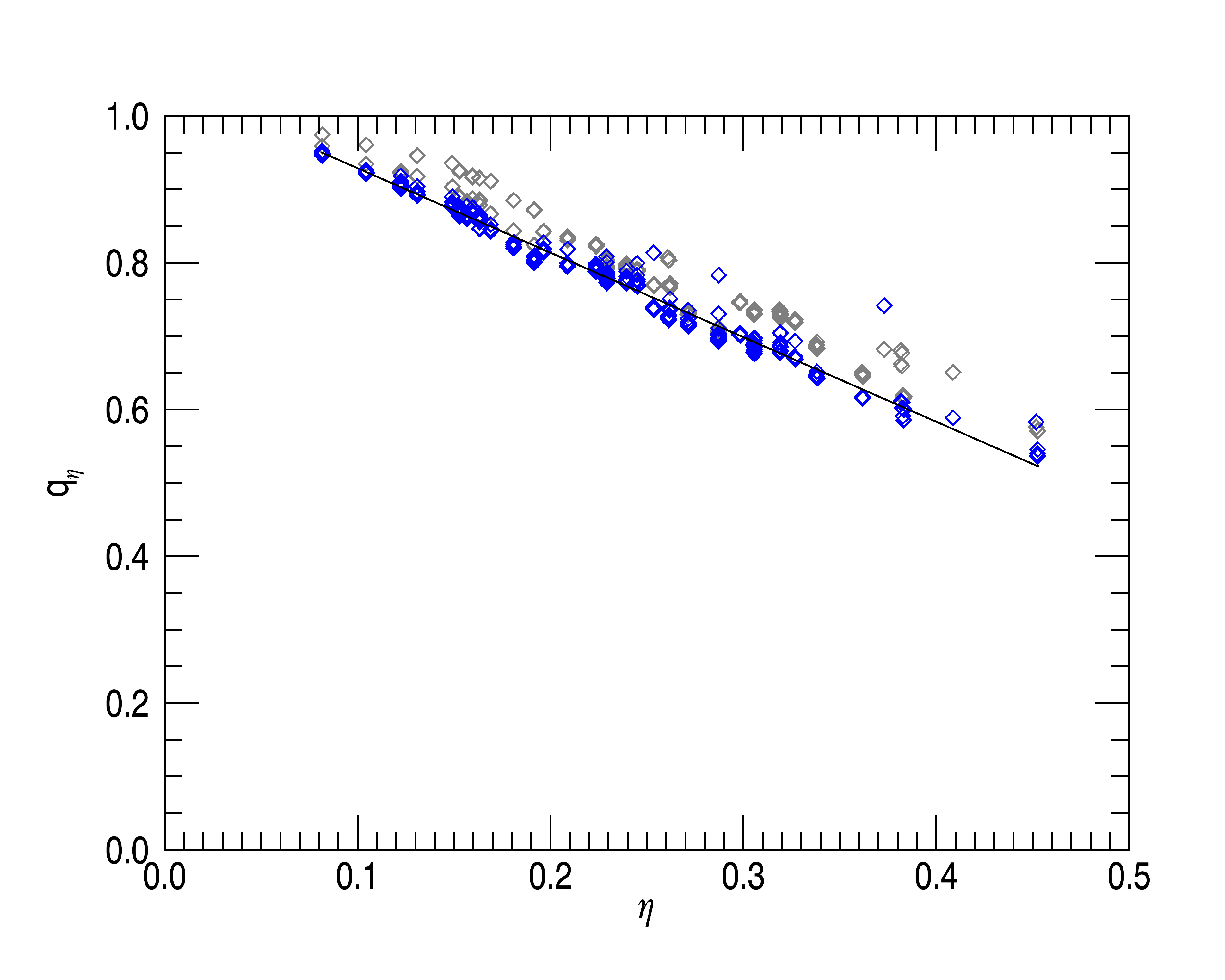}}

	\caption{The ratio of \fsignalmeasured{} to \fsignaltrue~for the same set of experiments as in Figure~\ref{fig:qcon_uncorrected_plot} after the application of the correction factor,\qcon{}, against  the tracer map filling factor,$\eta$. Blue diamonds correspond to experiments that satisfy the criterion $\qcon \geqslant 0.9 $, whereas grey points do not. The solid black line shows the best-fitting line to the blue points, which is used to calibrate the correction factor \qeta. }

	\label{fig:qcon_qzeta_plot}

\end{figure}

\begin{figure}
	\centering
	\subfloat{\includegraphics[width=\columnwidth]{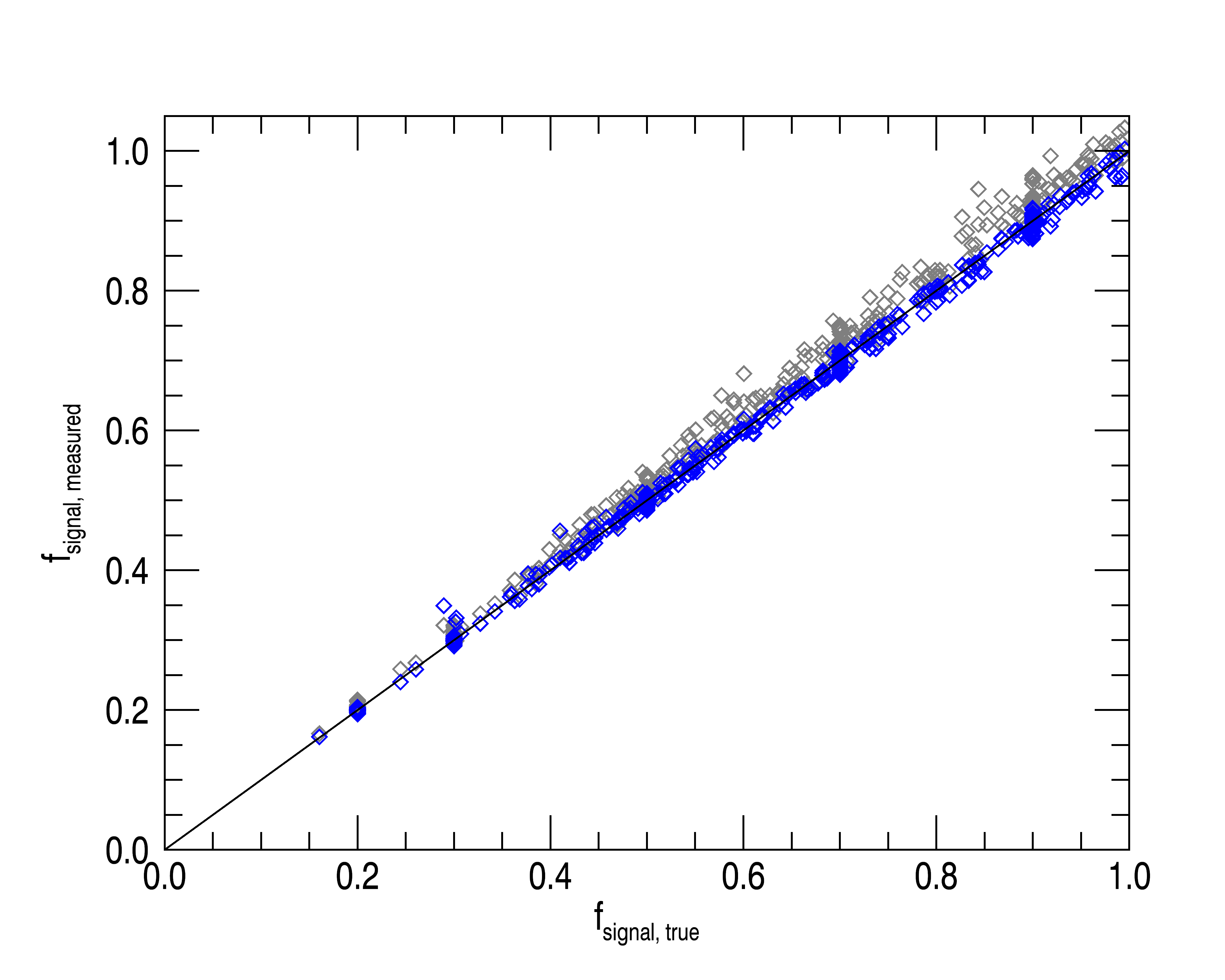}}

	\caption{The measured compact emission fraction, \fsignal{}, for the same set of experiments as in Figure~\ref{fig:qcon_uncorrected_plot} after the application of the correction factors \qcon{} and \qeta~against the true value of \fsignal. Blue diamonds correspond to experiments that satisfy the criterion $\qcon{} \geqslant 0.9 $, whereas grey points do not. The solid black line indicates the 1:1 relationship between \fsignalmeasured{} and \fsignaltrue. It can be seen that \fsignalmeasured{} is well-correlated with \fsignaltrue{} after correction by \qcon{} and \qeta{}, with the correlation somewhat worse for experiments where $\qcon{} \geqslant 0.9 $ (grey diamonds).}

	\label{fig:qcon_final_plot}

\end{figure}

We generate a set of simulated datasets with single pixel compact regions and a constant background to serve as a control dataset (`\qeta{} points'). As point sources are not extended in space, we expect no flux loss from the compact regions after the application of a filter in Fourier space (i.e. $\qeta=1$ and $\qcon{} = 1$). For Gaussian compact regions (`\qeta{} Gaussians'), we do expect flux loss due to overlapping regions. For all of these datasets, we set $\tstar = 10\, \myr{}$. Detailed testing of the \code by \citet{kruijssen18} shows that the condition $\zeta < 0.5$ should be satisfied for accurate measurements to be obtained from the code and that the \code code produces better results the lower the value of $|\log_{10}(\tstar / \tgas) |$ is. As we will use this dataset to fit for an empirical relationship between \qeta{} and $\eta$, we ensure good performance by considering a restricted range of \tgas{}, i.e. 10-50 \myr{} and a range of values of $\zeta$ up to the maximum recommended value. (i.e. $\zeta = 0.2, 0.3, 0.4, 0.5$). A summary of the parameters used to generate these datasets is given in Table~\ref{tab:all_parameter_space}. The value of \fsignal{} is randomly selected to be between 15\% and 100\% of the total flux in each image. For a full description of the method used to generate these images see Section~\ref{sec:test_images}.

Figure~\ref{fig:qcon_uncorrected_plot} shows the measured value of the image compact emission fraction against the true compact emission fraction, measured according to equation~\ref{eq:signal_fraction} with the application of Gaussian filters with varying values of $\nlambda \lambda$. This figure illustrates how, for images where the compact regions are single pixels, the true value of the compact emission fraction is recovered with equation~\ref{eq:signal_fraction}. However, for those images where the compact regions are extended Gaussians, the true value is not accurately recovered.

We correct these measurements for the flux loss from a single Gaussian, \qcon{}. The resulting measurements of \fsignal{} (according to Equation~\ref{eq:signal_fraction_qcon}) are shown in Figure~\ref{fig:qcon_qzeta_plot} as a function of $\eta$. It can be seen that, while the spread in measured values has been greatly reduced, we do not yet recover the true values of \fsignal{} and that $\fsignalmeasured{} / \fsignaltrue$ decreases with increasing $\eta$.

We then calibrate an empirical relationship between $\eta$ and the ratio of the measured flux fraction to the true flux fraction, after the flux fraction has been corrected for the flux lost from a single Gaussian region. We adopt a critical value of $\qcon{} = 0.9$, because, as can be seen in the right hand panel of Figure~\ref{fig:single_gauss_part1}, a small uncertainty in the measurement of $\nlambda{} \lambda / \fwhm{}$ leads to a small uncertainty in the measured value of \qcon{} above this threshold. By contrast, a small measurement uncertainty in $\nlambda{} \lambda / \fwhm{}$ below this threshold leads to a large uncertainty in \qcon{}, making this correction unreliable. Indeed, the experiments with $q_{\rm con}>0.9$ display a clear relation with $\eta$ in Figure~\ref{fig:qcon_qzeta_plot}, because they represent cases for which a small fraction of the compact emission is filtered out. We fit a linear relationship between the flux loss due to overlap, \qeta, and the evolutionary timeline normalised filling factor, $\eta$, truncated such that \qeta~may never be greater than unity:

\begin{equation}
\label{eq:qeta}
q_{\eta} = \min(A \eta + B, 1) .
\end{equation}

\noindent
The best-fitting slope and intercept for the Gaussian filter is listed in Table~\ref{tab:qeta_table} and is plotted in Figure~\ref{fig:qcon_qzeta_plot}.

\begin{table}
	\centering
	\caption{The best fitting parameters for the fitting function for  flux-loss in compact regions due to overlap between regions for the Gaussian filter, \qeta{} (see equation~\ref{eq:qeta})}
	\label{tab:qeta_table}
	\begin{tabular}{l|l}
		\hline
		Parameter & Gaussian \\
		\hline
		A & $-1.15 $  \\
		B & $1.04$ \\
		\hline
	\end{tabular}
\end{table}

Thus applying this empirical correction in addition to the correction for the flux lost from a single Gaussian region, the final measurement of the  compact emission fraction, \fsignal, in an image, $f(m,n)$ is:

\begin{equation}
\label{eq:signal_fraction_qeta}
\fsignal{} = \frac{1}{q_{\rm con} q_{\eta}}  \frac{\sum_{m,n} f'(m,n)}{\sum_{m,n} f(m,n)} .
\end{equation}

\noindent
The original measurements shown in Figure~\ref{fig:qcon_uncorrected_plot} are shown with these corrections applied in Figure~\ref{fig:qcon_final_plot}. The true value of \fsignal{} is then recovered, with some scatter for those cases where $\qcon \geq 0.9$. For cases where $\qcon \leq 0.9$, the upward scatter on the measured value of \fsignal~increases, biasing the measurement. For this reason we recommend the choice of the filtering-to-region separation length scale ratio, \nlambda{}, such that $\qcon \geq 0.9$. We can quantify this recommendation for \nlambda{} in terms of the map filling factor, $\zeta$, by substituting equations~\ref{eq:zeta_star} and \ref{eq:zeta_gas} into equation~\ref{eq:qcon} for \qcon{} and rearranging for \nlambda:

\begin{equation}
	\label{eq:nlambda_min}
	\nlambda = 2 \sqrt{\ln{2}}~\zeta_{\rm max}~c~ \left(\frac{\qcon - a}{d -\qcon}\right)^{\frac{1}{b}} ,
\end{equation}

\noindent
where  $\zeta_{\rm max} = \max{(\zetastar, \zetagas)}$ is the maximal filling factor for the pair of gas and stellar maps. For the Gaussian filter (values of the fitting parameters, $a$, $b$ and $c$ are shown in Table~\ref{tab:qcon_table}) at the maximally recommended value of the filling factor ($\zeta = 0.5$, from testing by \citealt{kruijssen18}) and given our recommendation that $\qcon \geq 0.9$ we would therefore recommend  $\nlambda  \geqslant 15$. For a lower value of the filling factor ($\zeta = 0.2$) this equation implies a recommendation of $\nlambda \geqslant 6$, allowing much tighter filtering. In the case that one wishes to filter out wavelengths shorter than allowed given these recommendations, a sharper filter, such as a second order Butterworth filter may be considered.

\newcommand{\qconwidth}{0.49\textwidth}

\section{Testing and validation}
\label{sec:testing}

We now test the performance of the method at separating diffuse and  compact emission in the generated images. We test this performance with regards to the recovered values of the three key fitting parameters of the \code code: the gas cloud lifetime, \tgas, the region overlap timescale, \tover,~and the mean separation length $\lambda$, and the image compact emission fraction, \fsignal.

We test the method on two test datasets, `Main set 1', for short to intermediate values of \tgas{} and `Main set 2', for intermediate to long values of \tgas{}. We make this division for computational reasons. In our simulated datasets, increasing the value of \tgas{} leads to an increase in the size of the maps in the datasets (see Equation~\ref{eq:pred_lambda}). In order to run datasets with large values of \tgas{} in a reasonable time, we reduce the number of Gaussian regions seeded into our reference stellar map (see Table~\ref{tab:parameters}) and thus the required number of Gaussian regions in the gas maps, leading to a reduction in the overall size of the maps. For these datasets, we set the reference timescale to $10\, \myr{}.$ (i.e. $\tstar{} = 10 \myr{}$). We consider a range of \tgas{} between 5 \myr{} and 100 \myr{} (i.e. $0.5-10~\tstar$). For \tover{}, the range of possible values for a dataset is between 0 \myr{} and the minimum of \tgas{} and \tstar{} (i.e. $0 \, \myr{} -\min(\tgas,10 \, \myr{})$). Initial applications of \code to nearby galaxies that are currently being undertaken \citep{Kruijssen18b,Hygate18,Chevance18,Ward18} have measured values of $\zeta $ between $ \sim 0.2 $ and $\sim 0.6$. We will thus consider values of $\zeta$ over a suitable range (i.e. $\zeta = 0.2, 0.3, 0.4, 0.5, 0.6$ and $0.7$). We note, however, that detailed testing of the \code by \citet{kruijssen18} shows that the condition $\zeta < 0.5$ should be satisfied for good performance of the \code code. For this reason, we focus our analysis on those experiments with $\zeta \leqslant 0.5$, keeping experiments with $\zeta$ = 0.5 as borderline cases.

Together, the selected evolutionary timelines and values of $\zeta$ translate to a range of $\lambda$  between 55 \pc{} for datasets with the highest value of the filling factor ($\zeta= 0.7$), and 420 \pc{} for datasets with the lowest value of the filling factor ($\zeta= 0.2$). For all images, the compact emission fraction, \fsignal, can vary between 10\% and 100\%. For each combination of compact region model (detailed in Section~\ref{sec:signal_component}) and diffuse model (detailed in Section~\ref{sec:backround_component}) and each considered value of the filling factor, $\zeta$, we generate 100 experiment datasets (50 each for main dataset 1 and main dataset 2).\footnote{For \fsignal, this equates to 200 measurements per combination as we make a measurement of the compact fraction in the stellar map and the gas map of each experiment dataset.} For main dataset 1 and main dataset 2, we generate 50 gas and stellar map pairs, for each combination of compact region model and each considered value of the filling factor, $\zeta$. We remove experiments with obviously visibly bad fits from the datasets with $\zeta$ between 0.2 and 0.5. This concerns 116 out 6400 experiments  ($\sim 2\%$ of the experiments).

This equates to a total of 100 experiments for each combination $\zeta$.\footnote{For \fsignal, this equates to 200 measurements per combination as we make a measurement of the compact fraction in the stellar map and the gas map of each experiment map pair.} For each map pair, we select a randomised evolutionary timeline, where \tgas~may vary between 5 \myr{} and 100 \myr{} (i.e. $0.5-10~\tstar$) and \tover{} may vary between 0 \myr{} and the minimum of \tgas{} and \tstar{} (i.e. $0 \, \myr{} -\min(\tgas,10 \, \myr{})$). The value of \fsignal~is randomly selected to be between 10\% and 100\% of the total flux of the image. For full details on the generation of experiment datasets see Section~\ref{sec:test_images} and for the full set of parameters used in this generation see Table~\ref{tab:all_parameter_space}.

\begin{figure*}

	\centering
	\subfloat{\includegraphics[width=\cwidth]{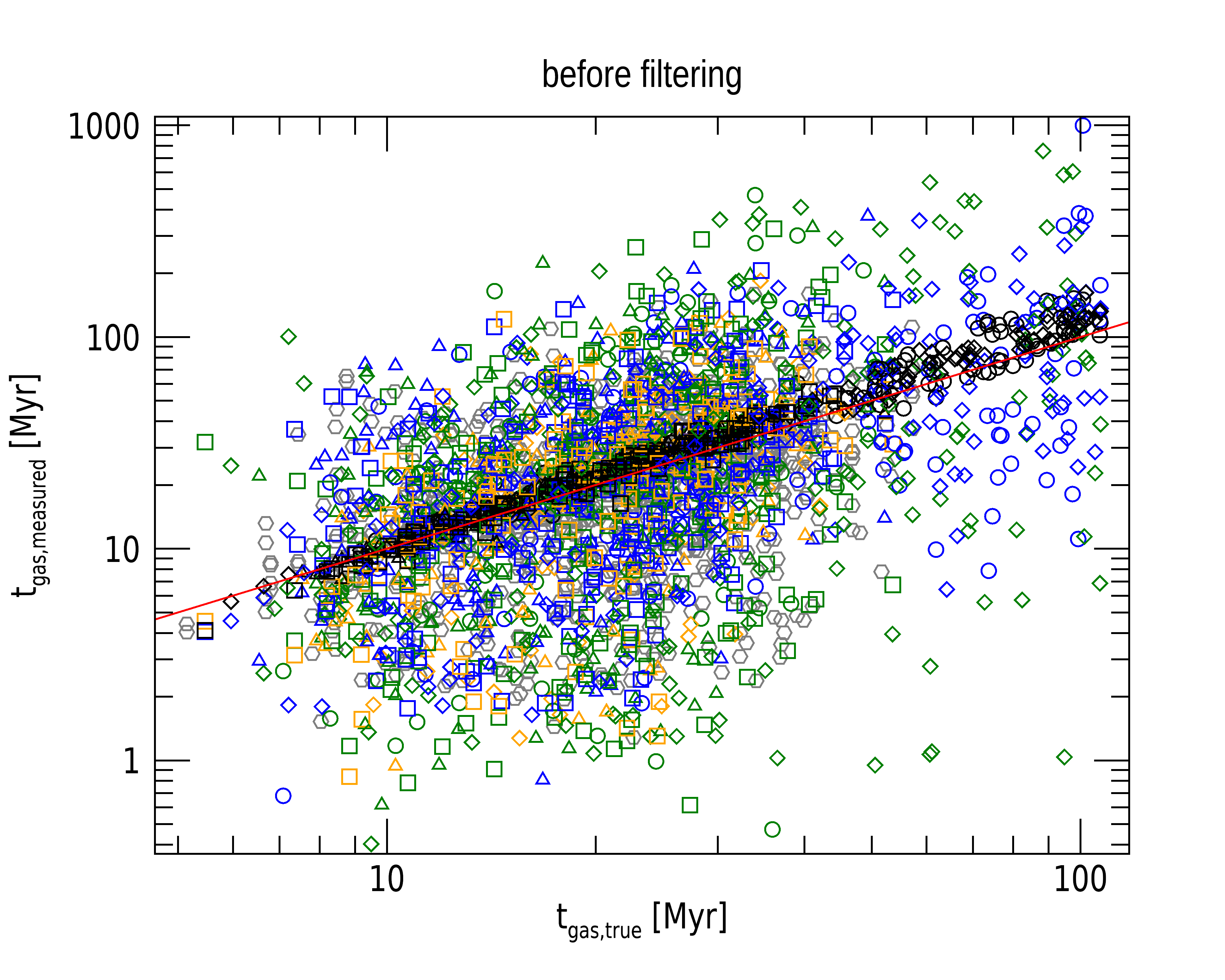}}
	\subfloat{\includegraphics[width=\cwidth]{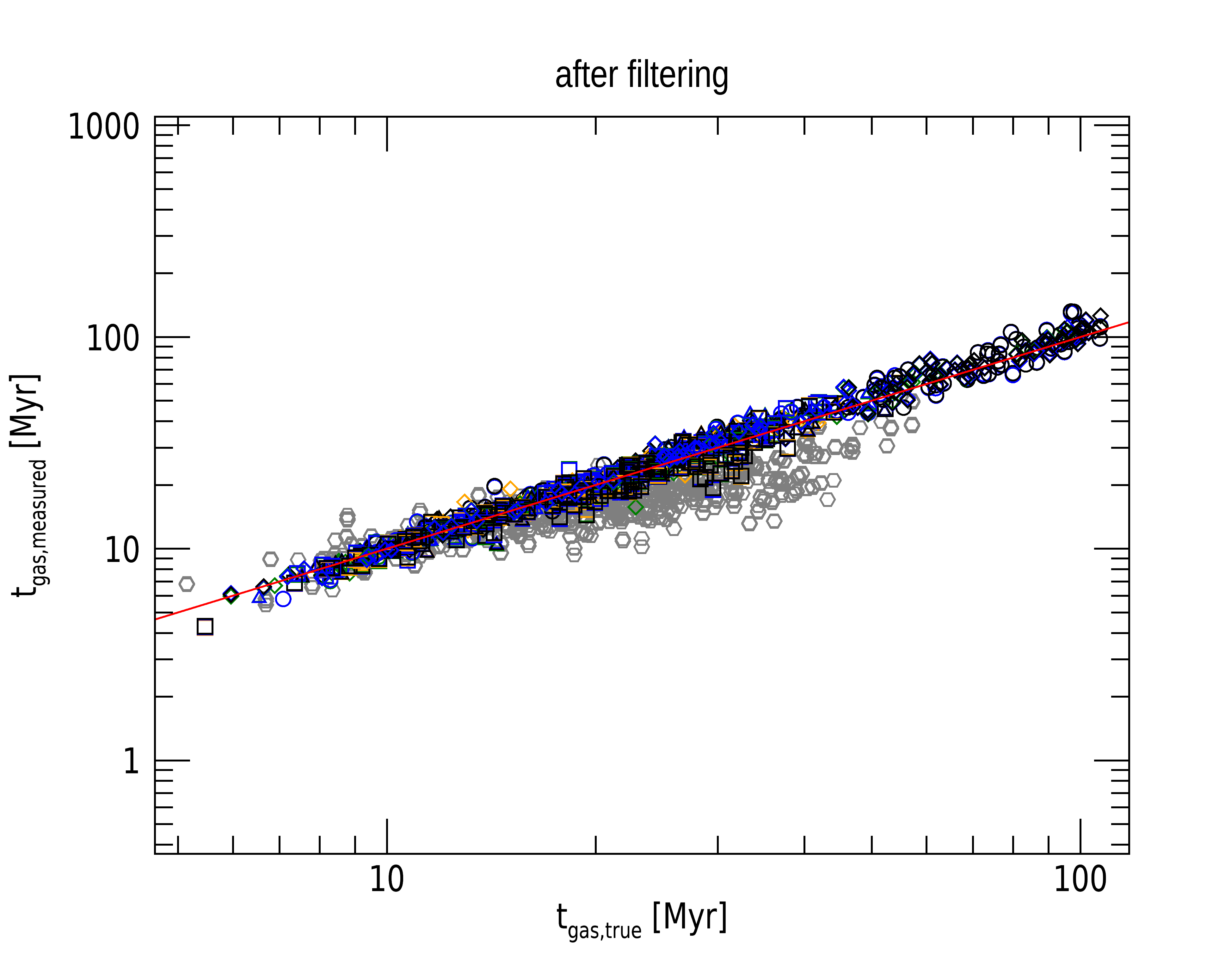}} \\
	\centering
	\subfloat{\includegraphics[width=\cwidth]{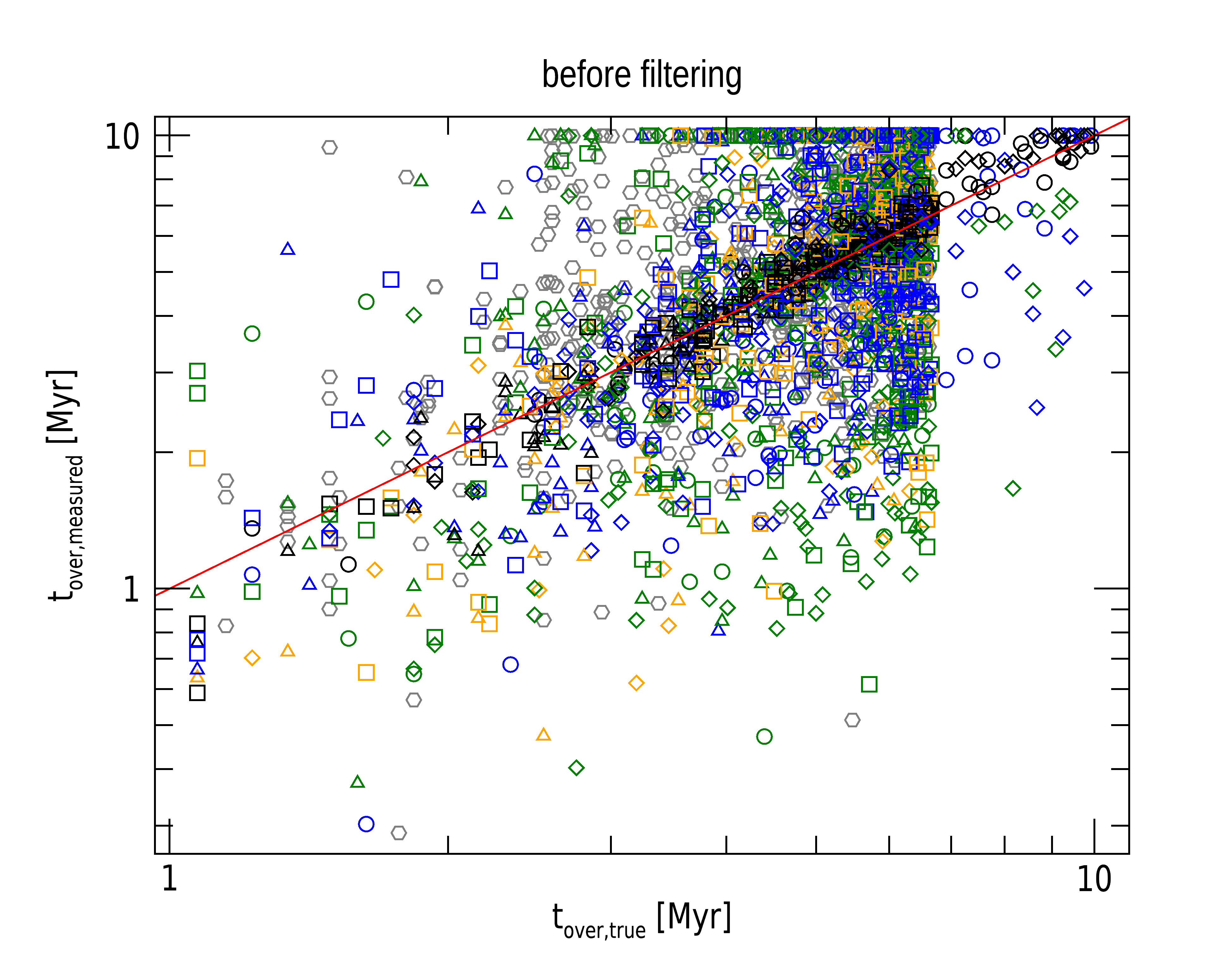}}
	\subfloat{\includegraphics[width=\cwidth]{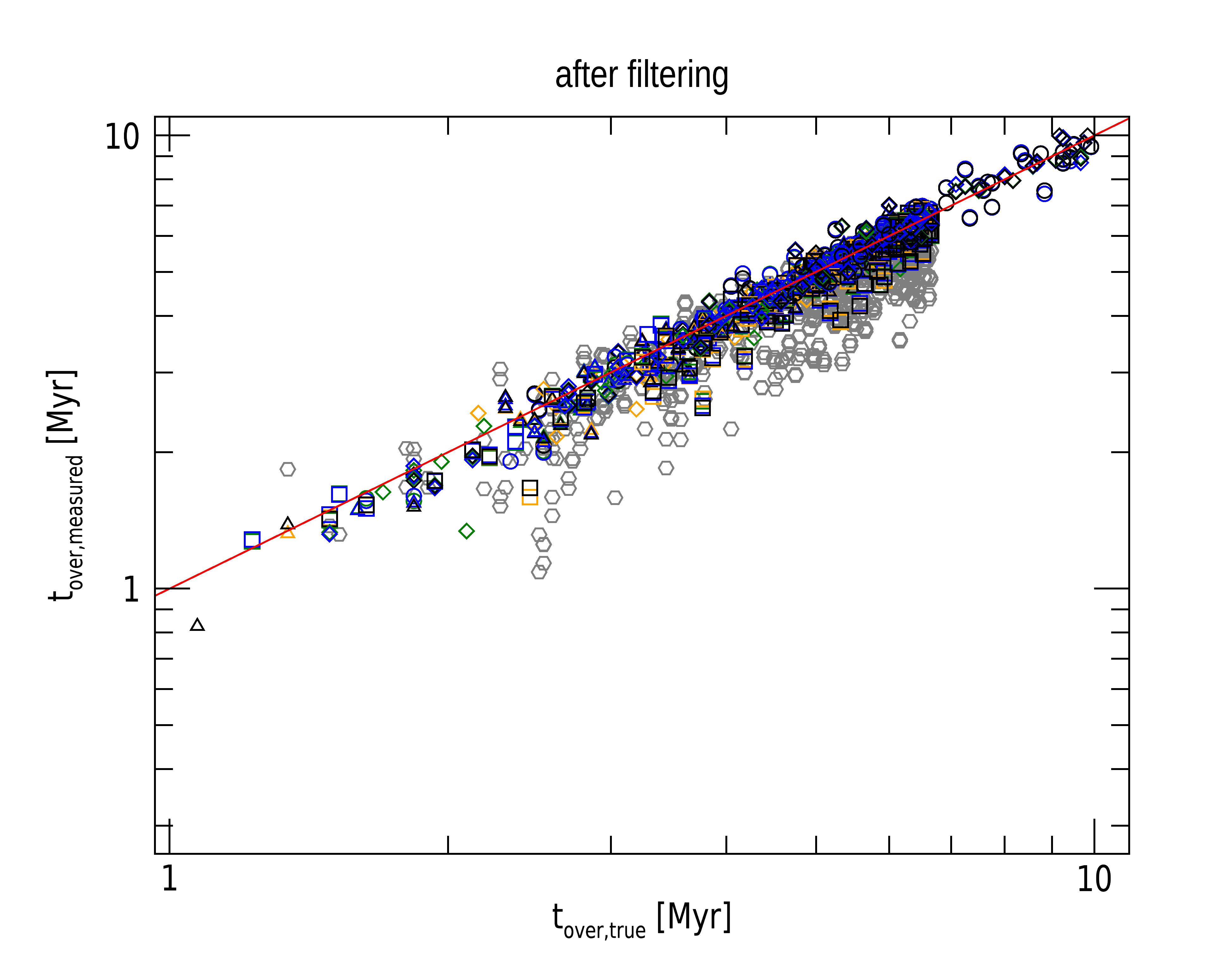}} \\
	\centering
	\subfloat{\includegraphics[width=\cwidth]{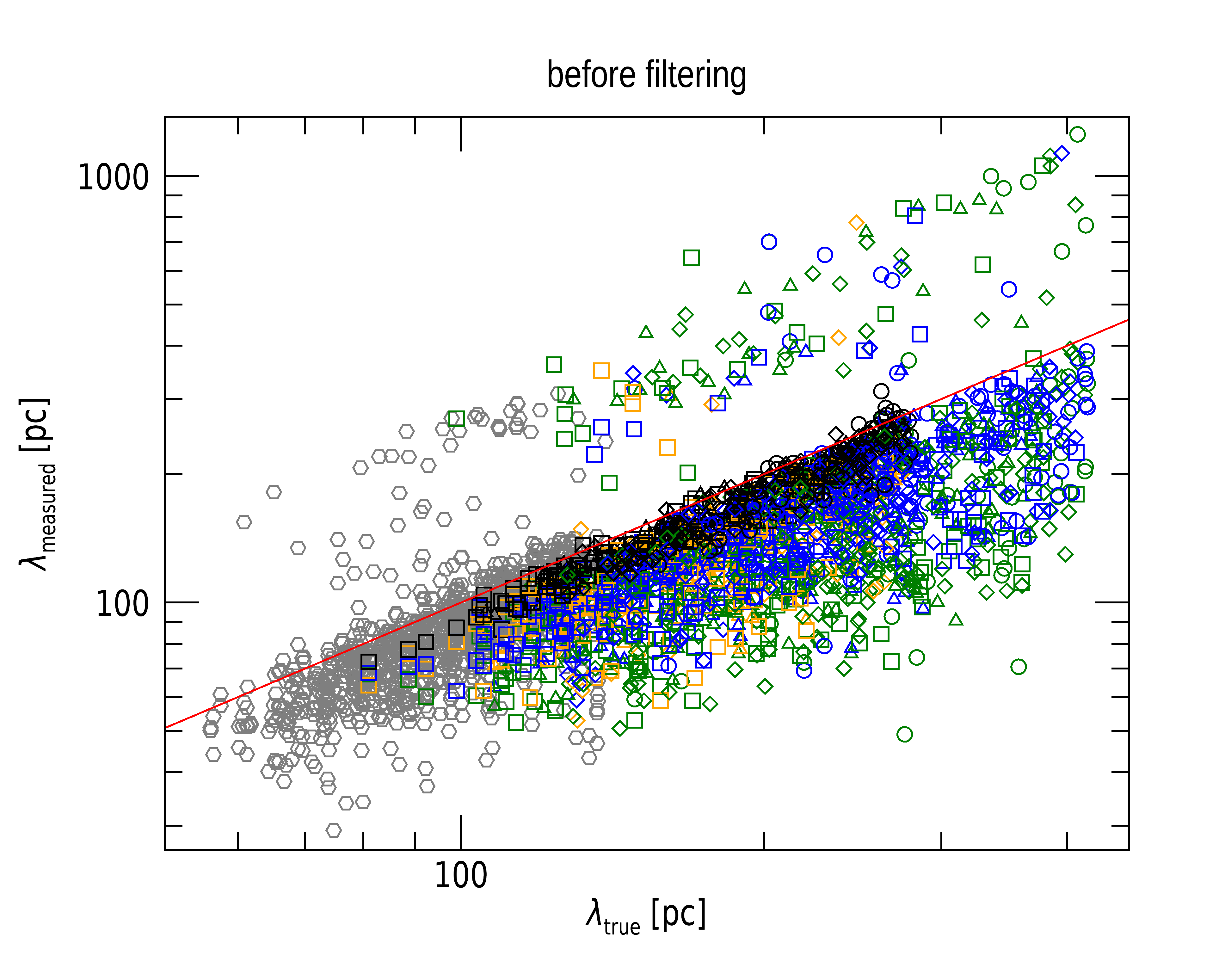}}
	\subfloat{\includegraphics[width=\cwidth]{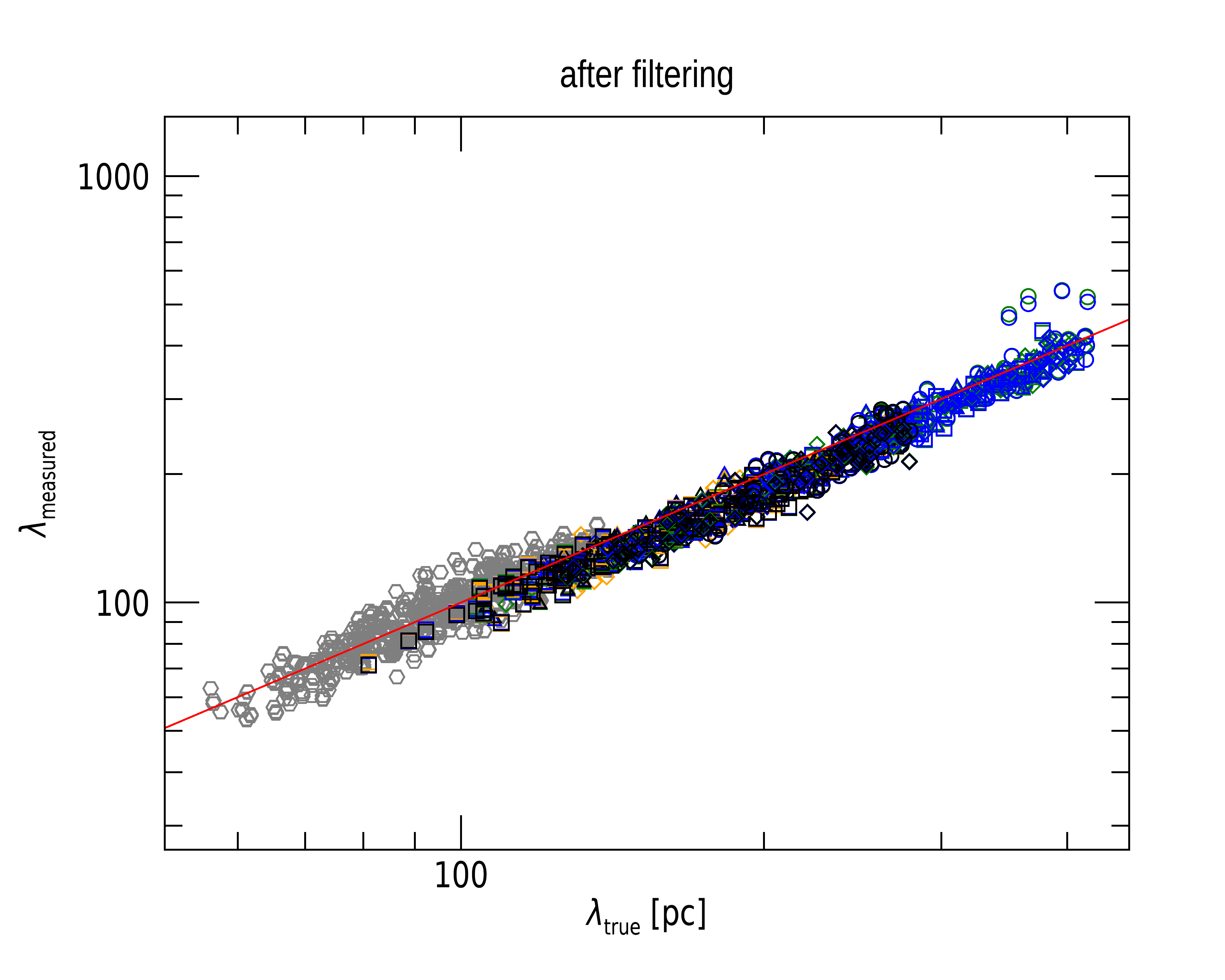}} \\

	\caption{The measured values of \tgas~(top row), \tover~(middle row) and $\lambda$ (bottom row) against the true value before filtering diffuse emission in Fourier space (left column) and after filtering (right column). The red line shows the 1:1 relation between the true and measured values for each of the quantities. Each combination of compact model and diffuse model is shown with a different symbol/colour combination, which we summarise in Table~\ref{tab:plot_symbols}. Experiments not meeting the filling factor criterion for the \code code (i.e $\zeta>0.5$) are shown as grey hexagons. This figure shows that a diffuse emission reservoir can significantly affect the measured values of these quantities and that the application of the presented Fourier filtering results in a much more accurate measurement.}

	\label{fig:main_heisenberg_plot}

\end{figure*}

\begin{figure*}

	\centering
	\subfloat{\includegraphics[width=\cwidth]{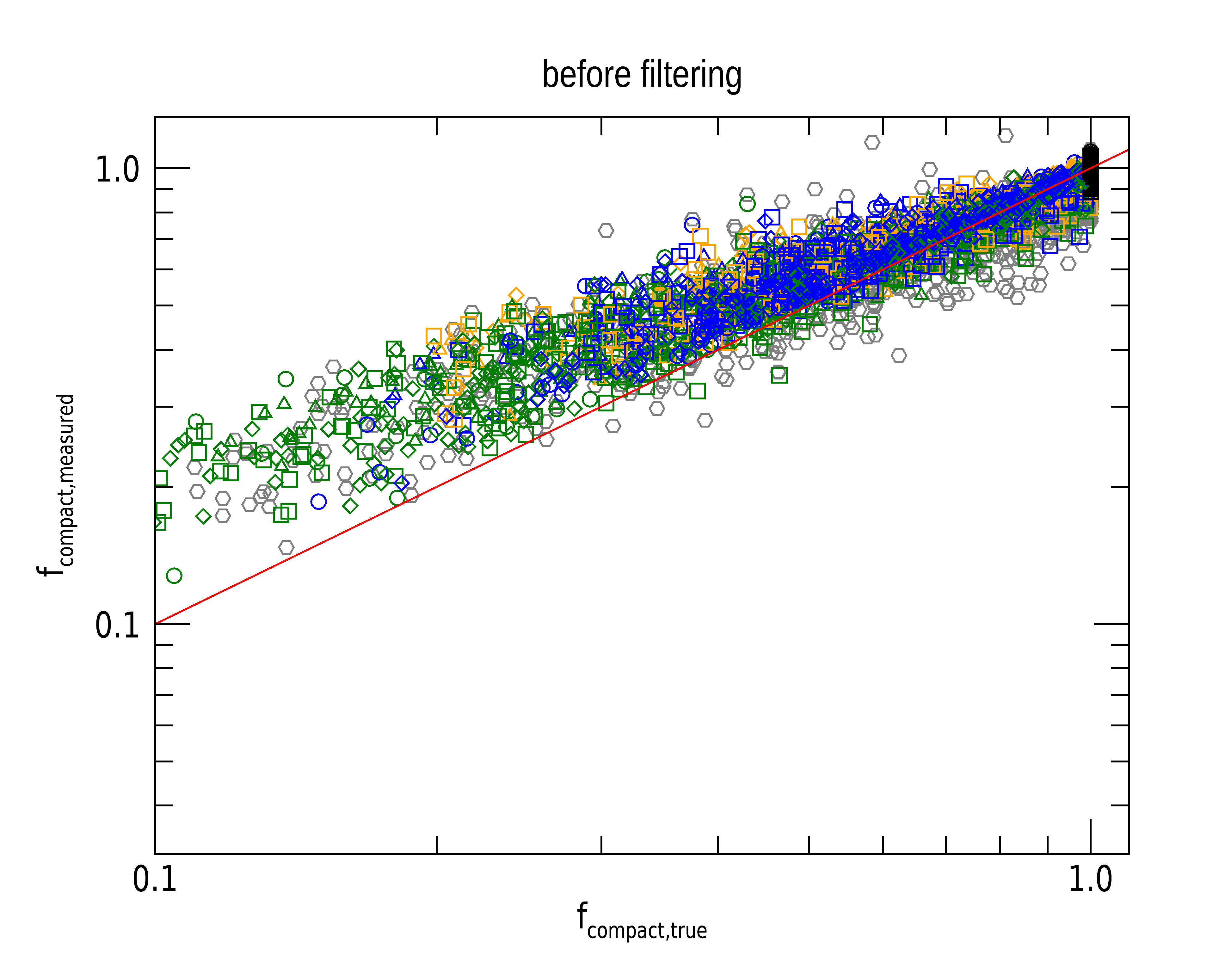}}
	\subfloat{\includegraphics[width=\cwidth]{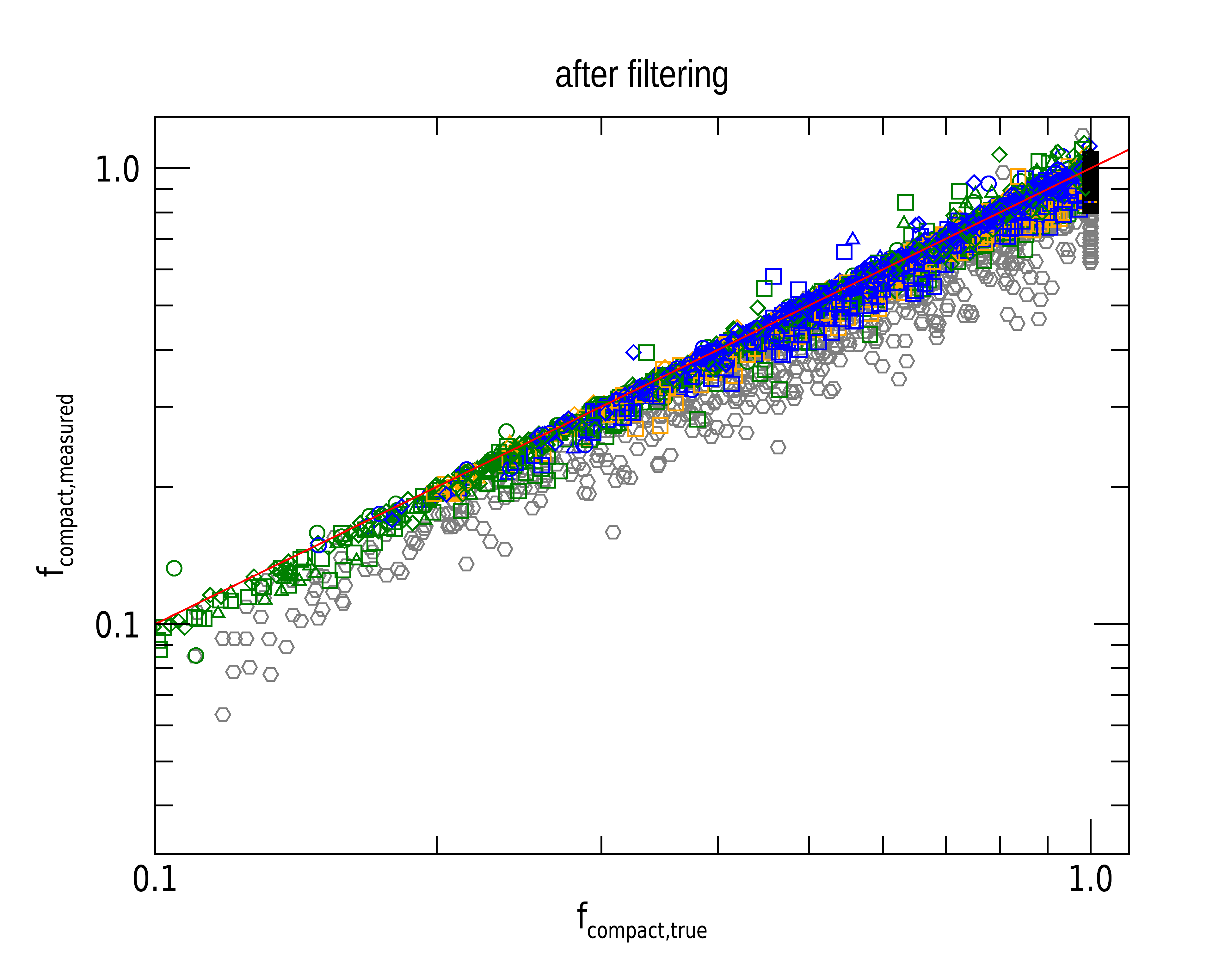}} \\

	\caption{The measured compact emission fraction, \fsignalmeasured{}, against the true compact emission fraction, \fsignaltrue{}. Each combination of compact model and diffuse model is shown with a different symbol/colour combination, which we summarise in Table~\ref{tab:plot_symbols}. Experiments not meeting the filling factor criterion for the \code code (i.e $\zeta>0.5$) are shown as grey hexagons. Left panel: values of \fsignalmeasured{} calculated on the basis of the value of $\lambda$ measured on the unfiltered map (i.e. at the end of the first iteration). Right panel: values of \fsignalmeasured{} calculated at the end of the iterative filtering process after the measured value of $\lambda$ has converged. After convergence, we recover values of \fsignalmeasured~that are tightly correlated around the 1:1 line with \fsignaltrue.}

	\label{fig:main_fractions_plot}

\end{figure*}

A summary of the measured quantities for these experiments, both before and after filtering, is shown in Figure~\ref{fig:main_heisenberg_plot} for \tgas, \tover~and $\lambda$ and in Figure~\ref{fig:main_fractions_plot}, for \fsignal. Experiments with unsatisfactory values of the filing factor ($\zeta=0.6, 0.7$) are shown as grey hexagons and those with satisfactory filling factor ($\zeta \leqslant 0.5$) are shown as coloured symbols. A key to the plotting symbols used for each combination of compact and diffuse model  is shown in Table~\ref{tab:plot_symbols}. A summary of the measured quantities for these experiments, both before and after filtering, is shown in Figure~\ref{fig:main_heisenberg_plot} for \tgas, \tover~and $\lambda$ and in Figure~\ref{fig:main_fractions_plot}, for \fsignal.

\definecolor{ndiffcolor}{HTML}{000000} 
\definecolor{cdiffcolor}{HTML}{007F00} 
\definecolor{gdiffcolor}{HTML}{0000FF} 
\definecolor{ediffcolor}{HTML}{FFA500} 
\definecolor{highzetacolor}{HTML}{7F7F7F} 

\begin{table}
	\centering
	\begin{minipage}{\columnwidth}
		\caption{The plotting symbols used in Figures~\ref{fig:main_heisenberg_plot} and \ref{fig:main_fractions_plot} for each combination of compact and diffuse models.}
		\label{tab:plot_symbols}\vspace{-1mm}
		\begin{tabular}{l| l l l l}
			\hline
			{} & \multicolumn{4}{l}{Diffuse Model} \\
			Compact Model & none & constant & large Gaussian & envelopes   \\
			\hline
			Uniform FWHM & $\textcolor{ndiffcolor}\Diamond$ & $\textcolor{cdiffcolor}\Diamond$ & $\textcolor{gdiffcolor}\Diamond$ & $\textcolor{ediffcolor}\Diamond$ \\
			Asymmetric FWHM & $\textcolor{ndiffcolor}\Circle$ & $\textcolor{cdiffcolor}\Circle$ & $\textcolor{gdiffcolor}\Circle$ & $\textcolor{ediffcolor}\Circle$ \\
			Spread FWHM & $\textcolor{ndiffcolor}\Box$ & $\textcolor{cdiffcolor}\Box$ & $\textcolor{gdiffcolor}\Box$ & $\textcolor{ediffcolor}\Box$ \\
			Flux spread FWHM & $\textcolor{ndiffcolor}\vartriangle$ & $\textcolor{cdiffcolor}\vartriangle$ & $\textcolor{gdiffcolor}\vartriangle$ & $\textcolor{ediffcolor}\vartriangle$ \\
			\hline
		\end{tabular} \\
		Experiments with unsatisfactory filling factor ($\zeta>0.5$) are displayed as grey hexagons ($\textcolor{highzetacolor}\hexagon$) regardless of the combination of diffuse and compact model
	\end{minipage}
\end{table}

For experiments with no added diffuse reservoir, we recover values of these quantities {close to the correct values} both before and after Fourier filtering. For \fsignal~this is unity, due to there being no diffuse reservoir in the images. This indicates that the distortions introduced into the images do not significantly affect the total flux of the  image or the correct identification of peaks (i.e. only few spurious peaks are introduced into the image by filtering) and thus in turn the measurement of these quantities. However, the addition of diffuse reservoirs does significantly affect the measured values of the quantities in the unfiltered images. After Fourier filtering, and the removal of the diffuse reservoir, we again measure values of these quantities that are close to the true values. The distribution of the recovered quantities is also not significantly different between experiments with added diffuse emission and without, with the exception of the measured  compact emission fraction, which is $\sim 1$ for all the experiments without an added diffuse background.

We do find a number of other elements that can bias the measurements. For instance, as $|\log_{10}(\tstar/\tgas) |$ increases, the measurement of \tgas~worsens, as described in \citet{kruijssen18}. For \tover, we note that the line of points in the middle left panel in Figure~\ref{fig:main_heisenberg_plot} (at $t_{\rm over, measured} = 10\, \myr{}$) is due to the fact that in all cases $\tover \leq \min(\tstar, \tgas)$, because it is physically impossible for regions to coexist with other regions for a time longer than their own lifetime (and $\tstar{} = 10$ \myr{} for all of our experiments).\footnote{In a minority of experiments, \tgas{} is shorter than \tstar{}. In this case, \tgas{} sets the upper limit for \tover }

For $\lambda$, we note that our prediction, as given in equation~\ref{eq:pred_lambda}, systematically overestimates the values measured in our experiments, by $\sim 10\% $ (see the bottom right panel of Figure~\ref{fig:main_heisenberg_plot}). As there is no significant offset between the values of $\lambda$ measured for experiments without added diffuse emission both before and after filtering we conclude that this is not an offset introduced by Fourier filtering. Instead this discrepancy is due to the geometric mean separation between regions that we calculate being an overestimate of the actual local mean separation of the regions. After all, the scatter introduced by the random positioning of the peaks causes them more often than not to be positioned closer to their near neighbours than the geometric mean separation length. This slight overestimation of $\lambda$ propagates into a corresponding, minor underestimation of the compact emission fraction in Figure~\ref{fig:main_fractions_plot}.

For those datasets with unsatisfactory values of the filling factor (i.e. $\zeta = 0.6, 0.7$), indicated as grey hexagons in Figures~\ref{fig:main_heisenberg_plot}~and~\ref{fig:main_fractions_plot}, the accuracy of the measurements decreases for larger values of \tgas{}. As the amount of blending in an individual tracer map ($\etastar$ and $\etagas$, see Equations~\ref{eq:eta_star}~and~\ref{eq:eta_gas}) scales with the lifetime of that particular tracer in units of the total duration of the timeline, the effect of blending in both maps is roughly equal when $\tgas \sim \tstar$. However, as \tgas{} increases, \etagas{} increases too (while \etastar{} decreases due to the increase of the total duration of the timeline), leading to more severe blending in the gas map and more inaccurate results overall.

\begin{figure*}

	\centering
	\subfloat{\includegraphics[width=\cwidth]{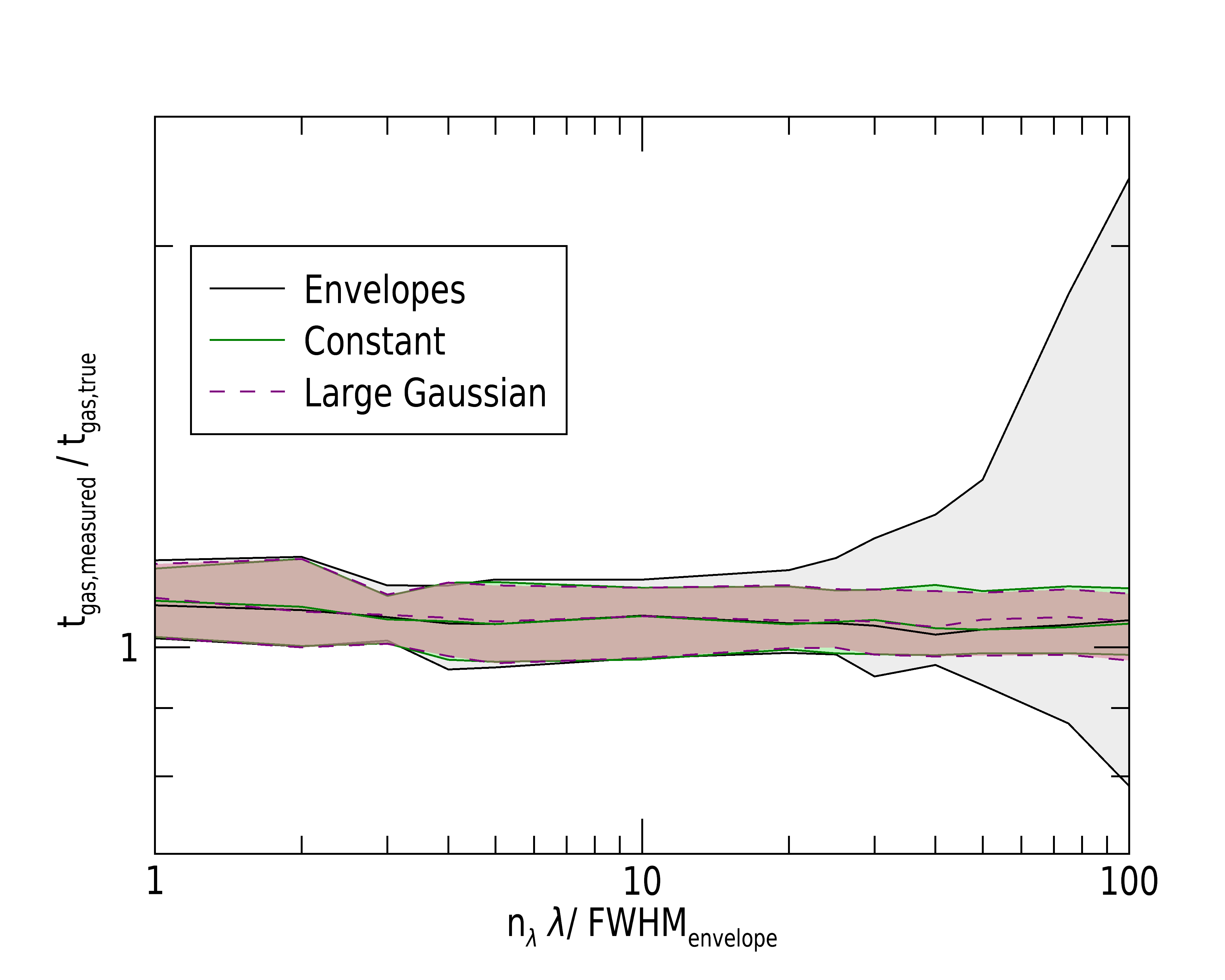}}
	\subfloat{\includegraphics[width=\cwidth]{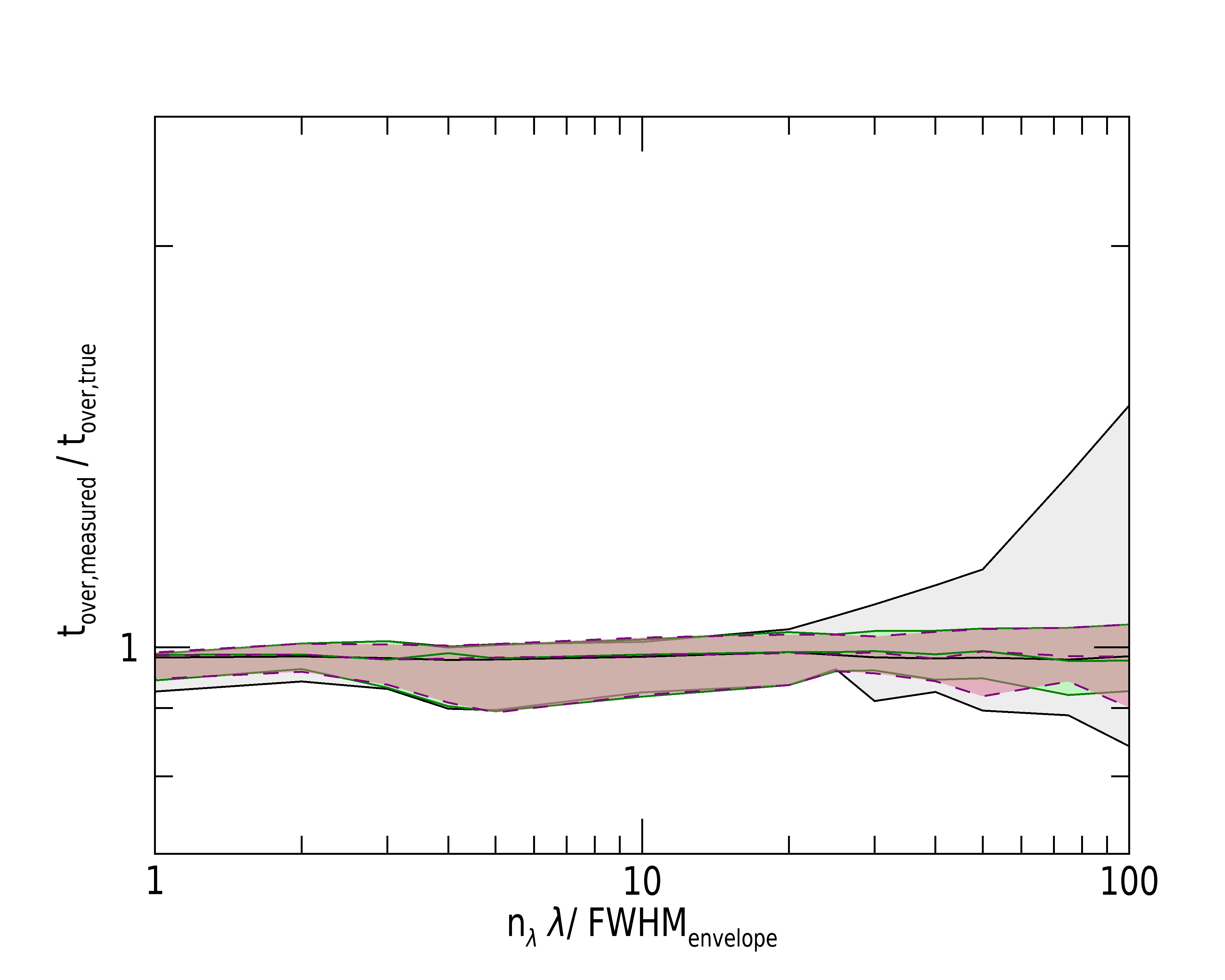}}  \\
	\subfloat{\includegraphics[width=\cwidth]{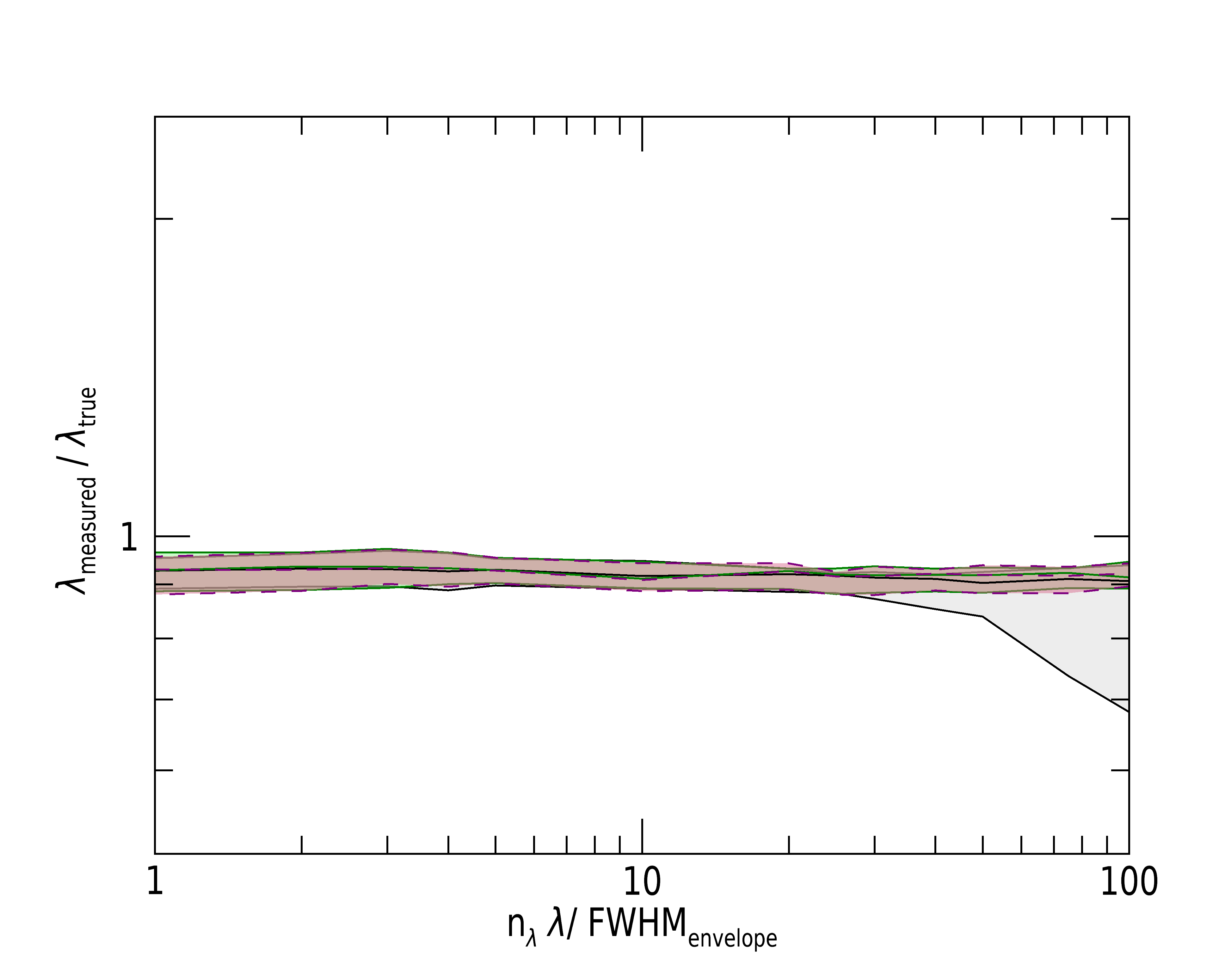}}
	\subfloat{\includegraphics[width=\cwidth]{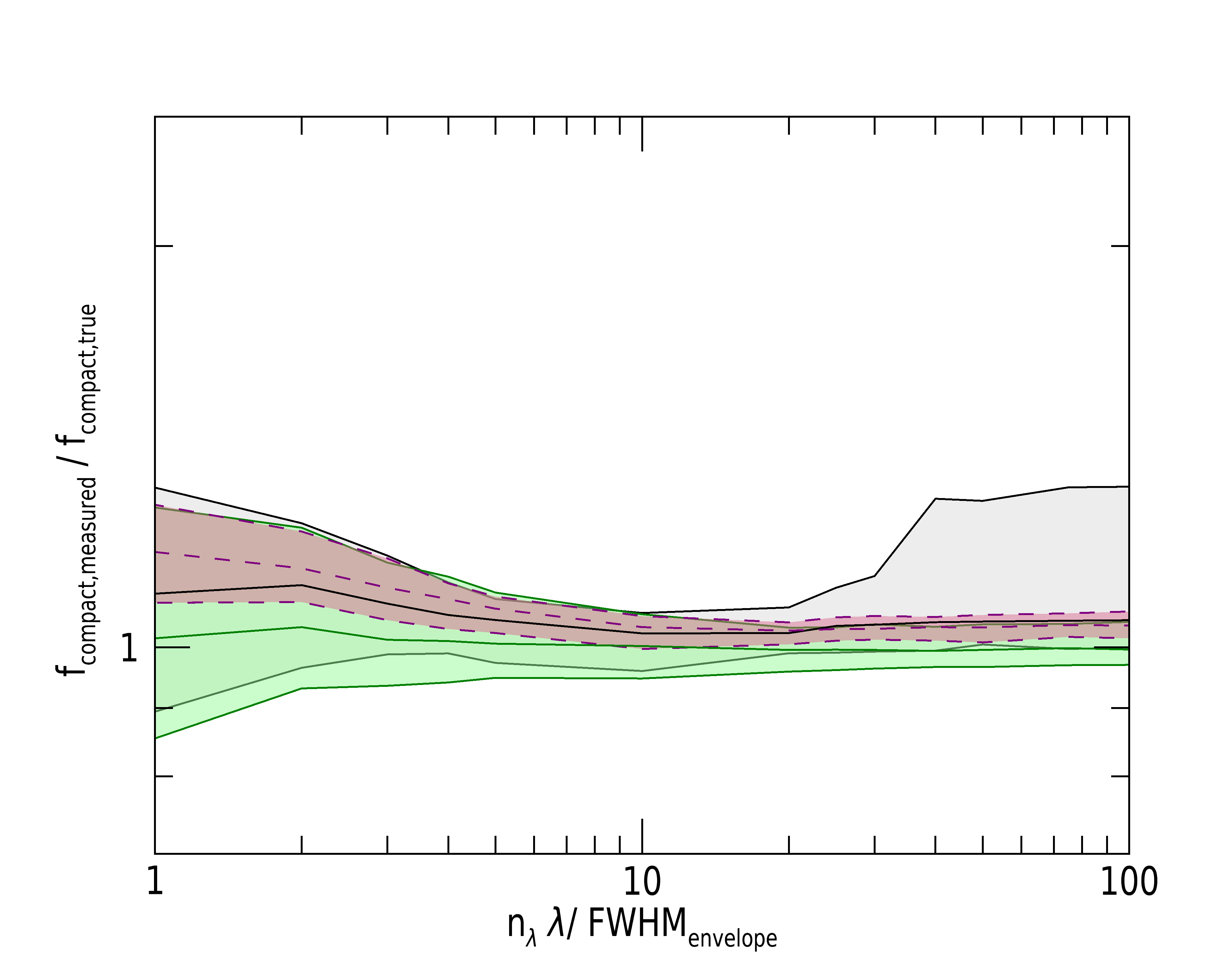}}  \\
	\caption{Measured values in units of the true value of  \tgas{} (top left), \tover{} (top right), $\lambda$ (bottom left), and  $\fsignal$ (bottom right) against the ratio of the critical length scale of the applied highpass filter normalised by the FWHM of the diffuse envelope, $\nlambda \lambda / {\rm FWHM_{\rm envelope}} $. Shaded regions show the $1\sigma$ region of measurements around the median for an example set of 10 image pairs with constant diffuse background (green), large-scale Gaussian background (purple) and envelope only (i.e. containing envelopes around the compact regions, but no constant or large-scale Gaussian background) background (grey). Both the constant and large-scale Gaussian backgrounds are well-removed across the entire considered range of the filtering-to-region separation length scale ratio, \nlambda{}. By contrast, the measurement scatter for the envelope-only model increases significantly at $n_\lambda>30\,{\rm FWHM}_{\rm envelope}/\lambda$, where the applied filter has not sufficiently removed the diffuse background. At $n_\lambda<3\,{\rm FWHM}_{\rm envelope}/\lambda$, the measured value of \fsignal{} is affected by the introduction of distortions to the image and the large amount of flux being removed from the compact regions.}

	\label{fig:ediff_cut_length_plot}

\end{figure*}

In summary, we see that the filtering method performs well regardless of the compact region model used, with no significant difference between the scatter of measured values between the different models after filtering. With respect to the diffuse background models, successful filtering of the constant diffuse model and the Gaussian diffuse model is relatively insensitive to the value of the filtering-to-region separation length scale ratio, \nlambda{}, chosen, due to the fact that a constant sheet of diffuse gas and a galaxy-scale Gaussian correspond to infinite and very large spatial wavelengths, respectively, whereas $\lambda$ is typically much smaller than a galactic radius. Galaxies with radii $\sim10~{\rm kpc}$ have a typical region separation length $\lambda \sim 100~{\rm pc}$ (as measured from initial applications of the method: \citealt{Kruijssen18b}; \citealt{Hygate18}; \citealt{Chevance18}; and \citealt{Ward18}), giving a factor of $\sim 100$ in size between the two. Thus the value of $\nlambda$ selected is relatively unimportant with regards to the filtering of galactic-scale components of diffuse emission. By contrast, the envelope diffuse model requires filtering with lower values of $\nlambda$, because the modelled envelopes are much closer in size to $\lambda$. Figure~\ref{fig:ediff_cut_length_plot} shows how the performance of the method at removing each of these diffuse models varies with the chosen value of \nlambda{}. This summarises the above discussion, i.e.~the value of \nlambda{} does not affect the filtering of galactic-scale emission, but in order to accurately filter diffuse envelopes of regions, we require $\nlambda{} < 30 \,{\rm FWHM}_{\rm envelope}/\lambda$. The criterion is easily satisfied, as in practice one would expect ${\rm FWHM}_{\rm envelope} \sim \lambda$ and the resulting $\nlambda{} < 30$ is a very generous condition.

Overall, we are able to accurately constrain the cloud lifecycle (\tgas~and \tover), the cloud separation length ($\lambda$) and the diffuse and compact emission fractions after filtering the diffuse emission in Fourier space, regardless of the significance of the added diffuse emission. We are able to do this for all of the compact region models and diffuse emission region models that we consider.

Finally, we note that the results presented in this section are based upon idealised datasets that contain no noise and are perfectly aligned in astrometry. In reality tracer maps will contain observational noise and two different tracer maps may be astrometrically offset from each other. We detail the impact that these two observational limitations have on the method in Appendix~\ref{confounding_obs_issues}. In Appendix~\ref{sec:astrometry}, we show that astronometric uncertainty less than one third compact region FWHM (i.e $\leqslant \fwhm/3$) must be achieved for meaningful constraints of \tover{} and in Appendix~\ref{sec:noise} we show that the use of low or intermediate signal to noise data will lead to biased results, unless a method to reduce the noise is used. 

\section{Conclusions}
\label{sec:conclusions}

We have presented a physically motivated method for the filtering and measurement of diffuse emission (such as the diffuse Warm Ionised Medium that is observed in H$\rm \alpha$ along with compact \htwo{} regions) in images by filtering in Fourier space. We have validated our method's performance within the \code code \citep{kruijssen18} by testing for the accurate recovery of the three key fitting-parameters measured by the \code code (\tgas, \tover~and $\lambda$). These tests are based on a range of models of compact regions and diffuse backgrounds, as detailed in Section~\ref{sec:test_images}. We demonstrate that the presence of diffuse emission can have a significant impact on the measurement of the measured cloud-scale quantities, but also demonstrate that our filtering method accurately alleviates this problem. A summary of these tests can be seen in Figure~\ref{fig:main_heisenberg_plot}. We have also validated its performance for measuring the fractions of diffuse and compact emission within tracer maps, which is summarised in Figure~\ref{fig:main_fractions_plot}. Finally, we report a number of criteria that need to be satisfied for the reliable application of the method, which we summarise here:

\begin{itemize}
	\item The observed compact regions need to be sufficiently resolved ($\fwhm \geqslant 2$ pixels) or to be point sources.
	\item The value of the filtering-to-region separation length scale ratio, $\nlambda{}$, should be chosen such that the flux-loss from compact regions due to the application of the Fourier filter is minimised, i.e. $\qcon \geqslant 0.9$. 
	\item For meaningful constraints on the value of \tover{}, the astrometric uncertainty between images should be less than one third the FWHM of the regions of interest (i.e $\leqslant \fwhm/3$), see Appendix~\ref{sec:astrometry}.
	\item In the case of low or intermediate signal to noise, the presence of noise will lead to biased results unless a noise-reduction method is employed such as lowpass Fourier filtering (See Appendix~\ref{sec:noise}).
\end{itemize}

We also note that common reduction techniques for astronomical data may reduce the amount of diffuse emission present in a tracer map. For example, the practice of filtering molecular gas tracer maps to exclude emission that is not associated in velocity space with H{\sc I} emission (see \citealt{2010A&A...512A..68G}) may remove some fraction of the diffuse emission from the final tracer map, depending on the filtering parameters chosen for this reduction process. The presented method will only measure the fraction of diffuse emission remaining in the map presented to it after any prior reduction process is complete.

In conclusion, we have demonstrated how, by filtering in Fourier space using the mean region separation length, $\lambda$, we can remove diffuse emission from tracer images. This allows the \code code to be applied to observationally constrain cloud scale ISM physics, such as cloud lifecycles, without the biasing impact of diffuse emission. In addition, by separating the diffuse and the compact emission within the images, we are able to measure their relative contribution to the total emission within the images. The method also produces maps of the compact and diffuse emission within each input tracer map that may be used for further analysis such as investigating spatial variations in the diffuse emission fraction within galaxies. This approach will provide critical astrophysical constraints on e.g.~the unresolved population of low-mass GMCs, and the ionised photon escape fraction.

\section*{Acknowledgements}

APSH and DTH are fellows of the International Max Planck Research School for Astronomy and Cosmic Physics at the University of Heidelberg (IMPRS-HD). JMDK and MC gratefully acknowledge funding from the German Research Foundation (DFG) in the form of an Emmy Noether Research Group (grant number KR4801/1-1). JMDK gratefully acknowledges funding from the European Research Council (ERC) under the European Union's Horizon 2020 research and innovation programme via the ERC Starting Grant MUSTANG (grant agreement number 714907). APSH warmly thanks Fabian Walter for extensive discussions on the content of this work, as well as more broadly on interferometric techniques.




\bibliographystyle{mnras}
\bibliography{bib_fourier} 

\begin{thebibliography}{}
\makeatletter
\relax
\def\mn@urlcharsother{\let\do\@makeother \do\$\do\&\do\#\do\^\do\_\do\%\do\~}
\def\mn@doi{\begingroup\mn@urlcharsother \@ifnextchar [ {\mn@doi@}
  {\mn@doi@[]}}
\def\mn@doi@[#1]#2{\def\@tempa{#1}\ifx\@tempa\@empty \href
  {http://dx.doi.org/#2} {doi:#2}\else \href {http://dx.doi.org/#2} {#1}\fi
  \endgroup}
\def\mn@eprint#1#2{\mn@eprint@#1:#2::\@nil}
\def\mn@eprint@arXiv#1{\href {http://arxiv.org/abs/#1} {{\tt arXiv:#1}}}
\def\mn@eprint@dblp#1{\href {http://dblp.uni-trier.de/rec/bibtex/#1.xml}
  {dblp:#1}}
\def\mn@eprint@#1:#2:#3:#4\@nil{\def\@tempa {#1}\def\@tempb {#2}\def\@tempc
  {#3}\ifx \@tempc \@empty \let \@tempc \@tempb \let \@tempb \@tempa \fi \ifx
  \@tempb \@empty \def\@tempb {arXiv}\fi \@ifundefined
  {mn@eprint@\@tempb}{\@tempb:\@tempc}{\expandafter \expandafter \csname
  mn@eprint@\@tempb\endcsname \expandafter{\@tempc}}}

\bibitem[\protect\citeauthoryear{{Baldwin}, {Phillips}  \&
  {Terlevich}}{{Baldwin} et~al.}{1981}]{Baldwin1981}
{Baldwin} J.~A.,  {Phillips} M.~M.,   {Terlevich} R.,  1981, \mn@doi [\pasp]
  {10.1086/130766}, \href {http://adsabs.harvard.edu/abs/1981PASP...93....5B}
  {93, 5}

\bibitem[\protect\citeauthoryear{{Blanc}, {Heiderman}, {Gebhardt}, {Evans}  \&
  {Adams}}{{Blanc} et~al.}{2009}]{Blanc2009}
{Blanc} G.~A.,  {Heiderman} A.,  {Gebhardt} K.,  {Evans} II N.~J.,   {Adams}
  J.,  2009, \mn@doi [\apj] {10.1088/0004-637X/704/1/842}, \href
  {http://adsabs.harvard.edu/abs/2009ApJ...704..842B} {704, 842}

\bibitem[\protect\citeauthoryear{{Bregman} \& {Pildis}}{{Bregman} \&
  {Pildis}}{1994}]{Bregman1994}
{Bregman} J.~N.,  {Pildis} R.~A.,  1994, \mn@doi [\apj] {10.1086/173587}, \href
  {http://adsabs.harvard.edu/abs/1994ApJ...420..570B} {420, 570}

\bibitem[\protect\citeauthoryear{{Cald{\'u}-Primo}, {Schruba}, {Walter},
  {Leroy}, {Sandstrom}, {de Blok}, {Ianjamasimanana}  \&
  {Mogotsi}}{{Cald{\'u}-Primo} et~al.}{2013}]{CalduPrimo2013}
{Cald{\'u}-Primo} A.,  {Schruba} A.,  {Walter} F.,  {Leroy} A.,  {Sandstrom}
  K.,  {de Blok} W.~J.~G.,  {Ianjamasimanana} R.,   {Mogotsi} K.~M.,  2013,
  \mn@doi [\aj] {10.1088/0004-6256/146/6/150}, \href
  {http://adsabs.harvard.edu/abs/2013AJ....146..150C} {146, 150}

\bibitem[\protect\citeauthoryear{{Cald{\'u}-Primo}, {Schruba}, {Walter},
  {Leroy}, {Bolatto}  \& {Vogel}}{{Cald{\'u}-Primo}
  et~al.}{2015}]{CalduPrimo2015}
{Cald{\'u}-Primo} A.,  {Schruba} A.,  {Walter} F.,  {Leroy} A.,  {Bolatto}
  A.~D.,   {Vogel} S.,  2015, \mn@doi [\aj] {10.1088/0004-6256/149/2/76}, \href
  {http://adsabs.harvard.edu/abs/2015AJ....149...76C} {149, 76}

\bibitem[\protect\citeauthoryear{{Chevance} et~al.,}{{Chevance}
  et~al.}{2018}]{Chevance18}
{Chevance} M.,  et~al., 2018, to be submitted

\bibitem[\protect\citeauthoryear{{Crocker} et~al.,}{{Crocker}
  et~al.}{2013}]{Crocker2013}
{Crocker} A.~F.,  et~al., 2013, \mn@doi [\apj] {10.1088/0004-637X/762/2/79},
  \href {http://adsabs.harvard.edu/abs/2013ApJ...762...79C} {762, 79}

\bibitem[\protect\citeauthoryear{{Dettmar}}{{Dettmar}}{1990}]{Dettmar1990}
{Dettmar} R.-J.,  1990, \aap, \href
  {http://adsabs.harvard.edu/abs/1990A%26A...232L..15D} {232, L15}

\bibitem[\protect\citeauthoryear{{Fabbiano}, {Heckman}  \& {Keel}}{{Fabbiano}
  et~al.}{1990}]{Fabbiano1990}
{Fabbiano} G.,  {Heckman} T.,   {Keel} W.~C.,  1990, \mn@doi [\apj]
  {10.1086/168778}, \href {http://adsabs.harvard.edu/abs/1990ApJ...355..442F}
  {355, 442}

\bibitem[\protect\citeauthoryear{{Gratier}, {Braine}, {Rodriguez-Fernandez},
  {Israel}, {Schuster}, {Brouillet}  \& {Gardan}}{{Gratier}
  et~al.}{2010}]{2010A&A...512A..68G}
{Gratier} P.,  {Braine} J.,  {Rodriguez-Fernandez} N.~J.,  {Israel} F.~P.,
  {Schuster} K.~F.,  {Brouillet} N.,   {Gardan} E.,  2010, \mn@doi [\aap]
  {10.1051/0004-6361/200911722}, \href
  {http://adsabs.harvard.edu/abs/2010A%26A...512A..68G} {512, A68}

\bibitem[\protect\citeauthoryear{{Haydon}, {Kruijssen}  \& {Longmore}}{{Haydon}
  et~al.}{2018}]{haydon18}
{Haydon} D.~T.,  {Kruijssen} J.~M.~D.,   {Longmore} S.~N.,  2018, \mnras~to be
  submitted

\bibitem[\protect\citeauthoryear{{Hoopes}, {Walterbos}  \& {Bothun}}{{Hoopes}
  et~al.}{2001}]{Hoopes2001}
{Hoopes} C.~G.,  {Walterbos} R.~A.~M.,   {Bothun} G.~D.,  2001, \mn@doi [\apj]
  {10.1086/322422}, \href {http://adsabs.harvard.edu/abs/2001ApJ...559..878H}
  {559, 878}

\bibitem[\protect\citeauthoryear{{Hygate}, {Kruijssen}, {Walter}, {Chevance},
  {Schruba}, {Longmore}  \& {Haydon}}{{Hygate} et~al.}{2018}]{Hygate18}
{Hygate} A.~P.~S.,  {Kruijssen} J.~M.~D.,  {Walter} F.,  {Chevance} M.,
  {Schruba} A.,  {Longmore} S.~N.,   {Haydon} D.~T.,  2018, to be submitted

\bibitem[\protect\citeauthoryear{{Kapala} et~al.,}{{Kapala}
  et~al.}{2015}]{Kapala2015}
{Kapala} M.~J.,  et~al., 2015, \mn@doi [\apj] {10.1088/0004-637X/798/1/24},
  \href {http://adsabs.harvard.edu/abs/2015ApJ...798...24K} {798, 24}

\bibitem[\protect\citeauthoryear{{Kreckel}, {Blanc}, {Schinnerer}, {Groves},
  {Adamo}, {Hughes}  \& {Meidt}}{{Kreckel} et~al.}{2016}]{Kreckel2016}
{Kreckel} K.,  {Blanc} G.~A.,  {Schinnerer} E.,  {Groves} B.,  {Adamo} A.,
  {Hughes} A.,   {Meidt} S.,  2016, \mn@doi [\apj]
  {10.3847/0004-637X/827/2/103}, \href
  {http://adsabs.harvard.edu/abs/2016ApJ...827..103K} {827, 103}

\bibitem[\protect\citeauthoryear{{Kruijssen} \& {Longmore}}{{Kruijssen} \&
  {Longmore}}{2014}]{KL14}
{Kruijssen} J.~M.~D.,  {Longmore} S.~N.,  2014, \mn@doi [\mnras]
  {10.1093/mnras/stu098}, \href
  {http://adsabs.harvard.edu/abs/2014MNRAS.439.3239K} {439, 3239}

\bibitem[\protect\citeauthoryear{{Kruijssen} et~al.,}{{Kruijssen}
  et~al.}{2018a}]{Kruijssen18b}
{Kruijssen} J.~M.~D.,  et~al., 2018a, submitted

\bibitem[\protect\citeauthoryear{{Kruijssen}, {Schruba}, {Hygate}, {Hu},
  {Haydon}  \& {Longmore}}{{Kruijssen} et~al.}{2018b}]{kruijssen18}
{Kruijssen} J.~M.~D.,  {Schruba} A.,  {Hygate} A.~P.~S.,  {Hu} C.-Y.,  {Haydon}
  D.~T.,   {Longmore} S.~N.,  2018b, \mn@doi [\mnras] {10.1093/mnras/sty1128},
  \href {http://adsabs.harvard.edu/abs/2018MNRAS.479.1866K} {479, 1866}

\bibitem[\protect\citeauthoryear{{Lacerda} et~al.,}{{Lacerda}
  et~al.}{2018}]{Lacerda2018}
{Lacerda} E.~A.~D.,  et~al., 2018, \mn@doi [\mnras] {10.1093/mnras/stx3022},
  \href {http://adsabs.harvard.edu/abs/2018MNRAS.474.3727L} {474, 3727}

\bibitem[\protect\citeauthoryear{{Leroy} et~al.,}{{Leroy}
  et~al.}{2012}]{Leroy2012}
{Leroy} A.~K.,  et~al., 2012, \mn@doi [\aj] {10.1088/0004-6256/144/1/3}, \href
  {http://adsabs.harvard.edu/abs/2012AJ....144....3L} {144, 3}

\bibitem[\protect\citeauthoryear{{Liu}, {Koda}, {Calzetti}, {Fukuhara}  \&
  {Momose}}{{Liu} et~al.}{2011}]{Liu2011}
{Liu} G.,  {Koda} J.,  {Calzetti} D.,  {Fukuhara} M.,   {Momose} R.,  2011,
  \mn@doi [\apj] {10.1088/0004-637X/735/1/63}, \href
  {http://adsabs.harvard.edu/abs/2011ApJ...735...63L} {735, 63}

\bibitem[\protect\citeauthoryear{{Mathis}}{{Mathis}}{1986}]{Mathis1986}
{Mathis} J.~S.,  1986, \mn@doi [\apj] {10.1086/163910}, \href
  {http://adsabs.harvard.edu/abs/1986ApJ...301..423M} {301, 423}

\bibitem[\protect\citeauthoryear{{Monnet}}{{Monnet}}{1971}]{Monnet1971}
{Monnet} G.,  1971, \aap, \href
  {http://adsabs.harvard.edu/abs/1971A%26A....12..379M} {12, 379}

\bibitem[\protect\citeauthoryear{{Oey} et~al.,}{{Oey} et~al.}{2007}]{Oey2007}
{Oey} M.~S.,  et~al., 2007, \mn@doi [\apj] {10.1086/517867}, \href
  {http://adsabs.harvard.edu/abs/2007ApJ...661..801O} {661, 801}

\bibitem[\protect\citeauthoryear{{Pety} et~al.,}{{Pety}
  et~al.}{2013}]{Pety2013}
{Pety} J.,  et~al., 2013, \mn@doi [\apj] {10.1088/0004-637X/779/1/43}, \href
  {http://adsabs.harvard.edu/abs/2013ApJ...779...43P} {779, 43}

\bibitem[\protect\citeauthoryear{{Rand}, {Kulkarni}  \& {Hester}}{{Rand}
  et~al.}{1990}]{Rand1990}
{Rand} R.~J.,  {Kulkarni} S.~R.,   {Hester} J.~J.,  1990, \mn@doi [\apjl]
  {10.1086/185679}, \href {http://adsabs.harvard.edu/abs/1990ApJ...352L...1R}
  {352, L1}

\bibitem[\protect\citeauthoryear{{Reynolds}, {Scherb}  \& {Roesler}}{{Reynolds}
  et~al.}{1973}]{Reynolds1973}
{Reynolds} R.~J.,  {Scherb} F.,   {Roesler} F.~L.,  1973, \mn@doi [\apj]
  {10.1086/152461}, \href {http://adsabs.harvard.edu/abs/1973ApJ...185..869R}
  {185, 869}

\bibitem[\protect\citeauthoryear{{Sembach}, {Howk}, {Ryans}  \&
  {Keenan}}{{Sembach} et~al.}{2000}]{Sembach2000}
{Sembach} K.~R.,  {Howk} J.~C.,  {Ryans} R.~S.~I.,   {Keenan} F.~P.,  2000,
  \mn@doi [\apj] {10.1086/308173}, \href
  {http://adsabs.harvard.edu/abs/2000ApJ...528..310S} {528, 310}

\bibitem[\protect\citeauthoryear{{Seon} \& {Witt}}{{Seon} \&
  {Witt}}{2012}]{Seon2012}
{Seon} K.-I.,  {Witt} A.~N.,  2012, \mn@doi [\apj]
  {10.1088/0004-637X/758/2/109}, \href
  {http://adsabs.harvard.edu/abs/2012ApJ...758..109S} {758, 109}

\bibitem[\protect\citeauthoryear{{Strickland}, {Heckman}, {Colbert}, {Hoopes}
  \& {Weaver}}{{Strickland} et~al.}{2004}]{Strickland2004}
{Strickland} D.~K.,  {Heckman} T.~M.,  {Colbert} E.~J.~M.,  {Hoopes} C.~G.,
  {Weaver} K.~A.,  2004, \mn@doi [\apjs] {10.1086/382214}, \href
  {http://adsabs.harvard.edu/abs/2004ApJS..151..193S} {151, 193}

\bibitem[\protect\citeauthoryear{{Thilker}, {Braun}  \& {Walterbos}}{{Thilker}
  et~al.}{2000}]{Thilker2000}
{Thilker} D.~A.,  {Braun} R.,   {Walterbos} R.~A.~M.,  2000, \mn@doi [\aj]
  {10.1086/316852}, \href {http://adsabs.harvard.edu/abs/2000AJ....120.3070T}
  {120, 3070}

\bibitem[\protect\citeauthoryear{{Thilker}, {Walterbos}, {Braun}  \&
  {Hoopes}}{{Thilker} et~al.}{2002}]{Thilker2002}
{Thilker} D.~A.,  {Walterbos} R.~A.~M.,  {Braun} R.,   {Hoopes} C.~G.,  2002,
  \mn@doi [\aj] {10.1086/344303}, \href
  {http://adsabs.harvard.edu/abs/2002AJ....124.3118T} {124, 3118}

\bibitem[\protect\citeauthoryear{{Thilker} et~al.,}{{Thilker}
  et~al.}{2005}]{Thilker2005}
{Thilker} D.~A.,  et~al., 2005, \mn@doi [\apjl] {10.1086/424816}, \href
  {http://adsabs.harvard.edu/abs/2005ApJ...619L..67T} {619, L67}

\bibitem[\protect\citeauthoryear{{Ward}, {Kruijssen}, {Chevance}, {Hygate},
  {Schruba}  \& {Longmore}}{{Ward} et~al.}{2018}]{Ward18}
{Ward} J.~L.,  {Kruijssen} J.~M.~D.,  {Chevance} M.,  {Hygate} A.~P.~S.,
  {Schruba} A.,   {Longmore} S.~N.,  2018, to be submitted

\bibitem[\protect\citeauthoryear{{Wilson} \& {Walker}}{{Wilson} \&
  {Walker}}{1994}]{WilsonWalker1994}
{Wilson} C.~D.,  {Walker} C.~E.,  1994, \mn@doi [\apj] {10.1086/174556}, \href
  {http://adsabs.harvard.edu/abs/1994ApJ...432..148W} {432, 148}

\bibitem[\protect\citeauthoryear{{Wood}, {Hill}, {Joung}, {Mac Low},
  {Benjamin}, {Haffner}, {Reynolds}  \& {Madsen}}{{Wood}
  et~al.}{2010}]{Wood2010}
{Wood} K.,  {Hill} A.~S.,  {Joung} M.~R.,  {Mac Low} M.-M.,  {Benjamin} R.~A.,
  {Haffner} L.~M.,  {Reynolds} R.~J.,   {Madsen} G.~J.,  2010, \mn@doi [\apj]
  {10.1088/0004-637X/721/2/1397}, \href
  {http://adsabs.harvard.edu/abs/2010ApJ...721.1397W} {721, 1397}

\bibitem[\protect\citeauthoryear{{Zhang} et~al.,}{{Zhang}
  et~al.}{2017}]{Zhang2017}
{Zhang} K.,  et~al., 2017, \mn@doi [\mnras] {10.1093/mnras/stw3308}, \href
  {http://adsabs.harvard.edu/abs/2017MNRAS.466.3217Z} {466, 3217}

\makeatother
\end{thebibliography}




\appendix

\section{Observational Considerations}
\label{confounding_obs_issues}

For the main body of the paper, we have tested diffuse emission filtering in Fourier space on idealised simulated datasets. In this appendix, we consider the impact of two features of astronomical observations on the validity of the method. We firstly consider the impact of astrometric offsets between two input images and then we consider the impact of noise in images.

\subsection{Systematic astrometric offsets}
\label{sec:astrometry}

\begin{figure*}
	\centering
	\subfloat{\includegraphics[width=\cwidth]{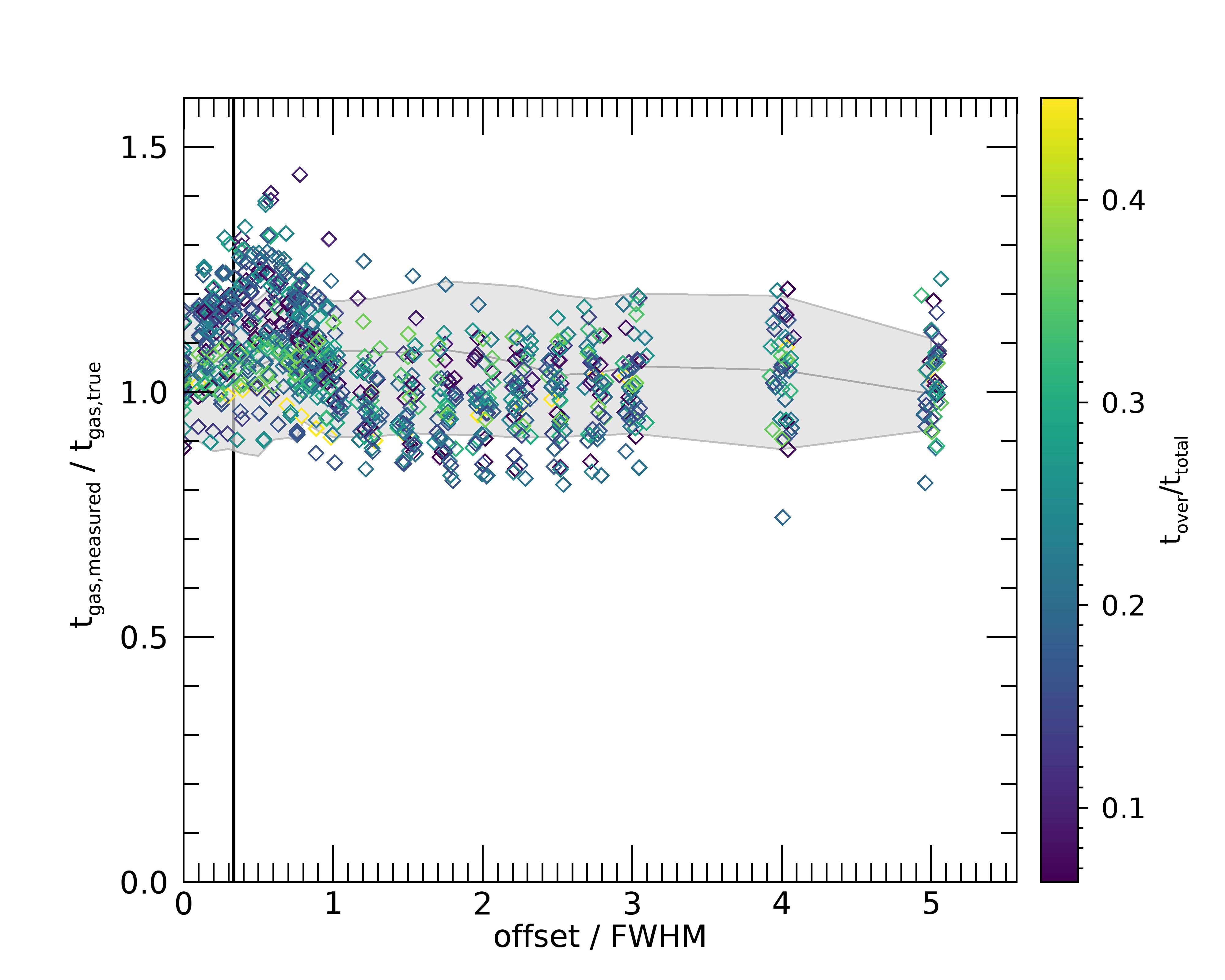}}
	\subfloat{\includegraphics[width=\cwidth]{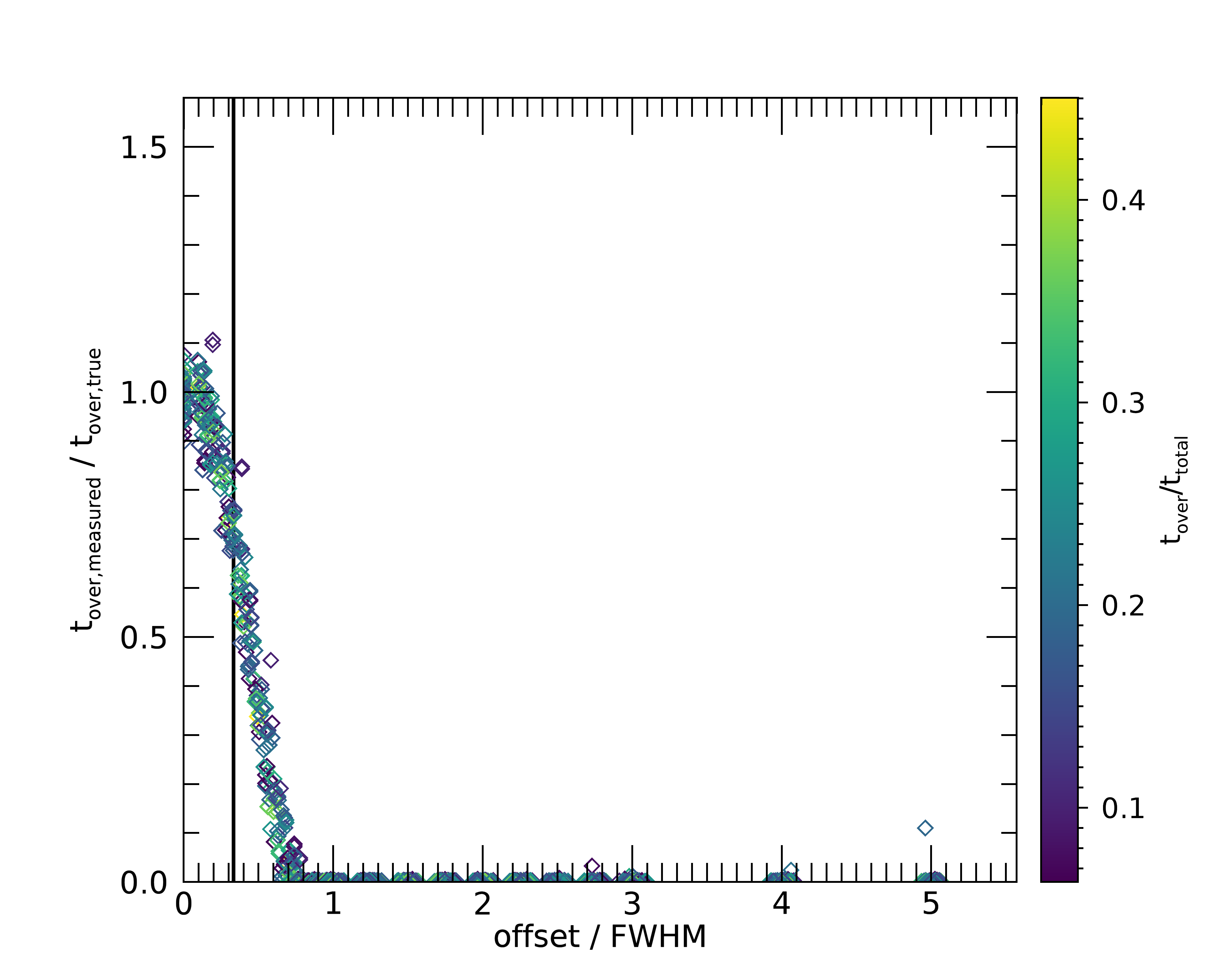}} \\
	\subfloat{\includegraphics[width=\cwidth]{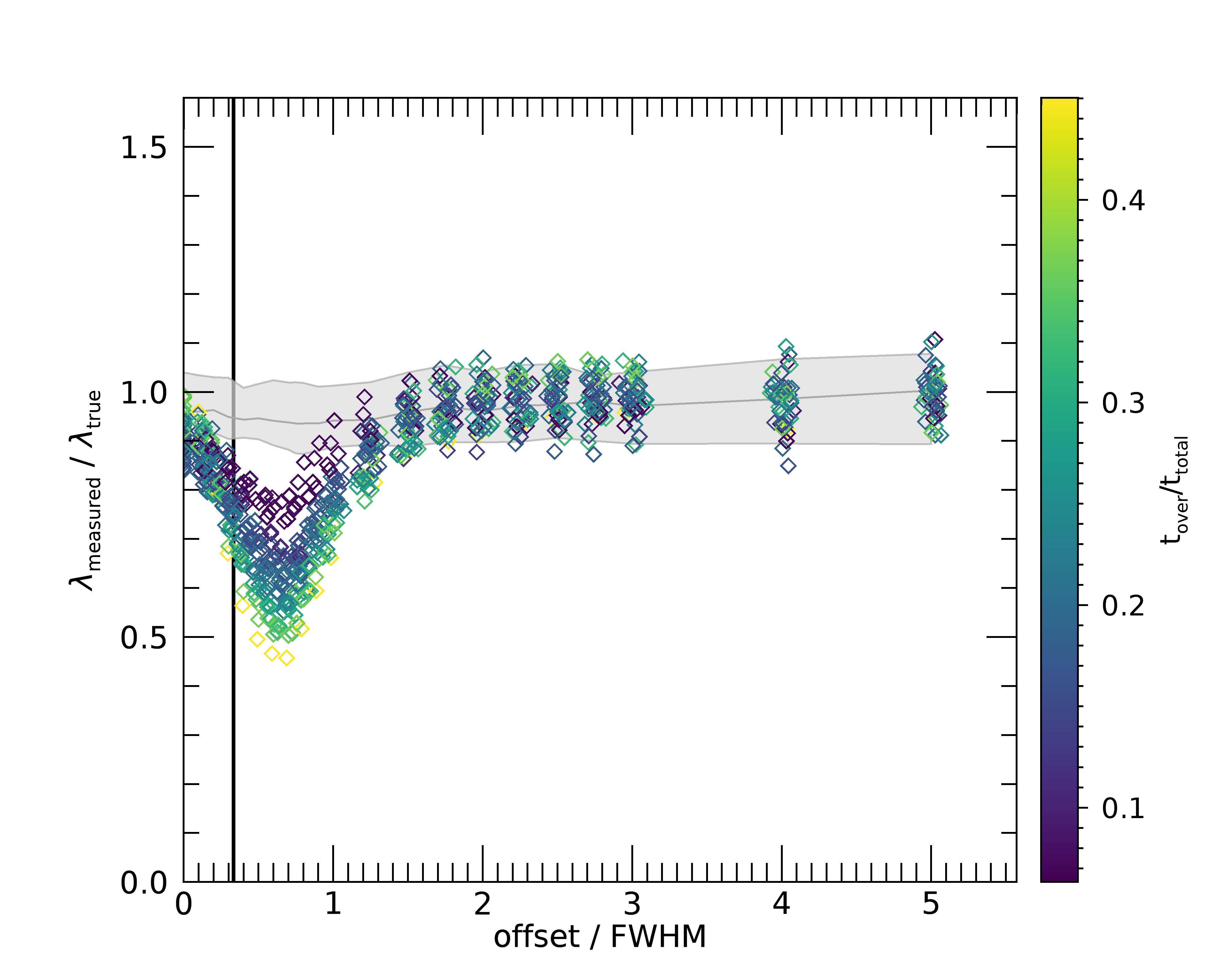}}
	\subfloat{\includegraphics[width=\cwidth]{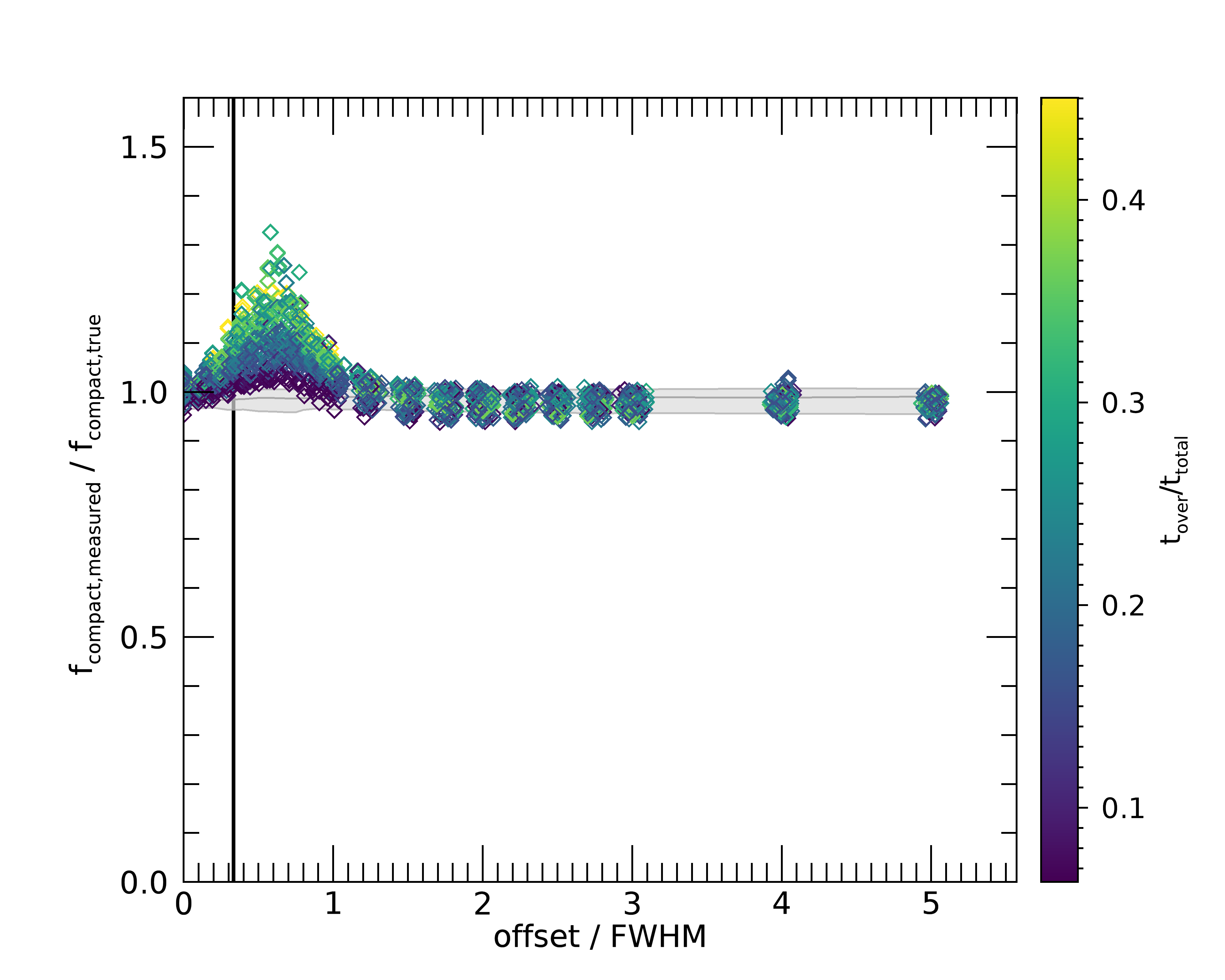}} \\
	\caption{Measured values in units of the true value of  \tgas{} (top left), \tover{} (top right), $\lambda$ (bottom left), and  $\fsignal$ (bottom right) against the astrometric offset introduced between the two input images in each experiment. The points are coloured to show the lifetime of the overlap phase relative to the total duration of evolutionary timeline ($\tover / \tau$). For each panel except the top right (showing \tover{}), a grey shaded area shows the $1\sigma$ region of measurements from experiments performed on control datasets with no regions in the overlap phase (i.e. $\tover=0$). The three dark grey lines show the 86th percentile, median and 14th percentile values from top to bottom. In each panel, the vertical line indicates an astrometric offset of ${\rm FWHM}/3$, below which the impact of the offset on the measured values is small. Measurements from the control datasets are not significantly affected by offsets of a few times the FWHM for any of the quantities, while in the datasets with a time overlap all the quantities are affected. Most notably, for \tover, measured values are scattered around the true value at zero offset. However, as the offset is increased, the measured value decreases sharply until reaching zero at offsets $\simeq 70\%$ of the FWHM of the compact regions. This in turn has an impact on the measured values of the other quantities.}
	\label{fig:offset_figure}
\end{figure*}

One issue that impacts the accuracy of parameters measured by the \code code is that of systematic astrometric offsets between the two tracer images used (i.e. the gas image and the stellar image). In order to test the impact of astrometric offsets, we generate test datasets, as described in Section~\ref{sec:test_images}, and then apply the presented method to them with a range of introduced offsets. We remove experiments with obviously visibly bad fits in the experiments with zero offset from consideration for all values of astrometric offset. This concerns 7 out of 50 experiments  ($\sim 14\%$ the experiments). Figure~\ref{fig:offset_figure} shows the measured values of the gas cloud lifetime, \tgas{}, the overlap phase lifetime, \tover{}, the mean separation length between regions, $\lambda$ and the  compact emission fraction, \fsignal{}, relative to their true values against introduced astrometric offset for these simulated datasets. In addition to the standard dataset, we also generate control datasets containing no regions in the overlap phase (i.e.~$\tover{}=0$). For the control datasets, there is no impact with increasing astrometric offset for these parameters up to astrometric offsets of a few times the FWHM of the compact regions. However, for the datasets containing overlap regions, the astrometric disassociation of the overlap regions affects the measured quantities. The most susceptible of the measured quantities to the effects of astrometric offset is the overlap time-scale, \tover. The gas and stellar `coexistence time-scale', \tover{}, is determined through the statistical spatial correlation between regions from one tracer with regions in the other tracer map. As the size of the astrometric offset increases with respect to the region size, these overlapping regions increasingly decorrelate with their counterparts in the other tracer map and cease to be co-spatial. For this reason, the measured value of \tover~drops sharply with increasing offset, reaching 0\% of the true value at offsets of $\simeq 75\%$ the FWHM of the compact regions.

In \citet{kruijssen18}, it was left undecided over what length scale feedback ejecta need to be displaced for the gas and stars to become decorrelated. Either the region separation length and region radius can be used in conjunction with the overlap time-scale to calculate quantities related to the feedback outflow (see e.g.~their equations 138 and 139). Figure~\ref{fig:offset_figure} demonstrates that the coexistence of regions becomes undetectable for a displacement of the order the region radius, hence the region radius should be used when calculating the feedback outflow-related quantities.

In order to provide meaningful constraints on the value of \tover, high astrometric precision is required. As a minimum, for a good estimate of \tover~(i.e. $t_{\rm over, meausred} \geqslant 75\%~t_{\rm over, true}$) one requires an astrometric offset that is less than 1/3 times the FWHM of the compact regions within the maps (such as molecular gas clouds or \htwo~regions). Uncertainty in the astrometry of a dataset introduces a systematic upwards uncertainty on $t_{\rm over}$ of the magnitude indicated by Figure~\ref{fig:offset_figure}.

For $\lambda$, the decorrelation between overlapping regions leads to the measurement of smaller separation lengths, with the minimum value of $\lambda$ measured at offsets of $\sim 0.7\, {\rm FWHM}$ corresponding to the point at which $\tover = 0$. After this point, the value of $\lambda$ increases, returning back to the true value at offsets of $\gtrsim 1.5\, {\rm FWHM}$. This is due to the fact that initially the regions become separated in both tracer maps, appearing to be nearby regions not in the overlap phase, thus decreasing $\lambda$. At larger offsets they become separated enough that they no longer decrease the value of $\lambda$. The impact in the measured value of $\lambda$ is greater the more significant a fraction of the total evolutionary timeline \tover~is.

The measured compact emission fraction, $\fsignal$, increases due to the decreasing value of $\lambda$, with again larger values of $\tover/\tau$ leading to a greater impact. As each image is filtered in Fourier space individually, a systematic offset between the astrometry of the two images has no impact on the filtering process directly. Instead it impacts the proper alignment of the compact regions during the application of the \code code and the proper measurement of $\lambda$, which sets the size of the filter. The reduced values of $\lambda$ cause an overestimate of the compact emission fraction in the images, due to an overcorrection for the presence of overlapping emission peaks ($q_\eta$, see Section~\ref{sec:signal_loss}).

The measured gas cloud lifetime, \tgas, can also increase as a result of the decreased value of $\lambda$. However, this is dependent on a negative covariance between the two quantities in the affected datasets. Those datasets that have greater measured negative covariance coefficients at zero offsets also have, in general, greater measured negative correlation coefficients and a larger increase in the measured value of \tgas~at offsets of $\sim 0.7 {\rm FWHM}$, where the effect on $\lambda$ is greatest. Thus observational datasets with uncertain astrometry and a larger covariance between $\lambda$ and \tgas~are most likely to be affected.

Overall, precision astrometry is important for the measurement of the region separation length, $\lambda$, crucial for measurement of the overlap time-scale, \tover{}, and  can have a small impact on measurements of the gas cloud lifetime, \tgas, and compact emission fraction, \fsignal. The recommended maximum astrometric uncertainty required for accurate measurements of $\lambda$ and \tover{} is 1/3 the FWHM of the regions of interest, such as molecular gas clouds or \htwo{} regions. This corresponds to the physical region size if they are resolved, or to the PSF size if they are barely resolved (or not resolved at all).

\subsection{Noise}
\label{sec:noise}

For the main results of the paper, we have so far not considered the impact of noise on our results. However, as noise is typically evenly distributed around zero in a well-prepared astronomical image (i.e. there is equal negative flux from noise as there is positive noise), the masking of all negative flux in an image as is done in the method (see Section~\ref{sec:post_processing} for details) will remove the negative noise flux and leave behind the positive noise flux. This may bias measurements made with the \code code, depending on the significance of the noise. As a method for mitigating this effect, we utilise lowpass filtering in Fourier space, predicated on the basis that flux from the noise is concentrated in the very low spatial wavelength (very high frequency)\footnote{We note that this assumption does not hold for interferometric images where each baseline, including those baselines that encode flux from large scale, contributes to the noise.}
part of Fourier space, whereas the flux from the compact regions and the diffuse emission is concentrated in intermediate to large wavelength portions of Fourier space. For this purpose we define a lowpass Gaussian filter to remove noise from the images:

\begin{equation}
	\label{eq:low_Gaussian_filt}
	\Psiuv = \exp \left( - \frac{(\Duv)^{2}}{2 \dcritlow^{2}}     \right).
\end{equation}

\noindent
The lowpass critical frequency, \dcritlow, is defined as:

\begin{equation}
\label{eq:d_crit_low}
\dcritlow  = \frac{\lpix}{\lpix{} n_{\rm pix}} = \frac{1}{n_{\rm pix}},
\end{equation}

\noindent
where $n_{\rm pix}$ is the number of pixels taken as the critical length scale over which to filter and \lpix{} is the pixel length scale.

In order to test the resilience of the method to the presence of noise, we generate test images as described in Section~\ref{sec:test_images}. However, in addition to the basic model of a compact region component $s(m,n)$ and background component $b(m,n)$ (see equation~\ref{eq:basic_image_model}), we add an additional noise component $k(m,n)$:

\begin{equation}
	f(m,n) = s(m,n) + b(m,n) + k(m,n) .
\end{equation}

\noindent
We consider two different models for the noise. Firstly uniform random noise:

\begin{equation}
	k(m,n) = \kappa_{\rm uniform}(m,n)
\end{equation}

\noindent
where the noise at position $(m,n)$ is a uniform random number $\kappa_{\rm uniform}$ such that  $\left\lbrace \nu_{\rm uniform} \in \mathbb{R} \ | \ -\nu_{\rm magnitude} \leqslant \nu_{\rm uniform} \leqslant \nu_{\rm magnitude} \right\rbrace$ and $\nu_{\rm magnitude}$ is a constant that sets the magnitude of the noise, which is determined in relation to the resultant signal to noise ratio  ($S/N$) of the generated image.

The second model we consider is Gaussian noise:

\begin{equation}
k(m,n) = \kappa_{\rm Gauss}(m,n),
\end{equation}

\noindent
where the noise at position $(m,n)$ is a Gaussian random number such that  $\left\lbrace \nu_{\rm Gauss} \in \mathbb{R}\right\rbrace$ selected from a distribution with standard deviation $\sigma(k)$.

\begin{figure*}
	\centering
	\subfloat{\includegraphics[width=\imgenwidth\textwidth]{Figures/Image_gen/gas_basefile.pdf}}
	\hfill
	\subfloat{\includegraphics[width=\imgenwidth\textwidth]{Figures/Image_gen/gas_diffonly.pdf}}
	\hfill
	\subfloat{\includegraphics[width=\imgenwidth\textwidth]{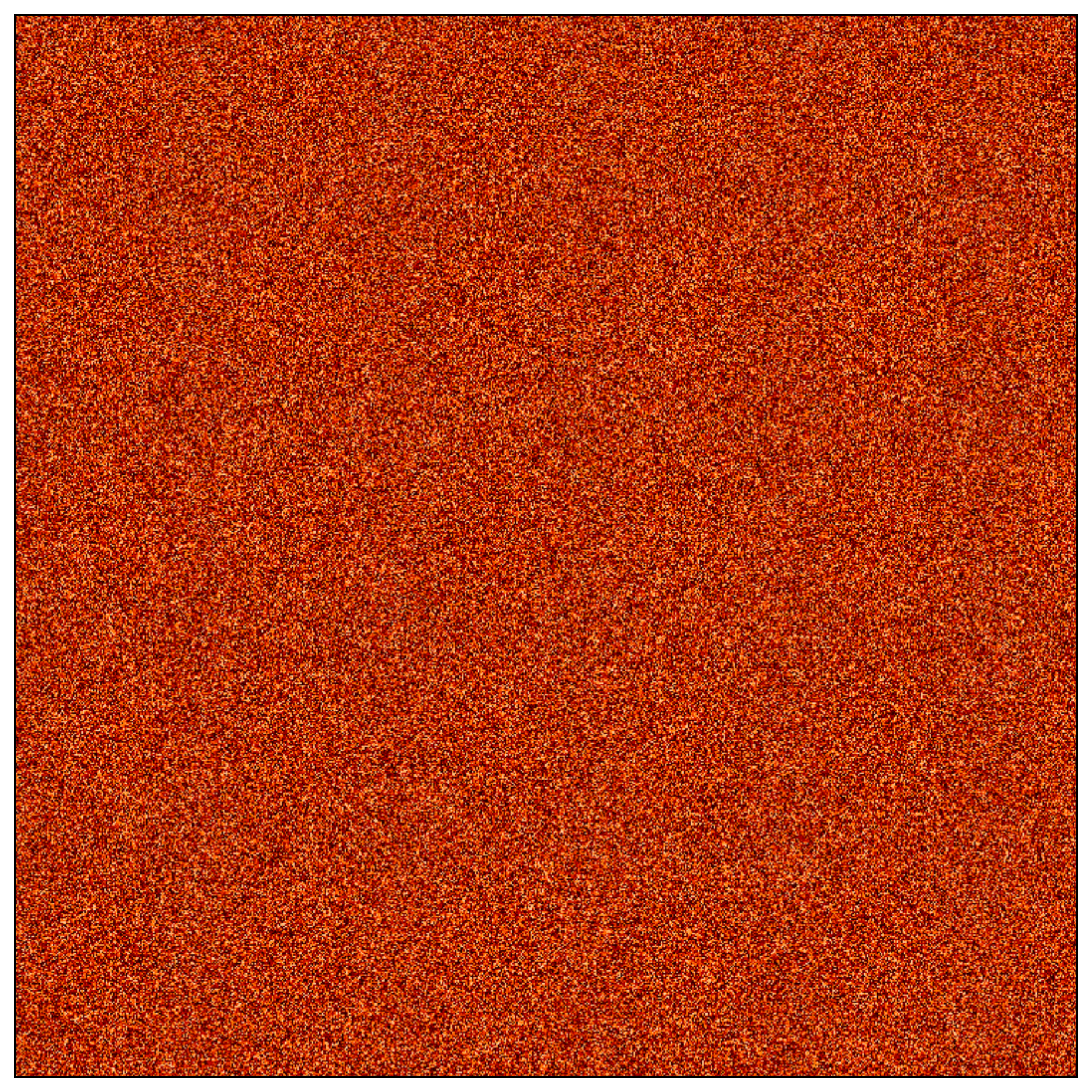}}
	\hfill
	\subfloat{\includegraphics[width=\imgenwidth\textwidth]{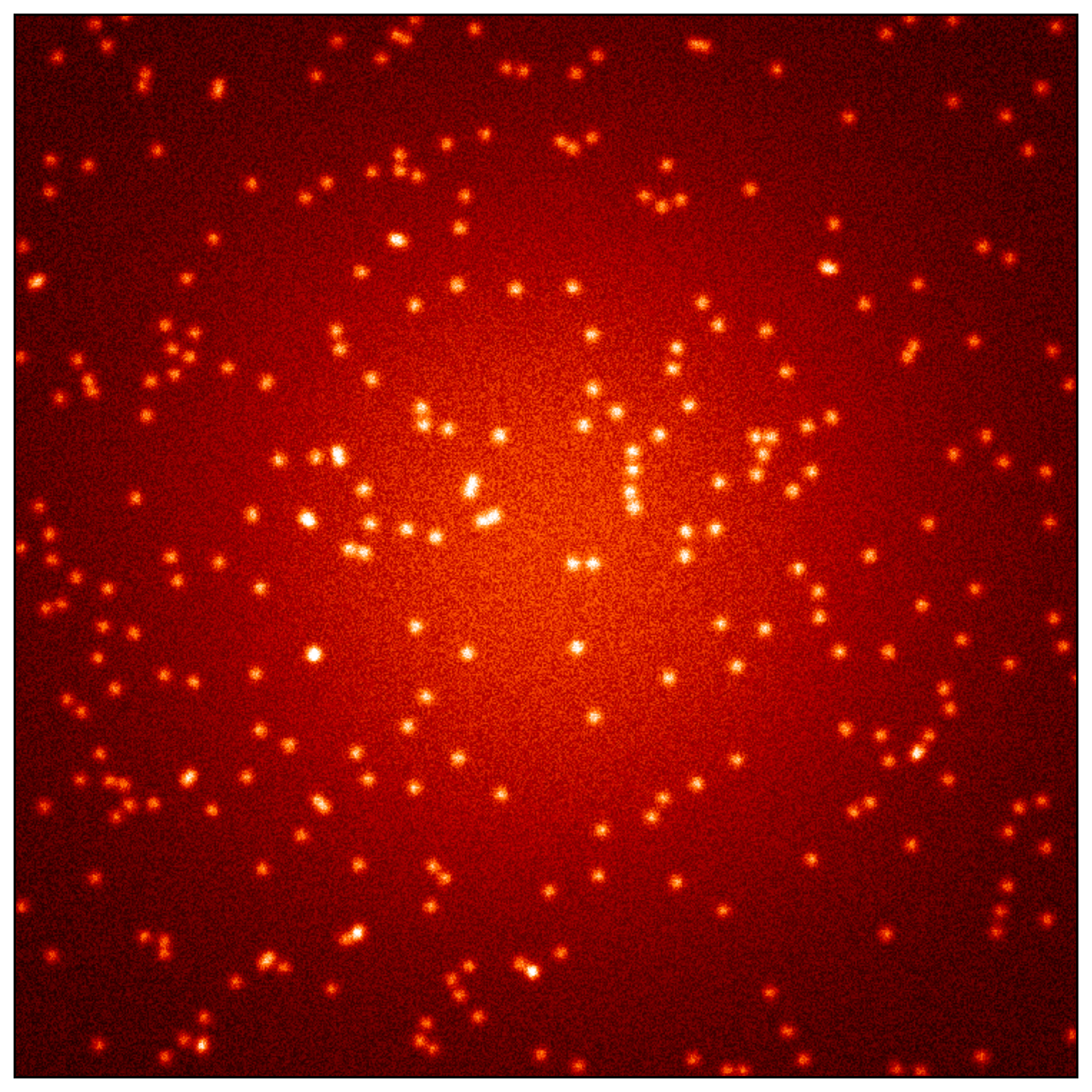}}

	\caption{Example generated image and its components.
		Left panel: the compact component. Note that the placement of Gaussians is entirely random such that they may overlap partially or entirely with other Gaussians, leading to pixels brighter than the peak brightness of a single Gaussian.
		Middle left panel: the background component, here a single large `galaxy scale' Gaussian function.
		Middle right panel: the noise component.
		Tight panel: the final generated image with all three components summed together.}\label{image_componets_with_noise}

\end{figure*}

For both models, we define the $S/N$ as the ratio of the peak brightness of the mean Gaussian function in the compact component, $s_{\rm peak}$, to the standard deviation of the noise component, $\sigma(k)$:

\begin{equation}
	S/N = \frac{s_{\rm peak}}{\sigma(k)}
\end{equation}
\noindent
An example image generated in this fashion is displayed in Figure~\ref{image_componets_with_noise}.

\begin{figure*}
	\centering
	\subfloat{\includegraphics[width=\cwidth]{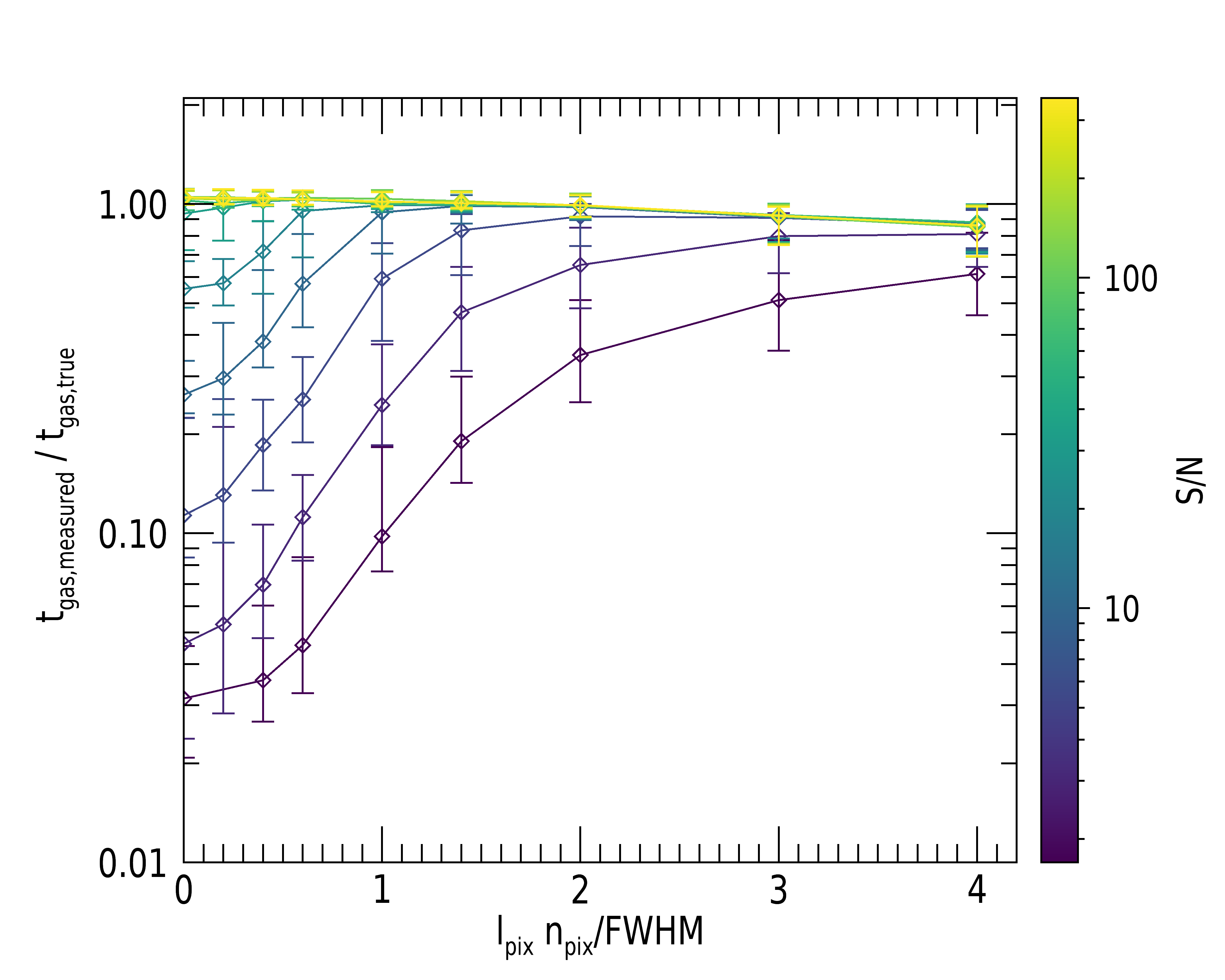}}
	\subfloat{\includegraphics[width=\cwidth]{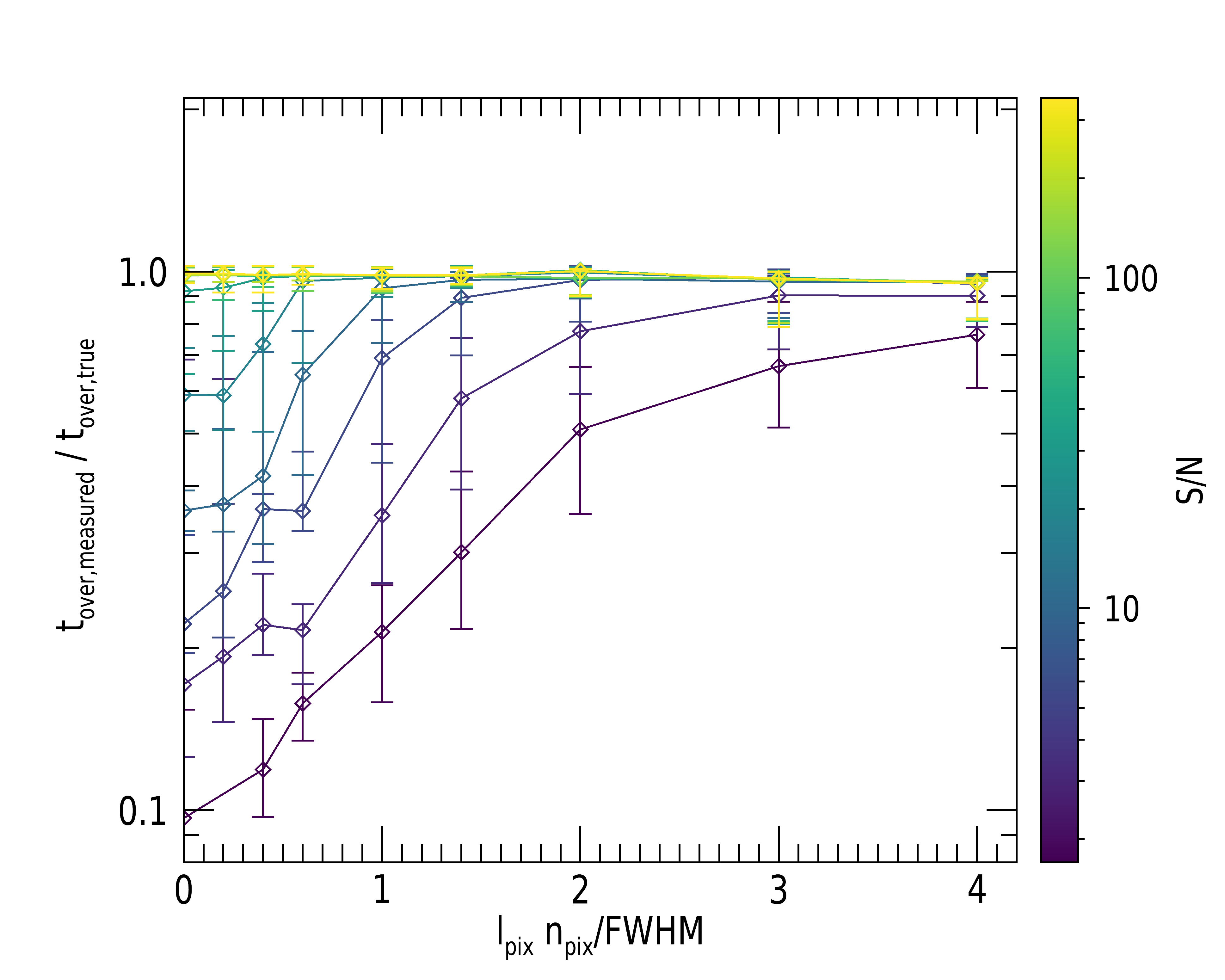}}  \\
	\subfloat{\includegraphics[width=\cwidth]{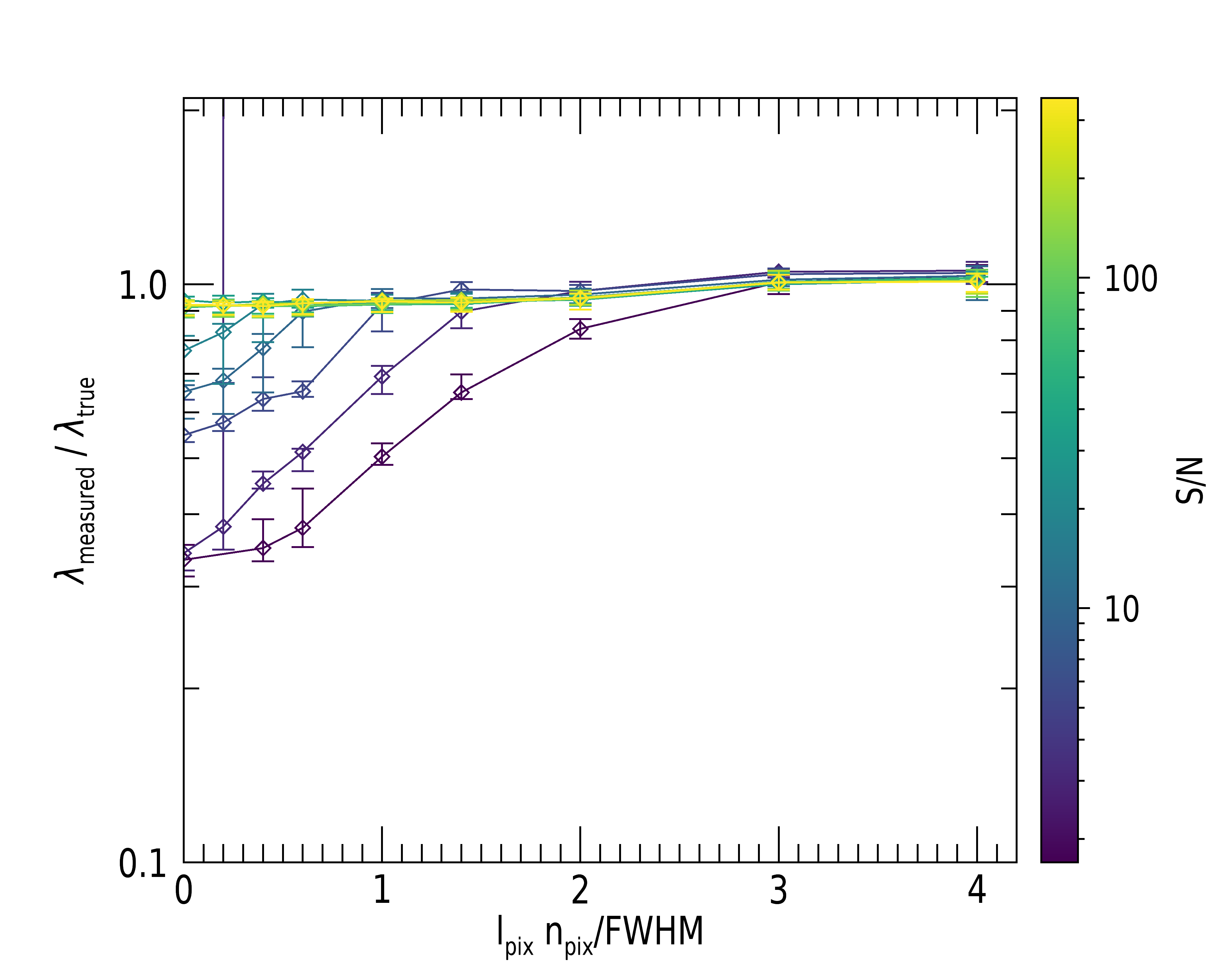}}
	\subfloat{\includegraphics[width=\cwidth]{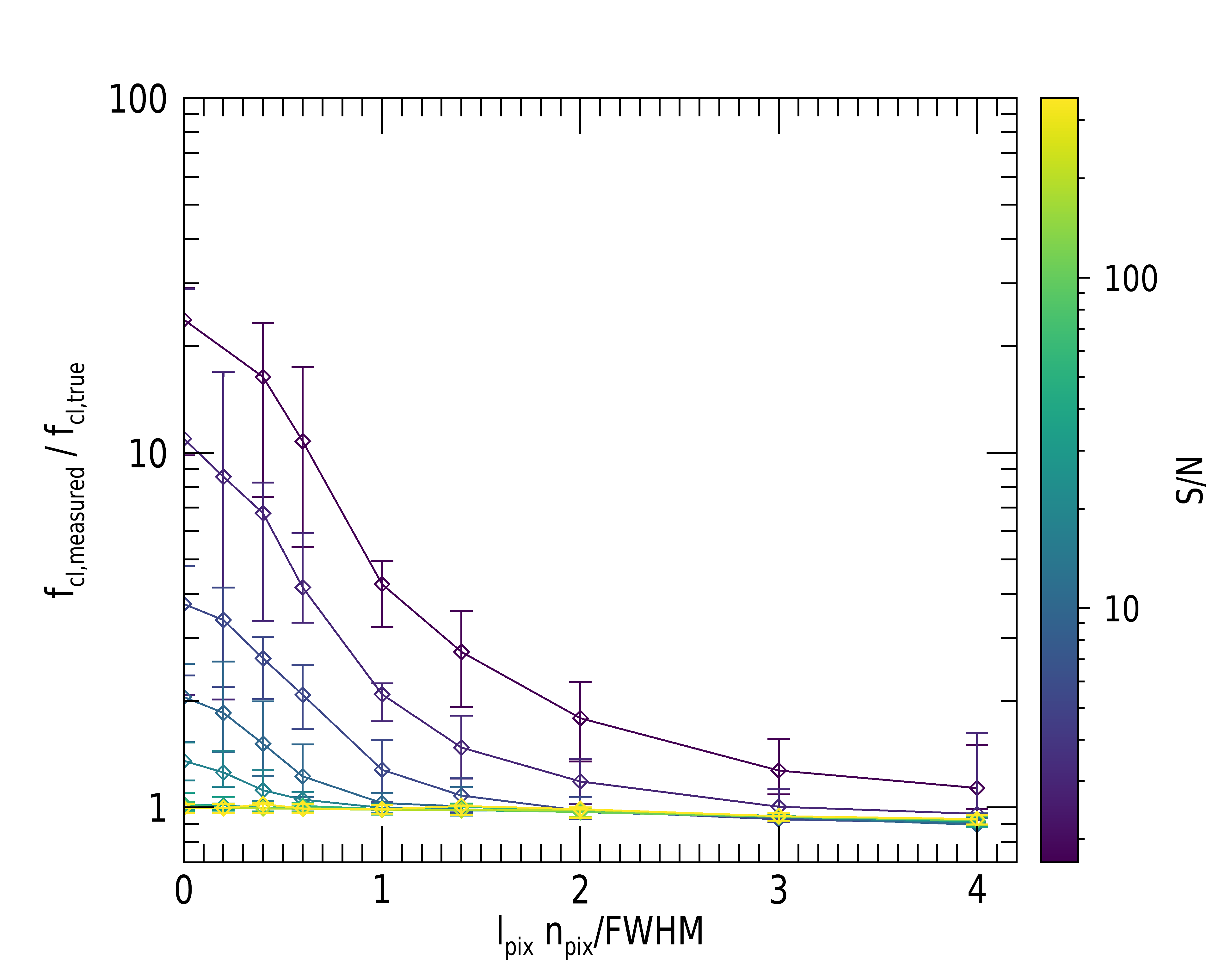}}  \\
	\caption{Measured values in units of the true value of \tgas{} (top left), \tover{}  (top right), $\lambda$ (bottom left), and $\fsignal$ (bottom right) for datasets with added uniform noise as a function of the critical length scale, $\lpix{} n_{\rm pix}$, of the applied lowpass filter in units of the compact region FWHM. The line colours indicate the $S/N$ of the map, with $10$ lines placed equally in logarithmic space in the range $\log_{10}{S/N}=0.230$--$2.544$. For each measured quantity, the presence of noise in the stellar map biases the result, with the magnitude of this effect decreasing with the application of lowpass filters with larger critical length scales. However, at larger critical length scales ($\lpix{} n_{\rm pix}/{\rm FWHM}>3$), the lowpass filtering introduces a small bias into the measured quantities.}
	\label{fig:noise_compare_figure}
\end{figure*}

\begin{figure*}
	\centering
	\subfloat{\includegraphics[width=\cwidth]{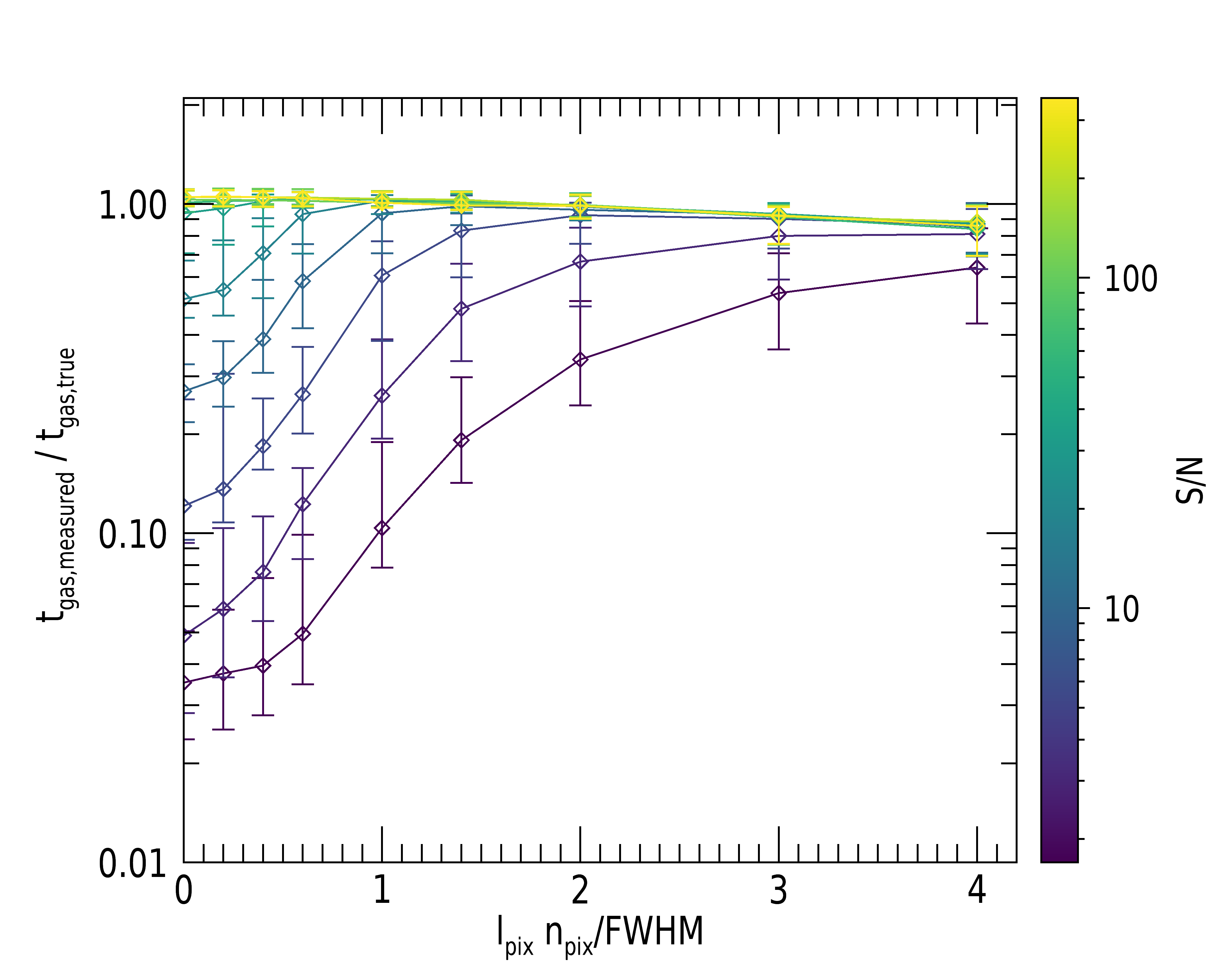}}
	\subfloat{\includegraphics[width=\cwidth]{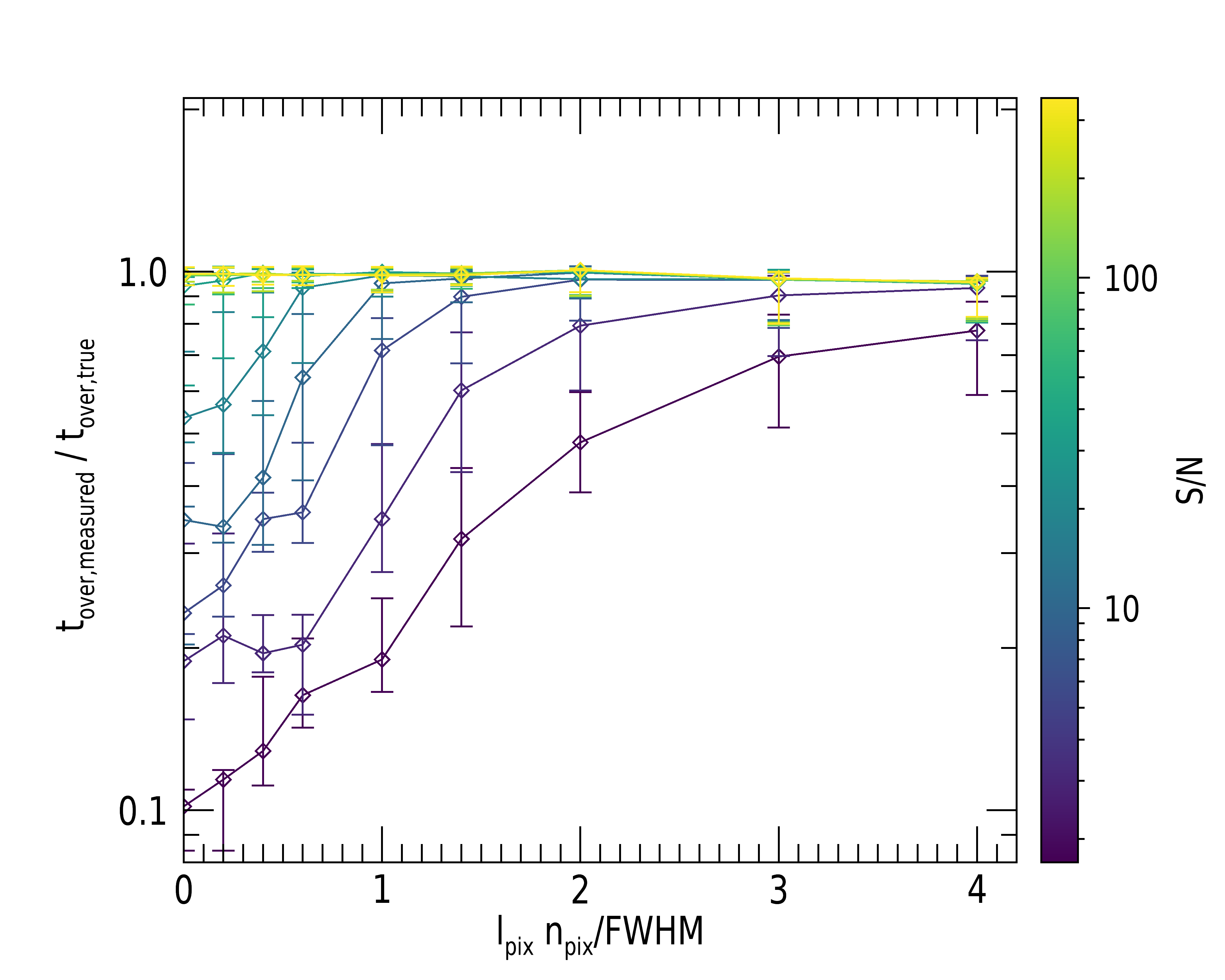}}  \\
	\subfloat{\includegraphics[width=\cwidth]{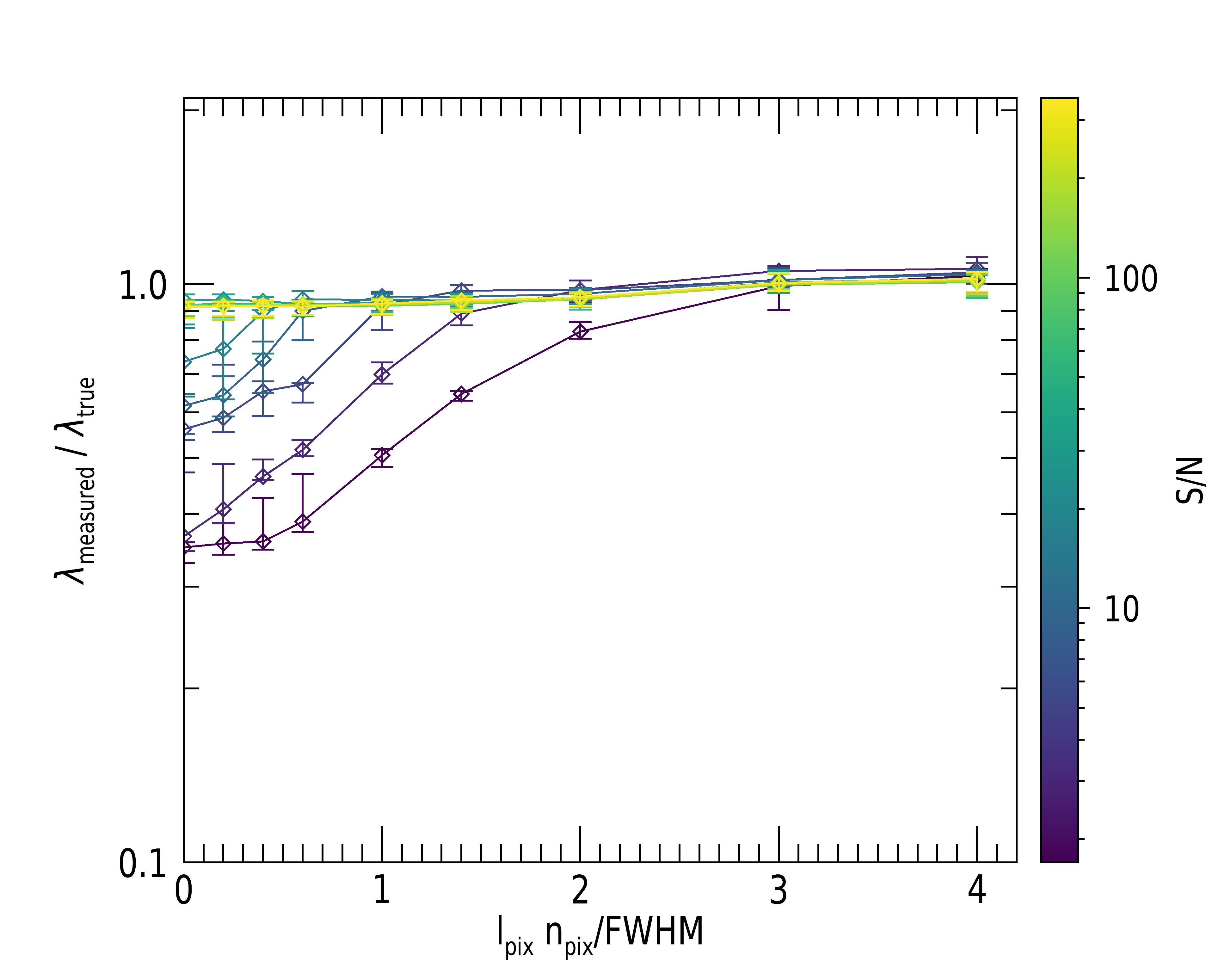}}
	\subfloat{\includegraphics[width=\cwidth]{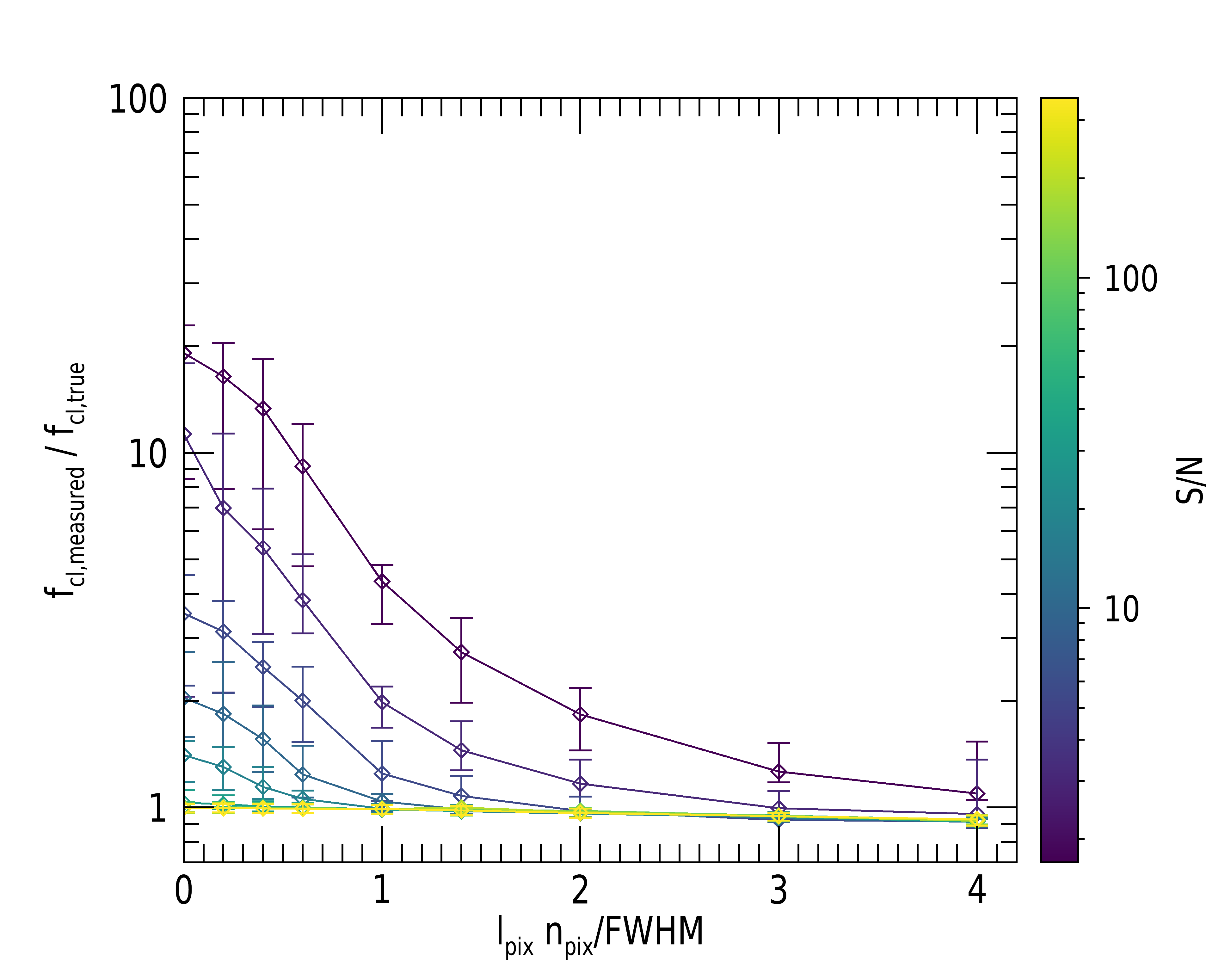}}  \\
	\caption{Measured values in units of the true value of \tgas{} (top left), \tover{}  (top right), $\lambda$ (bottom left), and $\fsignal$ (bottom right) for datasets with added Gaussian noise as a function of the critical length scale, $\lpix{} n_{\rm pix}$, of the applied lowpass filter in units of the compact region FWHM. The line colours indicate the $S/N$ of the map, with $10$ lines placed equally in logarithmic space in the range $\log_{10}{S/N}=0.230$--$2.544$. For each measured quantity, the presence of noise in the stellar map biases the result, with the magnitude of this effect decreasing with the application of lowpass filters with larger critical length scales. However, at larger critical length scales ($\lpix{} n_{\rm pix}/{\rm FWHM}>3$), the lowpass filtering introduces a small bias into the measured quantities.}
	\label{fig:gauss_noise_compare_figure}
\end{figure*}

We create simulated datasets where we add noise to the stellar map only, for simplicity of interpretation. We simulate ten different levels of noise spaced equally in logarithmic space in the range $\log_{10}{S/N}=0.230$--$2.544$. Before we apply the diffuse filtering method presented in this paper, we apply lowpass filters with a critical length scale from $0{-}4$ times the FWHM of the Gaussian regions in the image, with a critical length scale of zero resulting in no filtering. The results of these experiments are summarised in Figure~\ref{fig:noise_compare_figure}, for the uniform noise model and Figure~\ref{fig:gauss_noise_compare_figure}, for the Gaussian noise model. In the regime of high signal-to-noise ($S/N \geqslant 20$), there is very little impact on the measured parameters without any lowpass filtering. However, as the noise in the datasets increases in significance, the added noise causes two main effects. Firstly, spurious noise peaks may be identified as star formation peaks. Secondly, as negative pixels are masked, as described in Section~\ref{sec:post_processing}, the remaining positive noise in the map results in a similar effect as when a diffuse emission reservoir is present in the map. This excess flux mainly affects galactic scales, over which the mean flux density is lower than in emission peaks.

These effects lead to systematically incorrect measurements of \tgas~and \tover~at intermediate and low signal to noise. The measured value of the compact emission fraction in the stellar maps, \fcl, is also increased due to the positive noise flux remaining in the map, even after filtering the diffuse emission, because the noise resides at high frequencies in Fourier space. This makes the compact component appear more significant than in actuality. As we have not introduced noise into the gas maps, the only effect on the measured compact gas fraction is as a result of an incorrectly measured $\lambda$, and we therefore do not display these measurements. The measured value of $\lambda$, is primarily affected by the introduction of spurious noise peaks. This effect is mitigated at intermediate signal to noise by the fact that the \code code selects only peaks above some threshold of the noise level, allowing for the accurate recovery of $\lambda$. For lower signal to noise, where it is more likely that a noise peak could be of comparable brightness to the compact regions themselves and thus pass through this peak filtering process, the measured value of $\lambda$ systematically decreases.

Lowpass filtering reduces the significance of the noise in the images and thus reduces the impact on the measured parameters.  We see in Figures~\ref{fig:noise_compare_figure}~and~\ref{fig:gauss_noise_compare_figure} that a lowpass filter with critical length scale twice that of the FWHM of the compact regions\footnote{We note that applying a lowpass filter to the images suppresses the high-frequency information in the compact regions to a certain extent. This has the effect of `blurring' the map, leading the signal regions to appear larger, with a $\sim 10 \%$ increase in the measured value of \fwhm{} in comparison to the true value for a filter with a critical length scale twice that of the FWHM of the signal regions} allows accurate measurements of \tgas, \tover~and \fsignal~in the intermediate-to-high noise regime. We thus recommend that applications of the method to noisy data employ a lowpass filter prior to the application of the diffuse filtering method, in order to mitigate the impact of noise emission. However, we note that application of a lowpass filter will introduce bias into the measurements of \tgas{}, \tover{}, $\lambda$ and \fsignal{}. With a critical length scale of twice the FWHM of the signal regions or less this effect is small ($< 10\%$ for all quantities) increasing to a $\sim 10 \%$ systematic uncertainty at four times the FWHM ($< 40 \%$ for all quantities). Alternatively, another approach may be considered if the noise flux is evenly distributed in negative and positive flux, such as applying a signal-to-noise threshold after highpass filtering.

\bsp	
\label{lastpage}
\end{document}